\documentclass{aa} 

\usepackage{preamble}  

\begin{document} 

    \title{High-z gamma-ray burst detection by \textit{SVOM}/ECLAIRs: Impact of instrumental biases on the bursts' measured properties}

    \titlerunning{Impact of instrumental biases on the \textit{SVOM}/ECLAIRs high-z GRBs measured properties}

   \author{M. Llamas Lanza \inst{1} \and O. Godet \inst{1} \and B. Arcier \inst{1} \and M. Yassine \inst{1} \and JL. Atteia \inst{1} \and L. Bouchet \inst{1}}

   \institute{IRAP, Université de Toulouse, CNRS, CNES, UPS, (Toulouse), France\\
              \email{Miguel.Llamas.Lanza@gmail.com}
             }

   \date{February 23, 2024}

 
\abstract
{\glspl{grb} can be detected at cosmological distances and therefore can be used to study the contents and phases of the early Universe. 
The 4 -- 150 keV wide-field trigger camera ECLAIRs to fly on-board the \gls{svom} mission dedicated to study high-energy transient sky in synergy with multi-messenger follow-up instruments, is adapted to detect high-z GRBs.}
{Investigating the detection capabilities of ECLAIRs for high redshift \glspl{grb} and estimating the impacts of instrumental biases in reconstructing some of the source measured properties, focusing on GRB duration biases as a function of redshift.}
{We simulated realistic detection scenarios for a sample of 162 already observed GRBs with known redshift values as they would have been seen by ECLAIRs. We simulated them at redshift values equal and higher than their measured value. Then, we assessed whether they would be detected with a trigger algorithm resembling that on-board of ECLAIRs, and derived some quantities such as $\mathrm{T_{90}}$ for those that would have been detected.}
{We find that ECLAIRs would be capable of detecting GRBs up to very high redshift values (e.g. 20 GRBs of our sample are detectable within more than 0.4 of the ECLAIRs Field of View for $\mathrm{z_{sim} > 12}$). 
The ECLAIRs low-energy threshold of 4\,keV, contributes to this great detection capability, as it may enhance it at high redshift ($z > 10$) by over 10\% compared with a 15 keV low-energy threshold.
We have also shown that the detection of GRBs at high-z values may imprint tip-of-the-iceberg biases on the GRB duration measurements, which can affect the reconstruction of other source properties. 
}
  {}


   \maketitle

\glsresetall
\glsunset{fcfov}
\glsunset{pcfov}
\section{Introduction}
\label{sec:introduction}

\glspl{grb} are powerful transient events emitting large amounts of energy in X-rays and $\gamma$-rays \citep{Klebesadel1973_def_GRBs, zhang2007gamma, vedrenne_and_atteia_2009gamma, zhang2018physics}. They are characterised by a short, highly variable and intense phase called prompt emission, followed by a multi-wavelength long-lasting afterglow emission characterised by a rapid flux decay \citep{Costa_1997_Xray_afterglow, vanParadijs1997_Optical_afterglow, zhang_2006_afterglow_LC, Nousek_2006, laskar2023radio}.
They are thought to signal the catastrophic formation of a compact object, for example a black hole (BH) or a neutron star (NS), and the launch of powerful ultra-relativistic jets \citep[e.g.][]{piran2005physics, kumar2015physics} following either the merger of a binary NS or a binary BH-NS \citep{Eichler1989, tanvir2013kilonova, abbott2017ligo} or the core collapse of some massive stars, with masses $M \gtrsim 30 M_\odot$ \citep{woosley1993gamma, galama1999possible, Hjorth_2003, Woosley_2006, melandri2019grb}. 
These two different physical events are typically associated with short and long GRBs, respectively. 
Nevertheless, the overlap in the duration of these two populations defies their classification based solely on this parameter \citep{Bromberg_2013}. 
A new class has been suggested \citep[e.g.][]{gehrels2006new, gottlieb2023unified} for long-lasting GRBs produced by a binary merger \citep[see][]{valle2006enigmatic, rastinejad2022kilonova, levan2023grb, sun2023magnetar}. 
Moreover, the threshold used to differentiate the two populations (typically $\sim$2s) is contingent on the specific instrument, given that duration measurements are influenced by the instrument's sensitivity and operational energy range \citep{Bromberg_2013}.

Despite years of theoretical and observational effort aimed at unravelling the physical origin and mechanisms responsible for the prompt emission, the details are somewhat uncertain. 
It is believed that part of the gravitational energy released rapidly in a compact region during the catastrophic event, is converted into the production of a relativistic jet.
While some models rely on a thermally driven explosion as the mechanism behind the jet formation and acceleration, others prescribe the production of very strong magnetic fields which subsequently accelerate the outflow to relativistic speeds.
Other models suggest the production of part of the prompt emission below the photosphere.
It is likely that a wide range of combinations of the proposed models could be responsible for the energy transfer that generates the jet. 
Part of the jet kinetic energy is dissipated (e.g. by internal collisions) producing the high-energy radiation observed in the prompt emission. For more comprehensive explanations and the corresponding references pertaining the prompt emission physics, see e.g. the reviews by \citealt{pe2015physics, dai2017theory, zhang2018physics}.
In contrast, the underlying physics of the afterglow emission is better understood. This emission is likely a consequence of the deceleration of the ejecta and deposition of the remaining jet kinetic energy as it interacts with the surrounding medium \citep{meszaros1997optical}.

Due to their very high isotropic luminosities reaching up to $\mathrm{\sim10^{55}\,erg\,s^{-1}}$ \citep{Atteia_2017}, GRBs associated with the core collapse of massive stars can be observed at cosmological distances that make them among the most distant objects known in the Universe.
The current record holders for the farthest known GRBs are GRB 090423 with a spectroscopic redshift of $z \sim 8.2$ \citep{Tanvir_2009, salvaterra2009grb} and GRB 090429B with a photometric redshift estimation of $z \sim 9.4$ \citep{Cucchiara_2011}, corresponding respectively to $\sim$0.630 and $\sim$0.520\,Gyr after the Big Bang.
In addition to investigating the evolution of GRBs and their progenitors with redshift, the study of high-z GRBs provides a unique opportunity to investigate the contents of the early Universe. This includes the chemical content, dynamics, and interstellar medium of the host galaxy \citep{sparre2014metallicity, Saccardi_2023_ISM}, the intergalactic medium around the host \citep{tanvir2019TSO_early_universe}, and the distance of absorbing material to the GRB and existence of other absorption systems along the line of sight \citep[see e.g.][]{prochaska2008survey, vergani2009statistics, campana2012x}.
High-z GRBs could also be used to probe the end of the re-ionisation phase of the Universe by measuring the fraction of neutral hydrogen gas of the intergalactic media at their location \citep[see e.g.][]{totani2006reionization_implications_050904, lidz2021future}.
Likewise, GRBs associated with the core collapse of massive stars could serve as tracers of the star formation history, thus offering a complementary approach to estimating the star formation
rate (SFR) \citep{Greiner_2015}. This is of particular interest at very high-z (i.e. z$\gtrsim$6), since the SFR tracks the formation of the first galaxies, whose detection is challenging with traditional techniques \citep{tanvir2019TSO_early_universe}. 
It is worth noting however that the \textit{James Webb} Space Telescope \citep{gardner2006james}, with its remarkable infrared sensitivity, may overcome some of these challenges \citep{macpherson2013potential}.
Understanding the formation of the first galaxies may also shed light onto the formation of the earliest cosmic structures driven by dark matter \citep[see e.g.][]{yoshida2003early}.
Pushing the redshift limit further ($\mathrm{z\sim20-30}$) may enable the identification of GRBs arising from the first generation of stars, commonly known as population III stars \citep{bromm2006high, toma2016_GRB_pop3}.

However, the current sample of known high-z GRBs is rather small. This may be due to several factors, including 1) selection effects by the current high-energy instrumentation, for example only the bright tail of the total high-z GRB population is detected \citep[e.g.][]{Littlejohns_2013}, and some high-z GRBs may not be detected if their emission peaks below the energy ranges covered by the current instrumentation in the observer's frame; 2) the proportion of high-z GRBs in the current sample may be higher than those that were identified as high-z GRB since measuring their redshift is challenging. A spectroscopic redshift measurement requires an accurate and fast arcminute localisation and is limited by the sensitivity to detect the afterglow; 3) 
the population of GRB progenitors at earlier epochs (i.e. high-z) could be different to that at later epochs \citep[see e.g.][]{palmerio2021constraining}.

The upcoming French-Chinese mission \gls{svom} \citep{wei2016deep, atteia2022svom} is scheduled to start monitoring the high-energy transient sky from June 2024, with a specific emphasis on GRBs. 
It will operate in synergy with multi-messenger facilities including the ground-based gravitational wave interferometers, the Vera Rubin Observatory \citep{ivezic2019lsst}, and neutrino detectors.
One of the primary science drivers of the mission is to detect GRBs at high redshift \citep{godet2009monte}.
The wide-field high-energy camera ECLAIRs \citep{Godet_2014}, thanks to its low-energy threshold of 4\,keV, is optimised for this purpose and for the detection of peculiar populations of high-energy transients such as X-ray flashes, X-ray-rich GRBs, and low-luminosity transients in the local universe \citep[see][]{Arcier_2020}. 
In addition, the fast and accurate (down to arcsecond) source localisation by the onboard narrow-field instruments in both X-rays and optical will ease the identification of the afterglow, and hence the redshift measurement.
It is worth noting that the high sensitivity of the \gls{vt} \citep{fan2020visible} on board, will enable the detection of GRB afterglows up to redshift $\sim$6.5.
Complementing these capabilities, two \glspl{gft} will contribute to the follow up and identification of the afterglow.
The near-infrared capabilities of the CAGIRE camera \citep{nouvel2023cagire} on the French-Mexican \gls{gft} \citep[Colibri,][]{basa2022colibri, corre2018end} will play an important role in the afterglow identification.

The \gls{svom} observing strategy with a near anti-solar pointing will enhance the follow-up capabilities by large ground-based facilities. All of this will increase the likelihood to measure the GRB redshift.
As a result, it is expected that a substantial fraction of \gls{svom} GRBs ($\sim 2/3$) will have their redshift measured.
The sample of high-z GRBs is therefore expected to augment upon \gls{svom} launch \citep{wei2016deep, atteia2022svom}. 

As for any study of a source population, it is important to estimate how instrumental selection effects impact the source reconstructed properties. This is necessary to retrieve the intrinsic properties of a source population.
Such effects may play a significant role in explaining the observed variations among groups of GRBs, allowing us to differentiate between intrinsic features of the sources and instrumental biases.
While achieving this distinction is challenging, evaluating these instrumental effects may provide insights into the reliability and accuracy of the measurements, thereby enhancing our comprehension of the genuine nature of GRBs and their properties.
\citet{Littlejohns_2013} and \citet{moss2021_tip_iceberg} have studied selection effects on the GRB prompt emission duration measurements using the data from the \gls{bat} \citep{Barthelmy_2005} on board the \textit{Swift Neil Gehrels Observatory} \citep{Gehrels_2004}.
These selection effects could be due to several factors, such as the incidence angle of the photons on the detector plane, the background level and the number of working detectors of the instrument.
It is also likely that these selection effects will be more or less prevalent depending on the intrinsic properties of a class of GRBs. The redshift of the source is an important factor to consider. 

The motivation of this study is two-fold: to investigate the detection capabilities of ECLAIRs for high-redshift GRBs and to estimate the impacts of instrumental biases in reconstructing some of the source measured properties. In this paper we focus on the impacts on the burst duration as a function of redshift.
In Section \ref{sec:SVOMECLAIRs} we present the \gls{svom} mission and the ECLAIRs instrument. 
To perform this study, we first built a sample of GRBs with a redshift measurement (0.078--9.4) and retrieved their properties (see Section \ref{sec:Sample}). From the temporal and spectral information of the bursts, we simulated them as they would have been seen by ECLAIRs at their measured redshift and at higher redshifts (see Section \ref{sec:Method}).
In Section \ref{sec:results} we present the detection performance of ECLAIRs and the evolution of the computed duration of the bursts as a function of redshift.
To conclude, Section \ref{sec:Discussion} is devoted to the discussion.

\section{The SVOM mission and the ECLAIRs instrument}
\label{sec:SVOMECLAIRs}
\begin{figure}[htp]
\centering
\subfloat[Outline of the \gls{svom} mission, including the space and ground instruments.]{
  \includegraphics[width = \hsize]{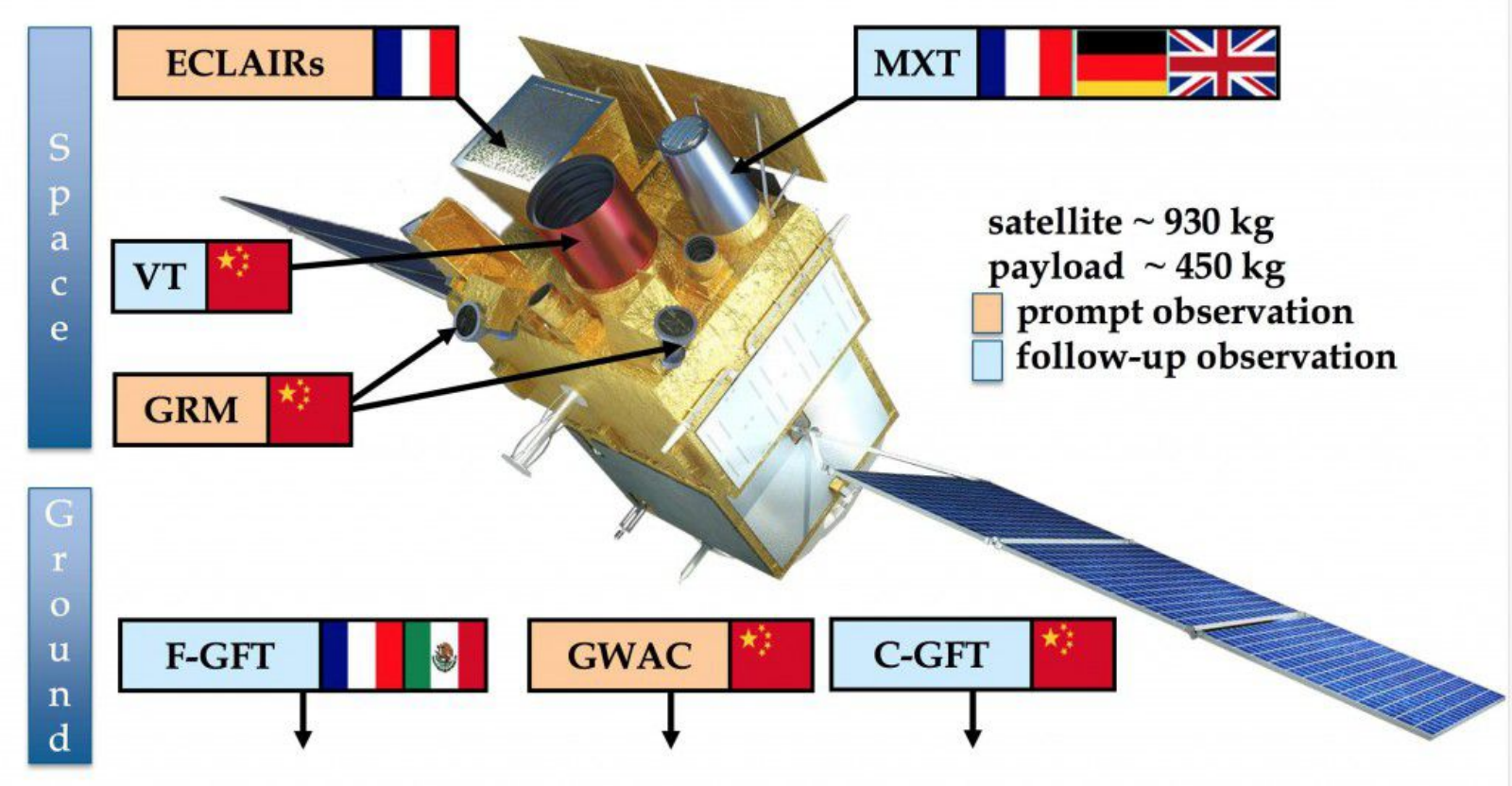}
}

\subfloat[Time and energy coverage of the instruments in the \gls{svom} mission.]{%
  \includegraphics[width = \hsize]{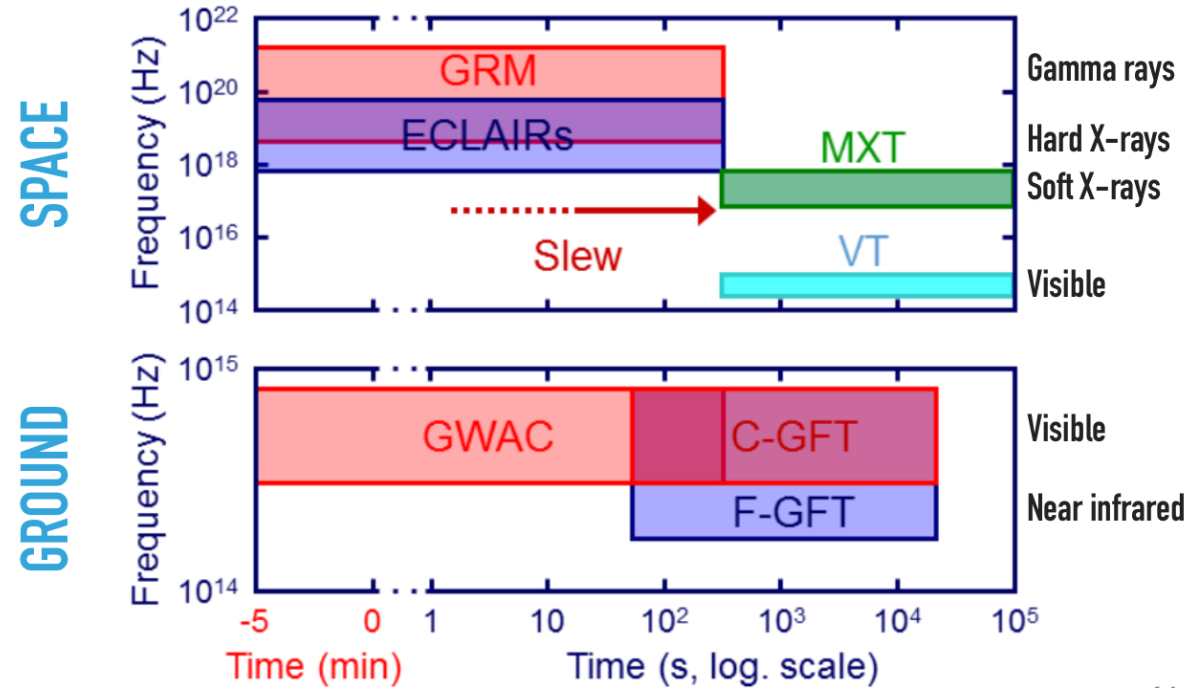}
}
\caption{The \gls{svom} mission}
\label{fig:svom_outline}
\end{figure}

The \gls{svom} satellite will be operated on a 96\,min low-Earth orbit with an inclination of 30\,\degree\, and an altitude of $\sim$625\,km.
Its onboard scientific payload comprises two wide-field and two narrow-field instruments, operating together with a well-defined ground segment (see Figure \ref{fig:svom_outline}).
The two wide-field instruments are the 4--150\,keV coded-mask camera ECLAIRs \citep{Godet_2014}
and the 15\,keV--5\,MeV \gls{grm} \citep{Dong_2009_GRM, wen2021calibrationGRM}.
The two narrow-field instruments are the 0.2--10\,keV \gls{mxt} \citep{Gotz16} and the \gls{vt} \citep{fan2020visible} observing simultaneously in the R and B bands.
On the ground \citep{chaoul2018svom}, two sets of Ground Wide Angle Cameras (GWACs) \citep{han2021automatic} located in two sites in China will enable the detection of the GRB prompt optical emission or any bright early afterglow emission in the optical.
Likewise, the two one-metre \glspl{gft} will perform follow-up observations of \gls{svom} transients and non-\textit{svom} targets; the \glspl{gft} are respectively located in China (Chinese \gls{gft}) and Mexico (French-Mexican \gls{gft}, Colibri), enhancing the possibility of immediate follow-up of \gls{svom} alerts.
The wide-field near-infrared imager CAGIRE on Colibri will enable the follow-up of GRBs up to $\mathrm{z\sim6.5}$ \citep{nouvel2023cagire}.

The attitude of \gls{svom} will be pre-defined, so that for most of the year the optical axis of the instruments will be pointed at an offset angle of 45\textdegree\;from the anti-solar direction. Likewise, it will include periods of avoidance of Sco X-1 and the Galactic plane within the ECLAIRs field of view (FoV).\glsunset{fov}
The pointing law has been optimised to ensure that most GRBs detected by \gls{svom} could be followed-up by large ground-based facilities to increase the number of GRBs with a measured redshift \citep{wei2016deep}.
Due to this pointing strategy, the Earth will transit through the ECLAIRs \gls{fov}, inducing significant modulation of the background count rate.
 
\begin{figure}
    \centering
    \includegraphics[trim={0 0 0 0},clip, width = \hsize]{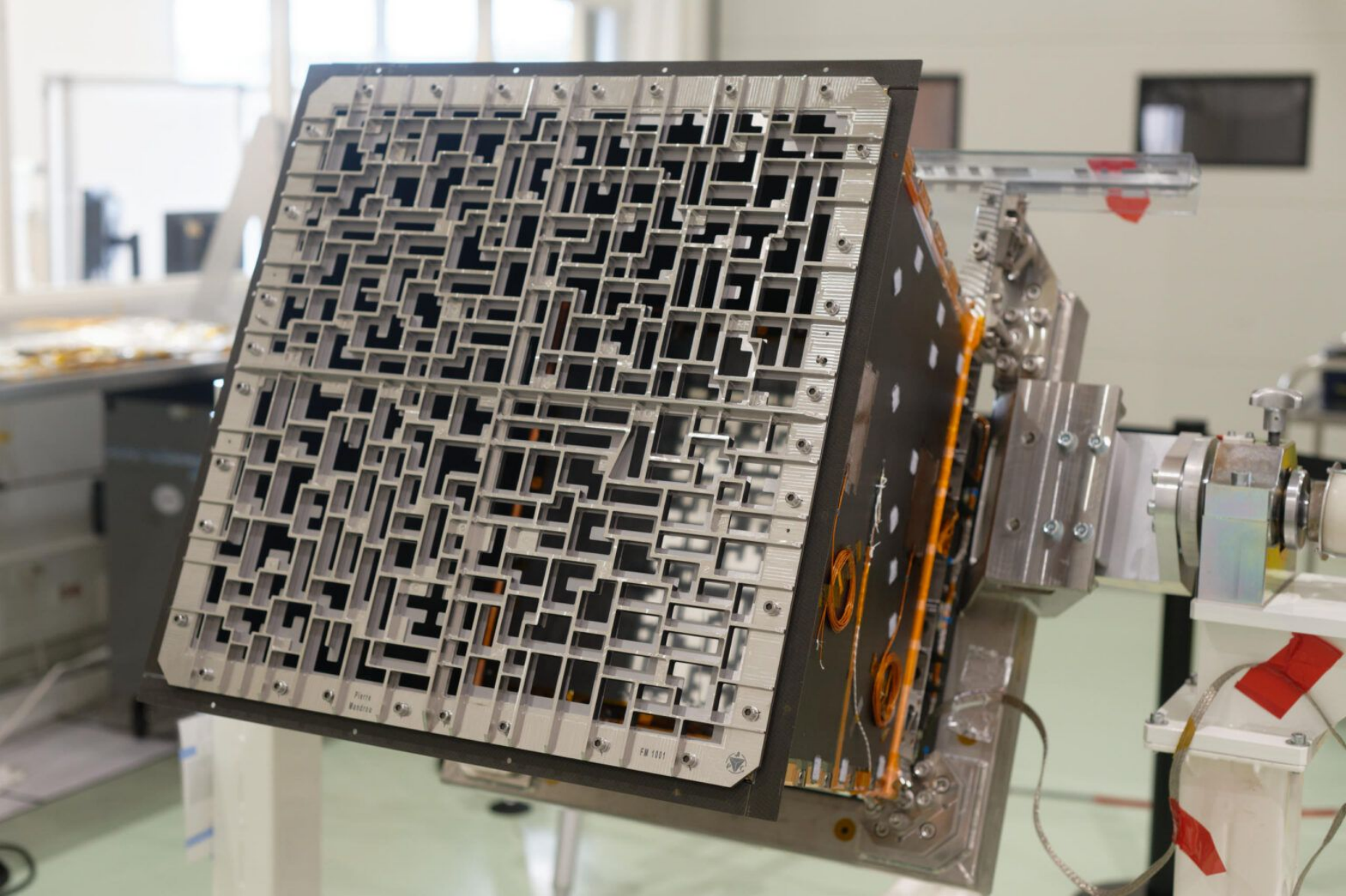}
    \caption{ECLAIRs flight model: view of the coded mask.}
    \label{fig:coded_mask_ECLAIRs}
\end{figure}

ECLAIRs is the wide-field ($\sim$2\,sr) coded-mask trigger camera of \gls{svom}. It is tasked with autonomously performing the detection and first localisation of a transient event with a typical accuracy better than 13$\arcmin$ at a 90\% confidence level.
The instrument consists of a pixelated array of $80\times 80$, $4\times 4$\,mm$^2$, and 1\,mm thick Schottky-type CdTe detectors 
totaling a geometrical area of $\sim$ 950\, cm$^2$ \citep{Remoue10, lacombe2013development, godet_2022_calibration}. A passive shield made of layers of lead and copper surrounds the detection plane to limit the background level. A coded-mask (see Figure \ref{fig:coded_mask_ECLAIRs}) with a $\sim$ 40\% transparency below 80\,keV is placed 46\,cm above the detector plane. 
The mask pattern consists of blocking and transparent elements to the high-energy radiation. It was designed so that its projection on the detection plane (called a shadowgram) is unique for a given source position within the ECLAIRs \gls{fov}.
Therefore, sky images can be reconstructed via deconvolution of the shadowgram with the mask pattern, enabling the localisation of sources in the \gls{fov}. The ECLAIRs \gls{fov} is thus divided into $199 \times 199$ sky pixels. 

The level of coding of the incoming source photons can vary depending on the position of the source within the \gls{fov}. When the source is located within the fully coded region of the FoV (\gls{fcfov}, 0.15\,sr), the source illuminates all of the plane modulated by the mask. Conversely, when the source is located within the partially coded region of the FoV (\gls{pcfov}), only a subset of the pixels are able to receive photons from the source.
ECLAIRs works in photon counting mode, that is, the readout electronics encode for each detected event its position on the plane, the detection time, and the deposited energy.

In order to optimise ECLAIRs detection sensitivity to certain populations of high-energy transients, including high-z GRBs, X-ray flashes, and X-ray rich GRBs, ECLAIRs presents a low-energy threshold (hereafter $\mathrm{E_{low}}$) of 4\,keV \citep{godet_2022_calibration}. 
This value is lower than that of previous similar high-energy instrumentation such as \textit{Swift}/\gls{bat}.
The key advantage of this low $\mathrm{E_{low}}$ value is that it increases the likelihood of detecting transients that may peak in the hard X-ray range.
This is the case of high-z GRBs since their measured emission in the observer's frame would peak at lower energy than if they were placed at lower z values, due to the reddening of the source emission.
Furthermore, this $\mathrm{E_{low}}$ value could aid in constraining the low-energy segment of the prompt emission spectra, contributing to a more precise understanding of the physics driving the prompt emission.

The ECLAIRs data processing unit makes use of the recorded events to search in near real time for the appearance of new transients within the instrument \gls{fov} \citep{dagoneau_thesis}. To do so, ECLAIRs uses two trigger algorithms \citep[see][for a detailed description]{schanne_2013_trigger_detailed}, envisioned for transients of different natures, mainly relative to their duration and intensity.
On short timescales, a count-rate trigger monitors count-rate excesses over four energy bands between 4 and 120\,keV, nine detector zones \citep[see Fig. 3 of][]{schanne_2013_trigger_detailed}, and 12 timescales ranging from 10\,ms to 20.48\,s. 
When the count rate exceeds a threshold in \gls{snr} (which depends on the energy band, timescale, and zone considered), a sky image is built on the corresponding timescale and energy intervals.
The trigger is then validated if a significant excess (e.g. \gls{snr} $\geq 6.5$) is found in the sky image. Longer transients are searched for via an image trigger on sky images built every 20.48\,s and stacked together on timescales up to $\sim$20\,min \citep{dagoneau2018ultralongGRBs}. The threshold used to validate an excess is also \gls{snr} $\geq 6.5$.

After the detection and localisation of a high-energy transient with ECLAIRs, alert messages carrying essential information on the detected events are broadcast to a network of very high-frequency (VHF) antennas spread around the Earth in the inter-tropical zone. In parallel, the satellite autonomously slews if possible, typically in less than five minutes, towards the source position provided by ECLAIRs. This enables the follow-up by the onboard narrow-field instruments \gls{mxt} and \gls{vt} to sample the afterglow emission and to sequentially refine the source localisation.
The wide-field \gls{grm} instrument extends the spectral coverage of the GRB prompt emission, up to $\sim$5\,MeV.
This enables the constraint of the spectral properties of the prompt emission and therefore the measurement of the energetics of the detected GRBs if the source redshift is measured.
In parallel, the \gls{svom} robotic ground telescopes are also able to re-point towards the obtained sky position to enhance the follow-up.


\section{Sample}
\label{sec:Sample}

\subsection{Sample selection}

The sample of \glspl{grb} selected for the analysis was built on previous prompt emission detection by \textit{Swift}/\gls{bat} and \textit{Fermi}/\gls{gbm} \citep{Meegan_2009, Goldstein_2012}.
For this, we selected all the \glspl{grb} with a redshift measurement, up to GRB 200829A, from the \textit{Swift} \gls{grb} website.\footnote{\textit{Swift} GRB table, available at \url{https://swift.gsfc.nasa.gov/archive/grb_table/}}
Additionally, some high-z \glspl{grb} also detected by \textit{Swift} but whose redshift information does not appear on this table were added based on the literature: GRB 050502 \citep[][photometric z = 5.2 $\pm$ 0.3]{Schulze_2015}, GRB 071025 \citep[][spectroscopic z $\approx$ 5.2, estimate]{Fynbo_2009}, GRB 080129 \citep[][spectroscopic z = 4.349]{Greiner_2009}, GRB 120923A \citep[][spectroscopic z = 7.8]{Tanvir_2018}, GRB 090429B \citep[][photometric z = 9.4]{Cucchiara_2011}.

In the cases where the redshift measurement is inaccurate \citep[denoted as tentative; see][]{Lien_2016_BAT_catalog}, where two or more discordant values are given, or where only an upper limit or range is available, the GRBs were not included. 
This reduced the sample from 385 to 368 GRBs.
When two or more close redshift measurements are given for the same GRB the mean value was taken, unless both spectroscopic and photometric measurements are given, in which case the spectroscopic were kept due to their better accuracy.
The \gls{gcn} circulars archive\footnote{All GCN circulars available at \url{https://gcn.gsfc.nasa.gov/gcn3_archive.html}} was used to complement the \textit{Swift} \gls{grb} redshift information when needed. 
More specific \textit{Swift}/\gls{bat} data, such as spectral parameters, burst duration parameters, and their 90\% confidence uncertainties, were retrieved from the Third \textit{Swift} \gls{bat} \gls{grb} Catalog \citep[][]{Lien_2016_BAT_catalog}.\footnote{The \textit{Swift}/BAT Catalog available at \url{https://swift.gsfc.nasa.gov/results/batgrbcat/}}


The \textit{Swift}/\gls{bat} data were complemented, when available, with \textit{Fermi}/\gls{gbm} data from the Fourth \textit{Fermi} \gls{gbm} Gamma-Ray Burst Catalog\footnote{The \textit{Fermi}-GBM Catalog 
available at \url{https://heasarc.gsfc.nasa.gov/W3Browse/fermi/fermigbrst.html}} \citep{von_Kienlin_2020}, since it provides more constrained spectra derived from a wider energy range (10--1000\,keV), and thus is useful to consider for the purpose of redshifting \glspl{grb}.
The spectra for \textit{Swift}/\gls{bat} GRBs (15--350\,keV) are typically described by a \gls{pl} or a \gls{cpl} \citep[see][]{Lien_2016_BAT_catalog}, while two additional models are used for \textit{Fermi}/\gls{gbm} \citep[see][]{Goldstein_2012}: the Band function \citep{Band} and the \gls{sbpl} \citep{Kaneko}.
In any case, the spectral data obtained from these catalogues and used in the present study correspond to time-integrated (or time-averaged) spectral data.
In Appendix \ref{appendix:TR_spectra} we compare the results of the analysis for three \glspl{grb} for which we used time-resolved spectral data from \citet{Yu_TR_spectra}.
This comparison shows that the qualitative impact of this simplification on the results is not significant.

When combining the two catalogues and to ensure that the detection from both instruments correspond to the same burst, the catalogues were merged based on the trigger time (within a margin of 300\,s) and the measured position (within a margin of 2\arcmin). All GRBs with z$<$3 that were not detected by both instruments were filtered out. However, no filtering was applied to \glspl{grb} above this redshift in order to maintain all high-z \glspl{grb} from both catalogues. 

The sample was further filtered by dropping those \glspl{grb} that lack complete spectral information from any of the catalogues.
Finally, the sample contains 162 \glspl{grb}, 20 of which have a less accurate redshift measurement (obtained via a photometric measurement or denoted as an estimation in the \textit{Swift}/\gls{bat} catalogue).
Appendix \ref{appendix:grb_table} contains a table with all GRBs in the sample.
The light curves used for our simulations are the 15--350\,keV background-subtracted 64\,ms binned $\mathrm{T_{100}}$ light curves from the Third \textit{Swift}/\gls{bat} \gls{grb} Catalog \citep{Lien_2016_BAT_catalog}.

To ensure compatibility between the spectral models derived from \textit{Fermi}/\gls{gbm} and the light curves used in the simulations, the normalisation factors for these models were recomputed. 
This normalisation factor, denoted as $A$, was computed as

\begin{equation}
    A = \frac{F_{E1\rightarrow E2}}{\int_{E1}^{E2} f_1(E) \cdot dE} \,\mathrm{ph\cdot cm^{-2} s^{-1} keV^{-1}} \,\,\,,
\end{equation}where $F_{E1\rightarrow E2}$ represents the total measured \gls{grb} photon flux over the energy range $E1$ -- $E2$, while $f_1(E)$ is the photon flux at energy E computed with $A=1$ (see Section 4 of \citealt{Goldstein_2012} for the details about the computation of $f(E)$ for each model). 
The primary interest in using \textit{Fermi}/\gls{gbm} data lies in the spectral shape information, which is contained in $f_1(E)$.
Re-computing the normalisation factor allows us to obtain a spectral model consistent with $F_{15\rightarrow 350}$, corresponding to the energy range of the \textit{Swift}/\gls{bat} light curves used. This is important for the first step of the simulations, which is the generation of the photon lists at various redshift values (see Section \ref{sec:simulations}).

\subsection{Sample properties}

The \gls{grb} duration (i.e. the duration of the prompt emission) is commonly described via the $\mathrm{T_{90}}$ parameter. This parameter indicates the duration of the time interval over which 90\% (i.e. from 5\% to 95\%) of the burst counts are recorded within the instrument's energy range.
Although it is measured in the observer frame ($\mathrm{T_{90}^{obs}}$), its rest-frame value, $\mathrm{T_{90}^{src}}$, can be computed as $\mathrm{T_{90}^{src} = T_{90}^{obs} / (1+z)}$.

\begin{figure}
    \centering
    \includegraphics[width=\hsize]{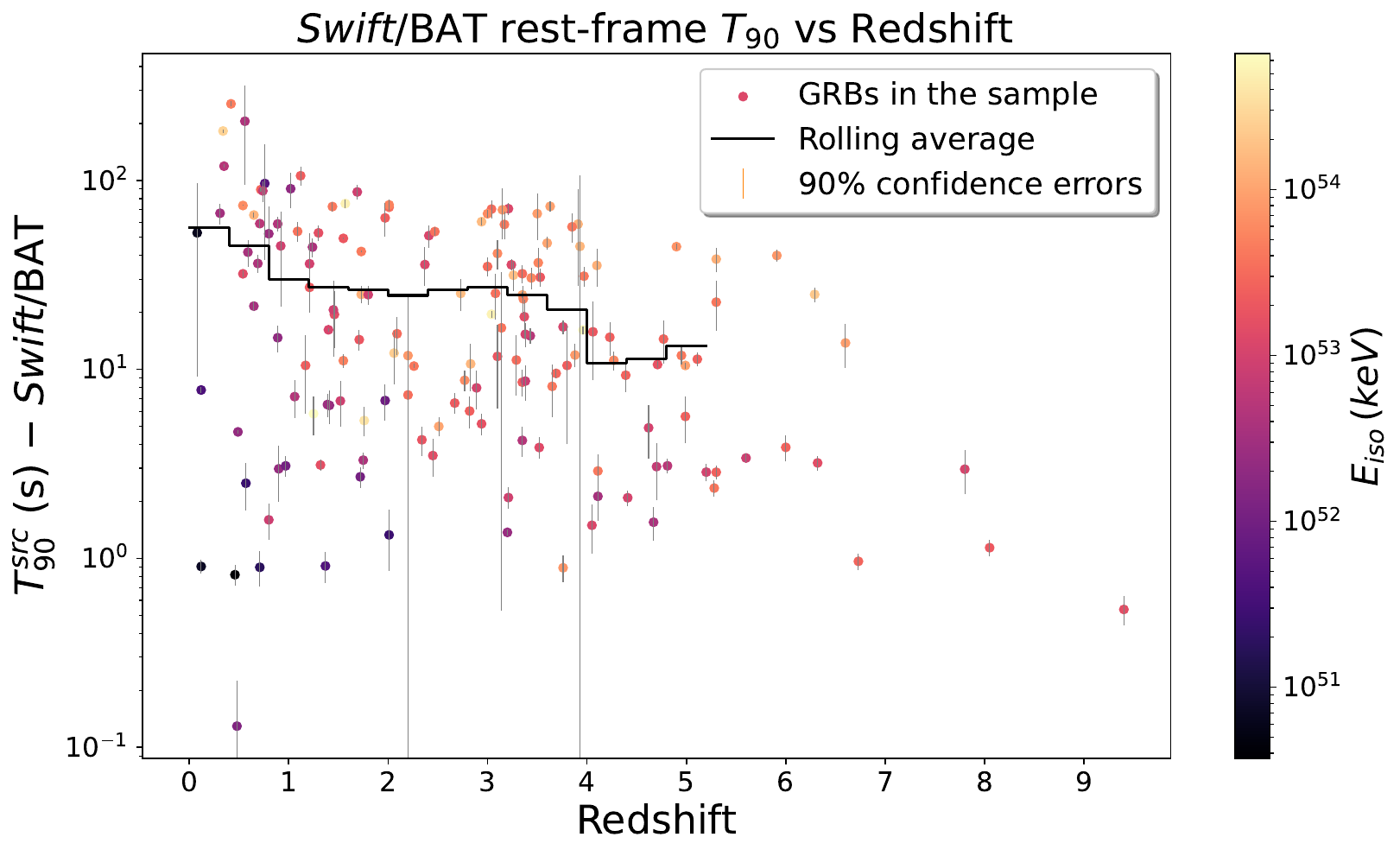}
    \caption{Rest-frame $\mathrm{T_{90}}$ as a function of redshift for the GRBs in the sample, colour-coded based on the bursts $\mathrm{E_{iso}}$. The solid line corresponds to the $\mathrm{T_{90}^{src}}$ rolling average up to $\mathrm{z\sim5}$ computed in z-bins of unity size, spaced by 0.4 in redshift.}
    \label{fig:T90vsRedshiftSample}
\end{figure}

The rest-frame $\mathrm{T_{90}}$ values are plotted against the measured redshift in Figure \ref{fig:T90vsRedshiftSample} for all GRBs in the sample. The errors are quoted at a 90\% confidence level. 
The solid line represents the $\mathrm{T_{90}^{src}}$ rolling average, computed for z-bins of unity size and spaced by 0.4 (not shown for $\mathrm{z>5}$ due to the sparsity of data).
The data points are colour-coded based on the rest-frame isotropic energy, $\mathrm{E_{iso}}$, which is computed as follows \citep[see e.g.][]{Zitouni_2014, Atteia_2017, Arcier_2020}:
\begin{align}
    \label{eq:eiso}
    E_{iso} = S_{E1\rightarrow E2}\cdot \frac{\int_{\frac{1}{1+z}}^{\frac{10^4}{1+z}} E\, f(E)\, dE}{\int_{E1}^{E2} E\, f(E)\, dE} \cdot \frac{4\pi {D_L(z)}^2}{1+z},\\
    D_L = \frac{(1+z)\cdot c}{H_0} \int^z_0 \frac{dz'}{\sqrt{\Omega_M(1+z')^3 + \Omega_L}}\nonumber.
\end{align}
Here $S_{E1\rightarrow E2}$ is the measured fluence over the energy range $E1$ -- $E2$ (in keV) given in the catalogues.
The spectral model parameters used here to compute the photon flux $f(E)$, including the normalisation factor, were those given in the corresponding catalogues. The ratio of integrals is the k-correction factor \citep{bloom2001prompt}, and extrapolates the fluence to a common energy range from 1\,keV to 10\,MeV.
The cosmological parameters used for computing the luminosity distance, $\mathrm{D_L}$, are those from the Planck collaboration: $\mathrm{\Omega_M = 0.315\pm0.007}$, $\mathrm{\Omega_L = 0.685\pm0.007}$, and $\mathrm{H_0 = 67.4\pm0.5\,km/s/Mpc}$ \citep{planck_2018_results}. The speed of light constant in vacuum is denoted by $c$.

The sparsity of the $\mathrm{z \gtrsim 6}$ GRB sample (see Fig. \ref{fig:T90vsRedshiftSample}) may hinder the identification of the intrinsic properties for this population of GRBs.
However, most GRBs with $\mathrm{z \gtrsim 6}$ present rather small $\mathrm{T_{90}^{src}}$.
We note that the $\mathrm{T_{90}^{src}}$ moving mean, depicted with a solid line, decreases with redshift.
Investigating the reasons behind this trend may help to decipher to what extent this is due to instrumental biases.
\section{Method}
\label{sec:Method}

\subsection{Simulations}
\label{sec:simulations}
\begin{figure*}[ht!]
    \centering
    \includegraphics[width=\textwidth, trim={0 9.5cm 0 0},clip]{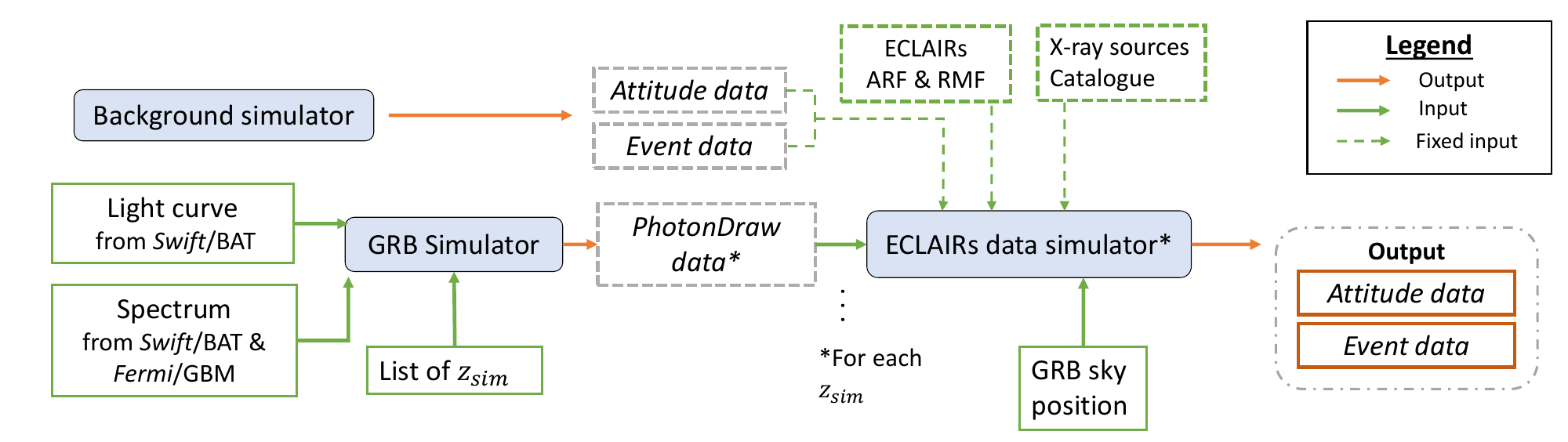}
    \caption{Flowchart of the simulation structure. The three simulator tools used in the analysis are shown as the blue bricks. This process is parallelised for all simulated \glspl{grb}, sky positions, and redshift values.}
    \label{fig:flowchart_method}
\end{figure*}

We simulated the detection by ECLAIRs of the \glspl{grb} in the sample using tools developed within the \gls{svom} collaboration, similarly to what was done by \citet{arcier2022detection}.
A flowchart showing the different steps of these simulations is displayed in Figure \ref{fig:flowchart_method}. The green arrows represent the input to each tool (blue boxes), while the orange arrows represent their outputs. The dashed green arrows represent inputs that are fixed for all the simulations.

\begin{figure}
    \centering
    \includegraphics[width = \hsize]{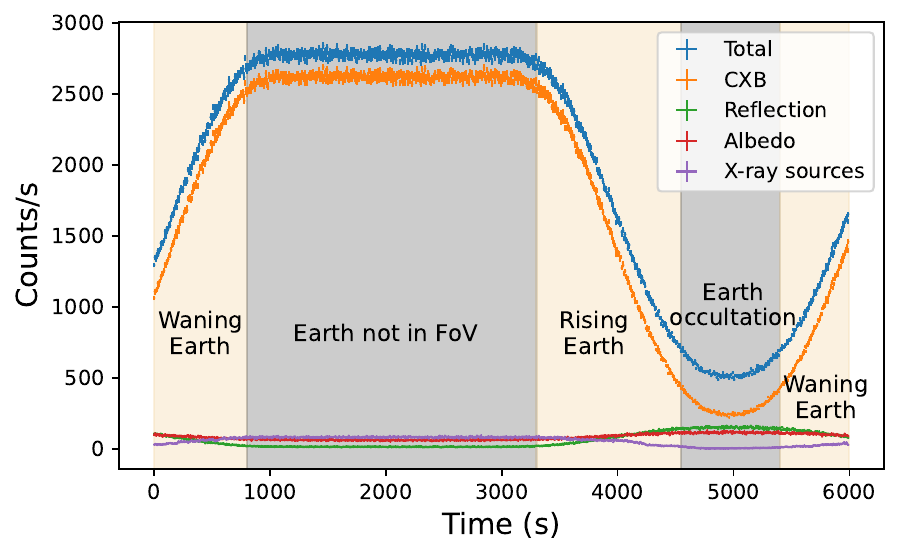}
    \caption{Count-rate evolution of the background measured by ECLAIRs for a standard \gls{svom} orbit, over the 4-120\,keV energy range and with a binning of 5.12\,s. The contribution of the main background components are displayed in different colours. }
    \label{fig:in-orbit_bkg}
\end{figure}

Firstly, the expected high-energy background along the \gls{svom} orbit is estimated via dynamical background simulations that consider the evolution of orbital parameters. It relies on a particle
interaction recycling approach (PIRA) based on a pre-computed database of particle--instrument interactions \citep{Sujay-bkg}. It accounts for the statistical fluctuations on the counts.
The dominant background contribution is the \gls{cxb} \citep{Churazov}. 
The reflection of \gls{cxb} photons on the Earth's atmosphere and the Earth's albedo, produced by the interaction of cosmic rays with the Earth's atmosphere \citep{sazonov_2007}, are additional components that can dominate the background counts when the Earth is passing through the instrument's \gls{fov}.
Figure \ref{fig:in-orbit_bkg} illustrates the 4--120\,keV and 5.12\,s binned background light curve for one orbit showing the background count-rate modulation caused by Earth's transit.

The GRB Simulator \citep{antierfarfar_thesis_2016} was used to build the list of photons emitted by each GRB, with their corresponding time and energy, hereafter called photon count data.
It takes as inputs the light curves and the spectral models from each GRB in the sample. 
It translates the source spectrum from its observed energy range into the \gls{svom}/ECLAIRs range and normalises the light curves in counts to recover the fluence measured in the ECLAIRs energy range.
This tool allowed us to simulate \glspl{grb} at redshift values ($\mathrm{z_{sim}}$) higher than their measured values ($\mathrm{z_{meas}}$) by taking into account cosmological effects such as time dilation, photon flux dimming, and energy shift.

\begin{figure}
    \centering
    \includegraphics[trim={0 0 1cm 1cm},clip, width = \hsize]{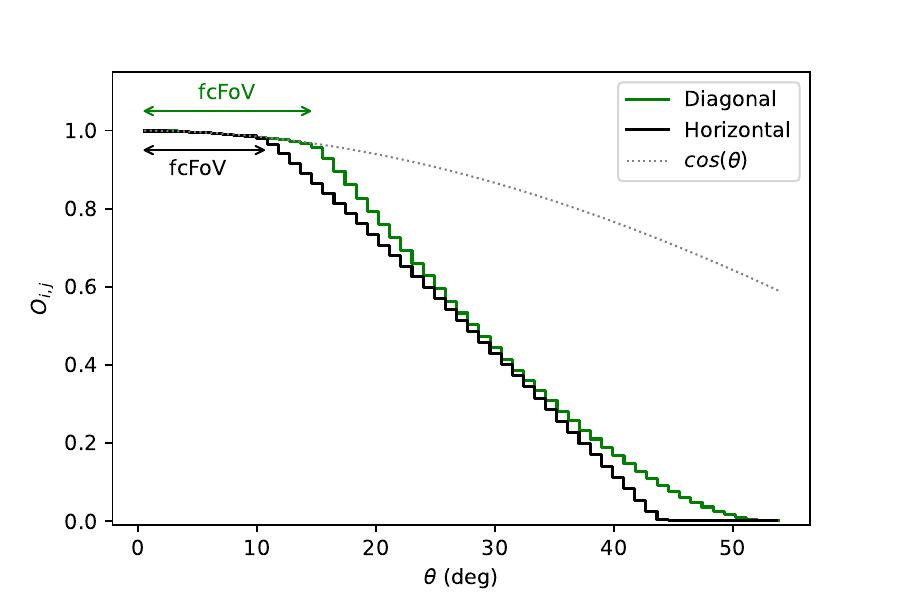}
    \caption{Mask response as a function of the source direction, for the pixels along the diagonal (green) and horizontal (black) cuts of the \gls{fov}.}
    \label{fig:mask_response}
\end{figure}

Finally, the ECLAIRs data simulator \citep{mate2021development} tool allowed us to place the GRBs anywhere along the orbit and in any position of the sky. It is a ray-tracing software that propagates photons one by one through ECLAIRs coded mask. 
The ray-tracing process considers the energy redistribution due to the transparency of the mask and the detector's efficiency, in relation to the energy of the incident photons.
This tool folds the light curves from the photon count data with the \gls{svom}/ECLAIRs ancillary response file (ARF) and redistribution matrix file (RMF) (see diagram in Fig. \ref{fig:flowchart_method}).
These response files (Yassine et al. in prep.) describe the on-axis detection efficiency (i.e. for sources in the centre of the \gls{fov}).
In the ray-tracing process, the mask response for off-axis sources is considered by applying an efficiency geometrical factor, denoted as $\mathrm{O_{i, j}}$, contingent on the sky pixel (i, j) of the incoming photons.
This factor can be regarded as a normalisation term of the curve that describes the effective area as a function of the energy given by the ARF.
Figure \ref{fig:mask_response} provides an illustration of the $\mathrm{O_{i, j}}$ values for the pixels along the diagonal and horizontal cuts of the \gls{fov}, where \(\theta\) is the angle between the source direction and the perpendicular to the detector plane (x-axis).
Pixels along the diagonal cut have i=j, while pixels along the horizontal cut have i = 100 (see Figure \ref{fig:positions_FoV} to visualise the pixel positions).
The extent of the \gls{fcfov} pixels with respect to $\theta$ is indicated for both cuts in their respective colours. 
At low $\theta$ values, corresponding to fully coded detections, the $\mathrm{O_{i, j}}$ factor is only described by a $\mathrm{cos(\theta)}$ component.
At larger $\theta$ values, the coding fraction dominates $\mathrm{O_{i, j}}$, illustrating how the detection efficiency is diminished in the \gls{pcfov}.

The output of this tool is composed of the attitude data for the simulated orbit and a photon list with the corresponding time, energy, and position of incidence on the detector plane, hereafter called the event data.
The contribution of known X-ray sources within the ECLAIRs FoV \citep{Dagoneau21} was also added to the event data along the orbit using this tool.
Although some sources can have a significant contribution within the ECLAIRs energy range, the standard orbit with extragalactic pointing considered in this work prevents strong sources within the \gls{fov} most of the time.

\subsection{Simulation set-up}
\label{sec:set-up}
All \glspl{grb} were simulated from $\mathrm{z_{meas}}$ up to $\mathrm{z_{sim}}$=15 by steps of 0.5 in z. For instance, a \gls{grb} with $\mathrm{z_{meas} = 4.3}$ would be simulated at $\mathrm{z_{sim}}$ = 4.3, 4.5, 5, ..., 15. 
The choice of this redshift range, extending up to 15, was made to investigate how ECLAIRs would detect GRBs at these remarkable distances, assuming a GRB population similar to those observed at lower redshifts. This decision is also relevant considering the concurrent operation of the \gls{svom} mission and the \textit{James Webb} Space Telescope, which detected several candidate galaxies at z$>$10 during its first year in operation \citep[e.g.][]{robertson2023identification}.
All the simulations were performed on the `flat' part of the background (i.e. when the Earth is not within ECLAIRs \gls{fov}; see Figure \ref{fig:in-orbit_bkg}).

The analysis was firstly carried out for GRBs within the \gls{fcfov} (i.e. within the central $20 \times 20$ sky-pixels, $\mathrm{\sim 575\,deg^2}$), to serve as a reference. 
One hundred trials were run for each simulation case with randomised positions and epochs within the above constraints to account for noise and source fluctuations. 
This also provided us with a measure of dispersion on the reconstructed GRB properties for each simulation case. 
A second analysis was performed with the GRBs placed in both the \gls{pcfov} and the \gls{fcfov}.
A grid of 100 positions, separated by ten sky pixels, in one quarter of the \gls{fov} was used as a reference to extrapolate the values to the rest of the \gls{fov}. This allowed us to compute the fraction of the \gls{fov} in which a \gls{grb} at a certain redshift can be detected.

\begin{figure}[ht!]
    \centering
    \includegraphics[width = \linewidth]{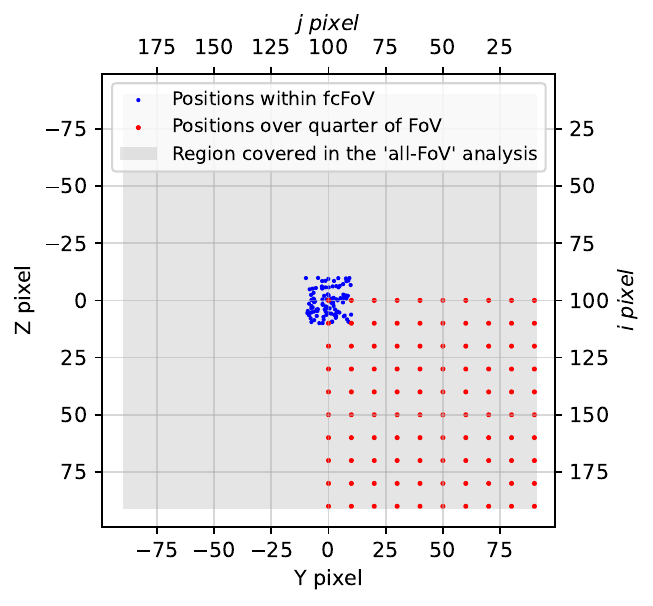}
    \caption{Representation of the positions within the \gls{fov} used in each simulation case. The grey-shaded region represents the part of the FoV for which the results of analysis in all the FoV are extrapolated.}
    \label{fig:positions_FoV}
\end{figure}

Figure \ref{fig:positions_FoV} shows the positions within the FoV selected for both analyses. The blue points represent the positions used for the \gls{fcfov} analysis, while the red points represent the grid of 100 positions for the analysis in all the \gls{fov}. The grey-shaded region shows the central $181 \times 181$ sky pixels within the FoV for which the results of the latter analysis are extrapolated.

\subsection{Signal-to-noise ratio computation}

The event data output from these simulations were used to compute the count and image \glspl{snr}, applying a suitable version of the count-rate trigger algorithm \citep{schanne_2013_trigger_detailed} presented in Section \ref{sec:SVOMECLAIRs}. 
Similarly to the flight algorithm, the \gls{csnr} was computed for 12 timescales exponentially spaced with base 2 from 10\,ms to 20.48\,s, four energy bands (4 -- 120, 4 -- 25, 15 -- 50, 25 -- 120 keV) and nine detector zones.
For each of these possible configurations, the light curve was binned with the corresponding timescale and the \gls{csnr} was computed for each time bin.
It was computed as $\mathrm{\gls{csnr} = C_{GRB} / \sqrt{\mathrm{C_{bkg}}}}$, where $\mathrm{C_{GRB}}$ corresponds to the counts associated with the GRB, and $\mathrm{C_{bkg}}$ to the background counts.
The count rate measured over an interval of three minutes before the \gls{grb} trigger time was linearly fitted to estimate the background level. This was then extrapolated to the time of interest (i.e. when the GRB occurred) to compute $\mathrm{C_{bkg}}$.
Then $\mathrm{C_{GRB}}$ was computed as the difference between the total counts and $\mathrm{C_{bkg}}$.

For each possible configuration the time bin with the largest \gls{csnr} value was chosen, and then the best \gls{csnr} of all those configurations was selected. 
This computation was also performed by shifting the time interval over which the \gls{csnr} was computed by half the width of the time bins, thus changing the binning.
This difference in binning accounts for possible better \gls{csnr} values.
Then, a sky image was reconstructed from the counts received within the energy band and time bin associated with the best \gls{csnr}.
If the transient was localised in this image with a significance exceeding a pre-set \gls{isnr} threshold of 6.5, it was considered a successful detection.
The onboard algorithm is slightly different in the sense that it selects all the excesses on the counts above a \gls{csnr} threshold and reconstructs the corresponding sky images to search for \gls{isnr} values above the \gls{isnr} threshold.
Here we just reconstructed the sky image and derived the \gls{isnr} for the best \gls{csnr} configuration over the time bins considered, assuming it would correspond to the highest \gls{isnr}, to minimise the computational cost of the simulations.

To assess whether a GRB could be detected at a certain redshift within the \gls{fcfov}, we used the median of the \gls{isnr} values obtained in the 100 trials, hereafter \gls{isnr}$\mathrm{_{med}}$.
The redshift horizon for a GRB, $\mathrm{z_{hor}}$, was computed as the maximum redshift at which the GRB is detectable with \gls{isnr}$\mathrm{_{med}}\geq$ 6.5.

Regarding the fraction of the \gls{fov} over which a GRB is detectable, we computed the \gls{isnr} values for 100 sky pixels within one quarter of the ECLAIRs \gls{fov}, including \gls{fcfov} and \gls{pcfov} (see Section \ref{sec:set-up} and Figure \ref{fig:positions_FoV}). These values, after extrapolation to the other three quarters, were interpolated to derive the expected \gls{isnr} value for each of the other sky pixels in between.
Because sky pixels do not all cover the same portion of the \gls{fov}, it was necessary to apply weights of \gls{fov} coverage by pixel in order to estimate the fraction of the complete \gls{fov} with \gls{isnr} $\geq$ 6.5.

\subsection{GRB duration measurements}

When a \gls{grb} is successfully detected, we derived its duration, $\mathrm{T_{90}^{obs}}$, using the standard \texttt{battblocks}\footnote{\url{https://heasarc.gsfc.nasa.gov/ftools/caldb/help/battblocks.html}} tool, from the \texttt{HEASOFT} v2.17 software.
It applies a Bayesian blocks algorithm \citep[see][]{Scargle_1998} for segmenting time-variable data into intervals based on Bayesian analysis.
Specifically, we input the 80\,ms binned light curve over the whole 4--120\, keV energy band.
This tool is commonly employed for ground analyses of \textit{Swift}/\gls{bat} data \citep[e.g.][]{Littlejohns_2013, Zhang_2013, Lien_2016_BAT_catalog, moss2021_tip_iceberg}.
The duration uncertainties are computed using $\pm 1 \sigma$ confidence bands around the cumulative light curve.
Since we computed the duration for 100 trials and we took the median, we also computed the 90\% confidence intervals for the median based on the values obtained in the trials.

The comparison of the $\mathrm{T_{90}^{obs}}$ median values computed for each GRB in the sample at $\mathrm{z_{meas}}$ within the ECLAIRs \gls{fcfov}, with the measured values from \textit{Swift}/\gls{bat} revealed a potential bias in the analysis.
It is worth noting, however, that this bias should not affect qualitatively the results presented in this paper (see Appendix \ref{appendix:LC_bias_in_t90} for more details).


\section{Results}
\label{sec:results}

As output of the simulations, for each GRB and $\mathrm{z_{sim}}$, we obtained the \gls{isnr}$\mathrm{_{med}}$ within the \gls{fcfov} as well as the fraction of the \gls{fov} over which the GRB would be detectable.
The maximum redshift horizon, $\mathrm{z_{hor}}$, was also obtained for each GRB, based on the \gls{isnr}$\mathrm{_{med}}$ within the \gls{fcfov}. 
For those $\mathrm{z_{sim}}$ values $\leq$ $\mathrm{z_{hor}}$, the corresponding $\mathrm{T_{90}}$ values were derived.
Table \ref{tab:results_table} in Appendix \ref{appendix:results_table} summarises some of these results for each GRB in the sample.

First, a subset of GRBs is chosen to present the results individually in Section \ref{sec:case_studies}. 
The choice of these \glspl{grb} is motivated by the shape and evolution of their light curves with increasing redshift. If we divide the GRB sample in different classes based on their $\mathrm{T_{90}}$ evolution with redshift, these GRBs can be representative of these diverse classes (see Section \ref{sec:t90_evolution_with_z}).
The results obtained for the whole sample are presented in the following sections.

\subsection{Case studies}
\label{sec:case_studies}
\begin{figure*}
    \centering
    \includegraphics[width = \textwidth, trim={0 7.5cm 0 0},clip]{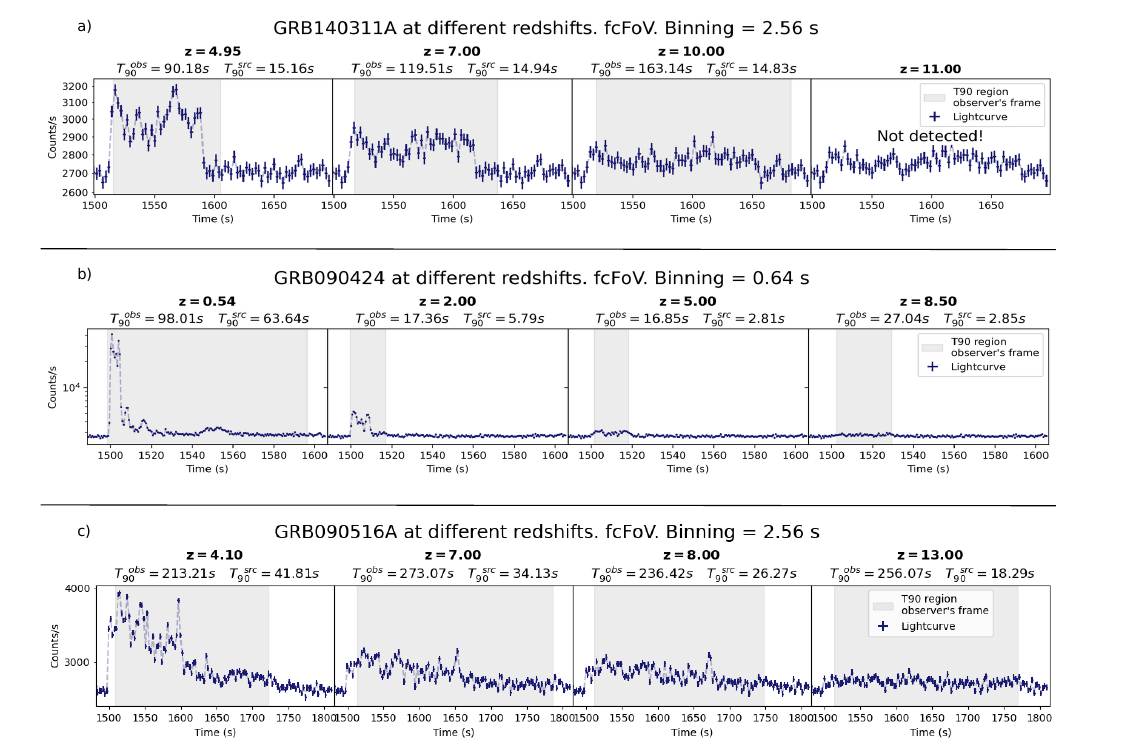}
    \caption{Evolution of the retrieved 4--120\,keV light curves at different $\mathrm{z_{sim}}$ for a specific position within the \gls{fcfov}, for the three GRBs described as case studies. GRB 140311A, GRB 090424, and GRB 090516A are shown in panels a), b), and c) respectively.
    For better visualisation, different binning is used for each GRB, both axes are fixed for all plots in the same panel, and the y-axis is shown in logarithmic scale in panels b) and c).
    The obtained $\mathrm{T_{90}}$ interval is shown with a grey shaded band on each light curve, and the corresponding values written on top, if the detection was successful. These light curves and values correspond to one trial, and hence they may differ from the values given in the text, which correspond to the median of 100 trials.}
    \label{fig:LC_evolution}
\end{figure*}

\begin{figure}
    \centering
    \includegraphics[width = \hsize]{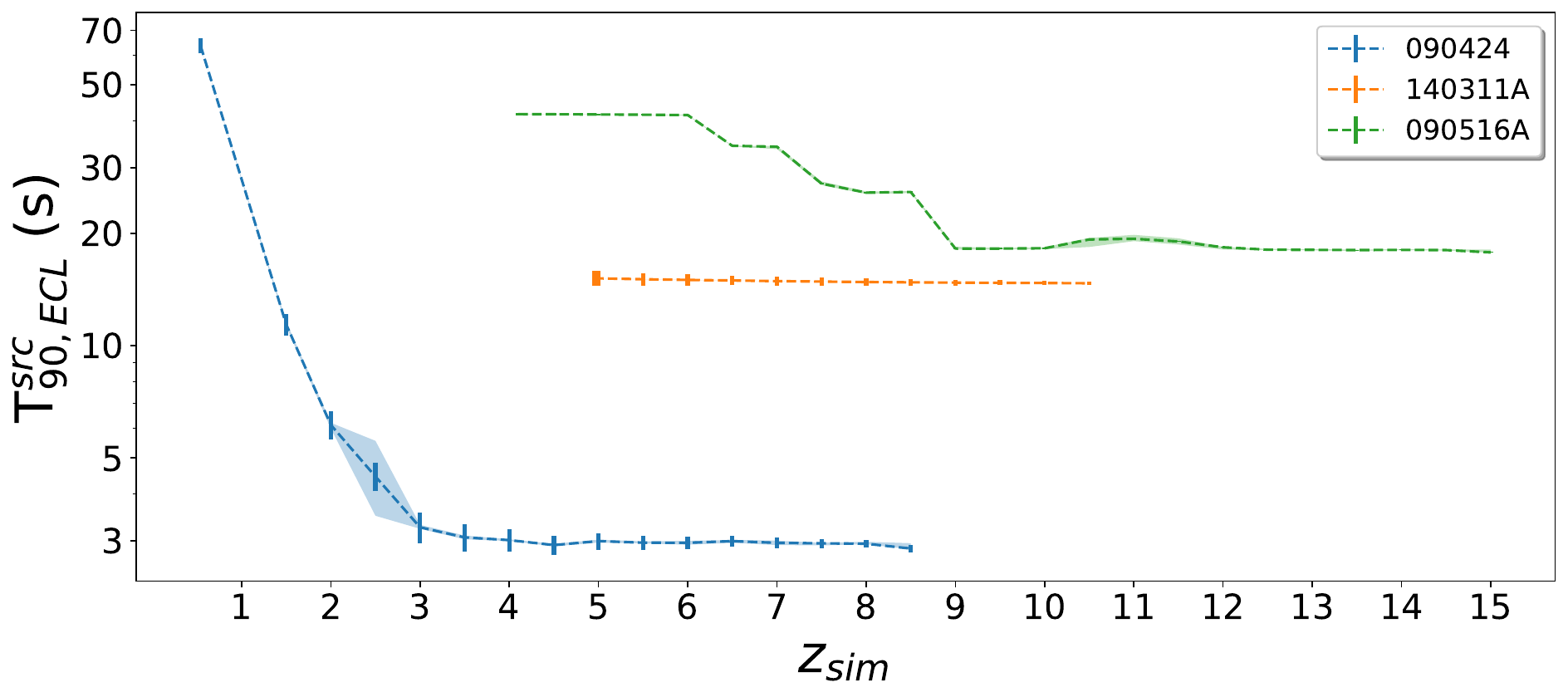}
    \caption{Evolution of $\mathrm{T_{90}^{src}}$ with respect to $\mathrm{z_{sim}}$ for GRB 140311A, GRB 090424, and GRB 090516A. The $\mathrm{T_{90}}$ values and error bars shown correspond to the median of 100 trials, derived with \texttt{battblocks}. The 90\% confidence interval of the median is shown with a shaded region around the plotted curves.}
    \label{fig:T90_ev_3GRBs}
\end{figure}

Together with the individual description of each GRB chosen for the case study, some plots are shown with the results of the three GRBs presented in this section.
Figure \ref{fig:LC_evolution} shows the light curves of these GRBs at different $\mathrm{z_{sim}}$ values, in one of the trials within the \gls{fcfov}. 
The obtained $\mathrm{T_{90}^{obs}}$ intervals for that trial are depicted with a grey-shaded band and the corresponding values are written at the top of each plot.
While the light curves and values shown in this figure correspond to one trial, the results mentioned in the text correspond to the median of 100 trials.
Figure \ref{fig:T90_ev_3GRBs} illustrates the evolution of the $\mathrm{T_{90}}$ value in the source rest frame ($\mathrm{T_{90}^{src}}$) as a function of $\mathrm{z_{sim}}$ for the three GRBs. The curves end at their redshift horizon, $\mathrm{z_{hor}}$. The median error bars correspond to those obtained with \texttt{battblocks}, which are the error values reported hereafter. The 90\% confidence interval of the $\mathrm{T_{90}^{src}}$ median value was computed for comparison, and is represented in this figure with a shaded region.

\begin{itemize}
    \item \textbf{GRB 140311A} was detected by \textit{Swift}/\gls{bat} yielding a $\mathrm{T^{obs}_{90}}$ value of 71.4 $\pm$ 9.5\,s
    within the 15--350\,keV band. Its 15--150\,keV time-averaged spectrum was best fitted by a \gls{pl} with an index of 1.67.
    Its light curve shows a weak peak of about 10\,s around the trigger time, and another weak peak of about 20\,s starting $\sim$40\,s after the trigger \citep{Krim_GCN15962}.
    There is no detection of this \gls{grb} by \textit{Fermi}/\gls{gbm}.
    Gemini-south \citep{Tanvir_GCN15961}, Gemini North \citep{Avanzo_GCN15964}, and the Nordic Observatory Telescope \citep{Chornock_GCN15966} observations found consistent spectroscopic values of redshift at $\mathrm{z_{meas}\sim 4.95}$.

    Our simulations yield that ECLAIRs would detect this GRB within its \gls{fcfov} at its $\mathrm{z_{meas}}$ value with an \gls{isnr}$\mathrm{_{med}}\sim$27 and a derived $\mathrm{T_{90}^{obs} = 90.1\pm4.17\,s}$. 
    The simulations at higher redshifts within ECLAIRs \gls{fcfov} show that it could be detected up to $\mathrm{z_{hor}\sim10.5}$, with \gls{isnr}$\mathrm{_{med}}$$\sim$6.9.
    Regarding the analysis over the entire \gls{fov}, GRB 140311A could be detected at its $\mathrm{z_{meas}}$ within $\sim$81\% of the \gls{fov}. This fraction decreases to below 50\% at $\mathrm{z_{sim} \sim 8.5}$.

    As seen in panel a) of Figure \ref{fig:LC_evolution}, the burst signature on the light curve is stretched in time and dimmed in intensity with increasing $\mathrm{z_{sim}}$ due to cosmological dilation. 
    The derived $\mathrm{T_{90}^{obs}}$ value increases accordingly over all the $\mathrm{z_{sim}}$ values at which it is detected. 
    As seen in Figure \ref{fig:T90_ev_3GRBs}, the $\mathrm{T_{90}^{src}}$ values for this GRB, in orange, stay relatively constant for all $\mathrm{z_{sim}}$.

    \medskip
    \item \textbf{GRB 090424} was a multi-peaked GRB first detected by \textit{Swift}/\gls{bat}, with $\mathrm{T_{90}^{obs}} = 49.5\pm 2.3\,s$. Its light curve shows very bright peaks during the first 5-6\,s trailing off into smaller peaks at around 8\,s, 15\,s, and 50\,s after the trigger \citep{Canizzo_GCN9223}.
    The coincident \textit{Fermi}/\gls{gbm} detection provided us with a Band function as the best-fit spectral model, with $\mathrm{E_{peak}^{obs}=177\,keV}$ \citep{Connaughton_GCN9230}. 
    The spectroscopic redshift measured for this \gls{grb} is relatively low at 0.544 (Gemini-South, \citealt{Chornock_GCN9243}; William Herschel Telescope, \citealt{Wiersema_GCN9250}).

    According to our simulations, this GRB could be detected by ECLAIRs at its $\mathrm{z_{meas}}$ within its \gls{fcfov} with \gls{isnr}$\mathrm{_{med}}$$\sim$237, and a $\mathrm{T_{90}^{obs}}$ duration of $\mathrm{98\pm4.56\,s}$.
    Its redshift horizon is $\mathrm{z_{hor}}$$\sim$8.5, with \gls{isnr}$\mathrm{_{med}}$$\sim$7.3. 
    The proximity of this GRB, together with its brightness make it detectable virtually anywhere within the \gls{fov} at its $\mathrm{z_{meas}}$ value.
    However, the detectable fraction of the FoV rapidly drops with increasing redshift.

    The evolution of the light curve with $\mathrm{z_{sim}}$ is shown in panel b) of Figure \ref{fig:LC_evolution}. The y-axes in all adjacent subplots are fixed in order to easily track the light curve evolution. In consequence, the burst signature on the right subplot at $\mathrm{z_{sim}}=8.5$, yet detectable by the trigger, is not easily distinguishable by eye.
    The evolution of the light curve with $\mathrm{z_{sim}}$ is consistent with the effect of cosmological dilation. 
    However, as opposed to the trend seen for the aforementioned GRB, the $\mathrm{T_{90}^{obs}}$ value rapidly decreases as the secondary peaks and tail of the emission are buried within the background noise. 
    Figure \ref{fig:T90_ev_3GRBs} reveals how the $\mathrm{T_{90}^{src}}$ value decreases during the first $\mathrm{z_{sim}}$, as this low-flux emission is lost. Above $\mathrm{z_{sim}}$$\sim$3, the $\mathrm{T_{90}^{src}}$ value stays roughly constant up to $\mathrm{z_{hor}}$.

    \medskip
    \item \textbf{GRB 090516A} was detected by \textit{Swift}/\gls{bat} with
    $\mathrm{T_{90}^{obs} = 210\pm 65\,s}$, later refined in the catalogue to $\mathrm{T_{90}^{obs} = 181\pm 41\,s}$.
    The time averaged \textit{Swift}/\gls{bat} spectrum is best fit by a \gls{pl} with an index of 1.84 \citep{Palmer_GCN9374}.
    The coincident \textit{Fermi}/\gls{gbm} detection \citep{McBreen_GCN9413} provided us with a \gls{cpl} function as the best-fit spectral model. The light curve consists of about five overlapping pulses lasting at least 150\,s.
    A redshift value of 4.1 was measured spectroscopically and photometrically using the Very Large Telescope \citep{Deugarte_GCN9383}, the Gamma-Ray Burst Optical/Near-Infrared Detector \citep{Rossi_GCN9382}, and the Stardome Observatory \citep{Christie_GCN9396}.

    The simulations within the ECLAIRs \gls{fcfov} at $\mathrm{z_{meas}}$ yield \gls{isnr}$\mathrm{_{med}}$$\sim$66 and $\mathrm{T_{90}^{obs} = 212.27\pm1.36\,s}$.
    This GRB is intrinsically bright enough that it could be detected at $\mathrm{z\gtrsim 15}$ within the \gls{fcfov}.
    Similarly to GRB 090424, the decrease in $\mathrm{T_{90}^{src}}$  as a function of $\mathrm{z_{sim}}$ is evident but gradual over several $\mathrm{z_{sim}}$ values (see Figure \ref{fig:T90_ev_3GRBs}).
    At $\mathrm{z_{meas}}$ the GRB is detectable over $\sim$98\% of the \gls{fov}. The detectable fraction of the \gls{fov} remains above 60\% up to $\mathrm{z_{sim} \sim 10}$, and rapidly drops below 30\% beyond $\mathrm{z_{sim} \sim 13}$.

\end{itemize}

\subsection{Sample detectability within the \gls{fcfov}}
\label{sec:detectability_fcfov}

We have shown as an example three GRBs that could be detected by ECLAIRs up to very high redshifts, higher than any GRB detected so far.
However, it is fundamental to check whether these are special cases or if there are more such GRBs in our sample.

\begin{figure*}
    \centering
    \includegraphics[width = 0.9\hsize]{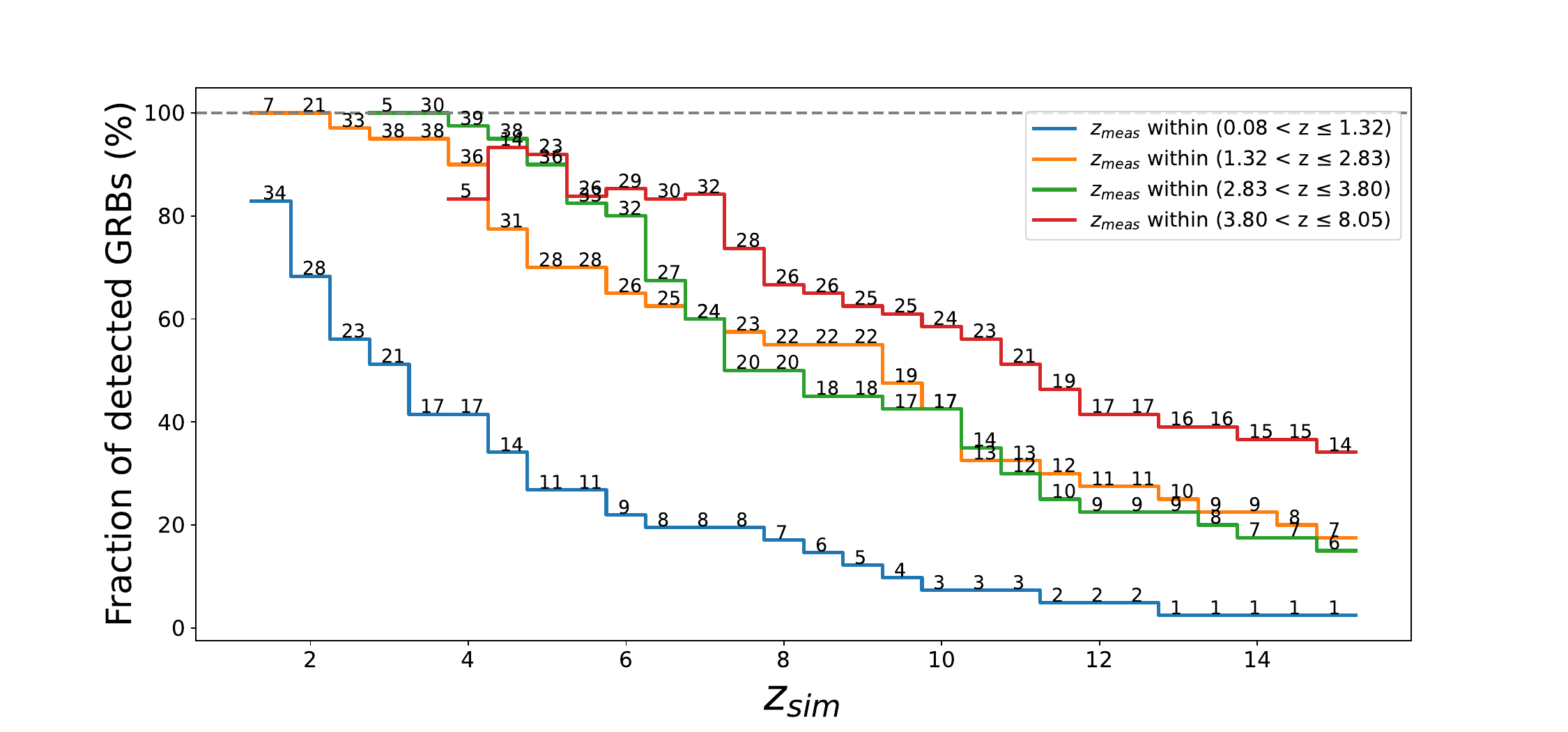}
    \caption{Percentage of detected GRBs within ECLAIRs \gls{fcfov} at each simulated redshift for the whole sample, separated into subsets based on their $\mathrm{z_{meas}}$. The numbers shown at the top of the histogram correspond to the number of detected GRBs in each redshift.}
    \label{fig:Detectability_per_z_in_zbins}
\end{figure*}

We divided our sample into four subsets of $\sim40$ GRBs, according to their $\mathrm{z_{meas}}$ value, into the following ranges: [0.078 -- 1.33], [1.33 -- 2.86], [2.86 -- 3.83], and [3.83 -- 9.4]. The lowest and highest redshift subsets contain 41 GRBs each, while the two intermediate subsets comprise 40 GRBs each.
Figure \ref{fig:Detectability_per_z_in_zbins} illustrates the percentage of GRBs detected within the \gls{fcfov} for each of these groups plotted as a function of $\mathrm{z_{sim}}$.
The number of detected GRBs out of the total number of simulated GRBs within each group is indicated above each bar in the histogram.
We recall that GRBs were not simulated at $\mathrm{z_{sim}}$ values lower than their corresponding $\mathrm{z_{meas}}$ value. Therefore, the number of GRBs corresponding to the first $\mathrm{z_{sim}}$ values for each curve in Figure \ref{fig:Detectability_per_z_in_zbins} might be small, while the percentage is high (except for the lowest-z subset represented by the blue curve where all GRBs have $\mathrm{z_{meas}}$<1.5, which is the lowest-$\mathrm{z_{sim}}$).

This figure illustrates that several GRBs could be detected at redshifts higher than 9 or 10, a few even detectable up to $\mathrm{z_{sim}\,\ge 15}$.
We can also see that the percentage of detected GRBs at a certain redshift is typically smaller for those GRBs which originally had a low redshift.
In particular, the percentage of detected GRBs from the lowest-z subset ([0.078 -- 1.33]) is significantly smaller than for the rest of GRBs, at any given redshift. This is due to the lower intrinsic brightness of the GRBs in this subset compared to the other ones.

\subsection{Impact of the low-energy threshold}
\label{sec:lowE_threshold}
\begin{figure}
    \centering
    \includegraphics[width = \hsize]{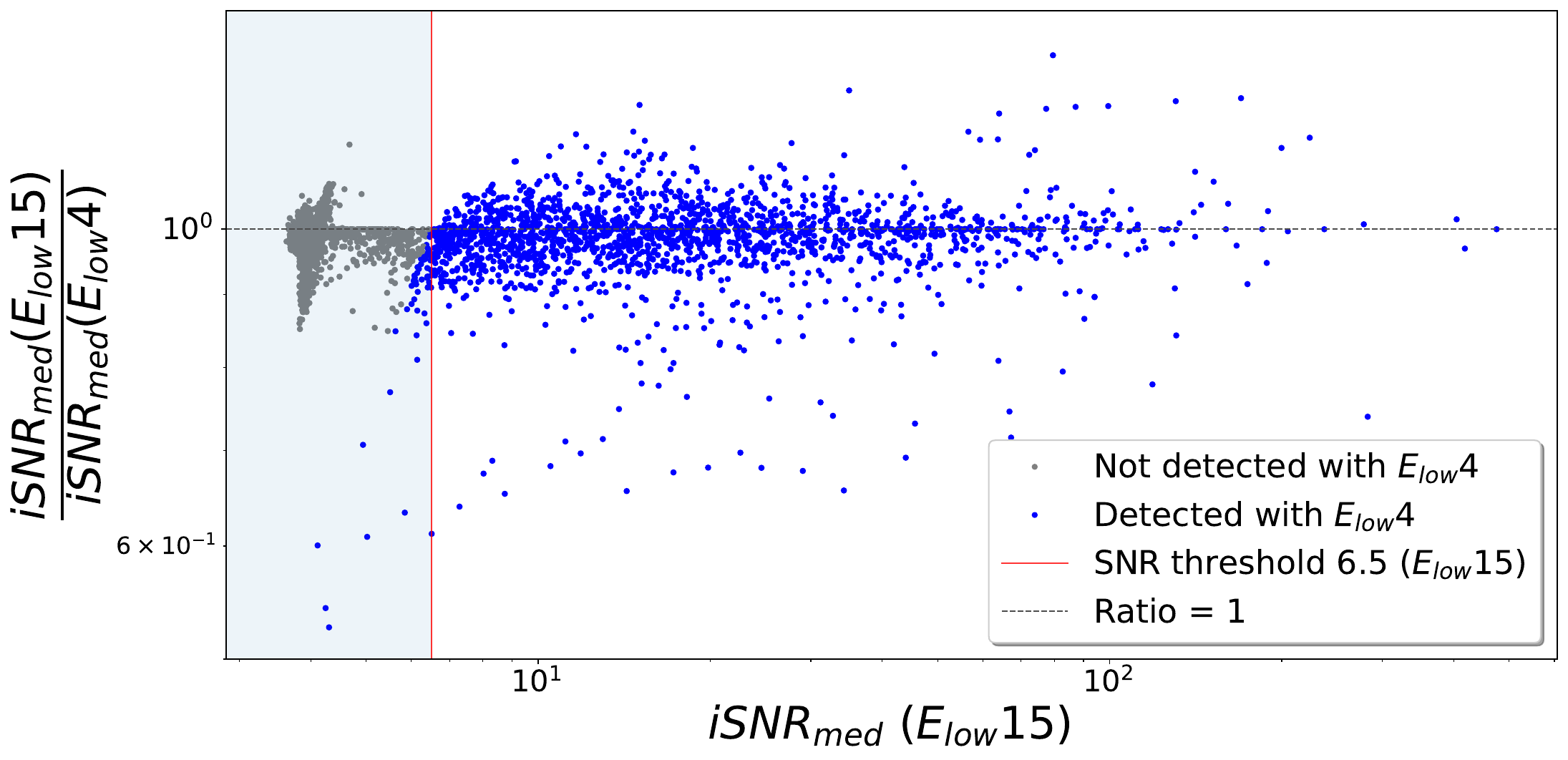}
    \caption{Comparison of the \gls{isnr}$\mathrm{_{med}}$ obtained with $\mathrm{E_{low}}$ values of 15 and 4 keV, for all GRBs, and redshift values simulated within the \gls{fcfov}. The blue and grey points represent respectively all successful and not successful detections with $\mathrm{E_{low}}$ = 4 keV. The red vertical line represents the \gls{isnr} threshold above which the successful detections with $\mathrm{E_{low}=15\,keV}$ lie.}
    \label{fig:iSNR_comparison_Elow}
\end{figure}

\begin{figure}
    \centering
    \includegraphics[width = \hsize]{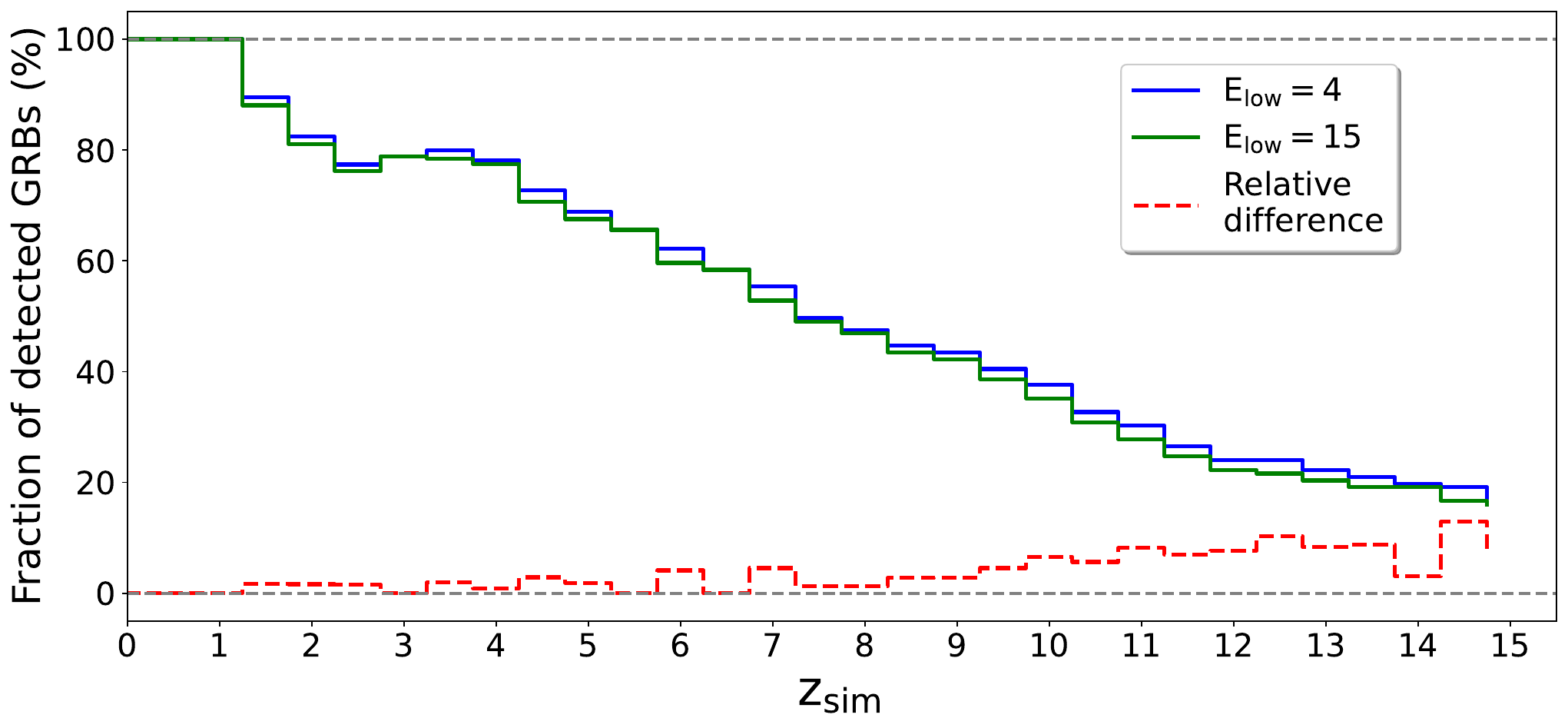}
    \caption{Comparison of the percentage of detected \glspl{grb} (\gls{isnr}$\mathrm{_{med}}$ $\geq$ 6.5) within the \gls{fcfov} at each $\mathrm{z_{sim}}$ for $\mathrm{E_{low}}$ values of 15 and 4\,keV. The dashed red line represents the relative difference between the percentage of both scenarios, with respect to the 4\,keV scenario.}
    \label{fig:Detectability_elow_comparison}
\end{figure}

The low value achieved for the low-energy threshold ($\mathrm{E_{low}}$) of ECLAIRs is a significant characteristic of the instrument.
To shed insight into its importance, a comparison was made between the previous analysis results with $\mathrm{E_{low}=4\,keV}$ and an analysis using an $\mathrm{E_{low}}$ of 15\,keV, similar to \textit{Swift}/\gls{bat}.
The comparison of these two scenarios is illustrated in Figures \ref{fig:iSNR_comparison_Elow} and \ref{fig:Detectability_elow_comparison} for all the simulation cases (i.e. all GRBs and $\mathrm{z_{sim}}$) within the \gls{fcfov}.

Figure \ref{fig:iSNR_comparison_Elow} shows the ratio of the \gls{isnr}$\mathrm{_{med}}$ obtained with $\mathrm{E_{low}=15\,keV}$ to that with $\mathrm{E_{low}=4\,keV}$ plotted against the \gls{isnr}$\mathrm{_{med}}$ with $\mathrm{E_{low}=15\,keV}$. 
Among the 4157 simulation cases, the ratio is less than unity in 2657 cases (63\%), equal to unity in 367 cases (9\%), and greater than unity in 1131 cases (22\%).
Notably, there are 57 cases (blue points on the left of the vertical red line in Figure \ref{fig:iSNR_comparison_Elow}) for which the detection is successful with $\mathrm{E_{low}=4\,keV,}$ but not with $\mathrm{E_{low}=15\,keV}$, while only two cases demonstrate successful detection with $\mathrm{E_{low}=15\,keV}$, but not with $\mathrm{E_{low}=4\,keV}$.

Figure \ref{fig:Detectability_elow_comparison} shows the percentage of GRBs detected within the \gls{fcfov} by bins of 0.5 width in redshift, for both scenarios. The relative loss shown by the dashed red line was computed as the difference between the percentage for both scenarios divided by the percentage for the 4\,keV scenario. This provides an idea of the extent of the detections that would be lost by setting $\mathrm{E_{low}}$ to 15\,keV instead of 4\,keV. This relative loss is larger for greater $\mathrm{z_{sim}}$ values, escalating from nearly 0\% below $\mathrm{z_{sim} \sim 2}$ up to $\sim10\%$ beyond $\mathrm{z_{sim} \sim 10}$.
However, these values should be taken cautiously due to the limited number of GRBs at these high redshifts, rendering the results less statistically robust. 

\subsection{Detectability outside the \gls{fcfov}}
\label{sec:detectability_all_fov}

\begin{figure}[H]
    \centering
    \includegraphics[width = \hsize]{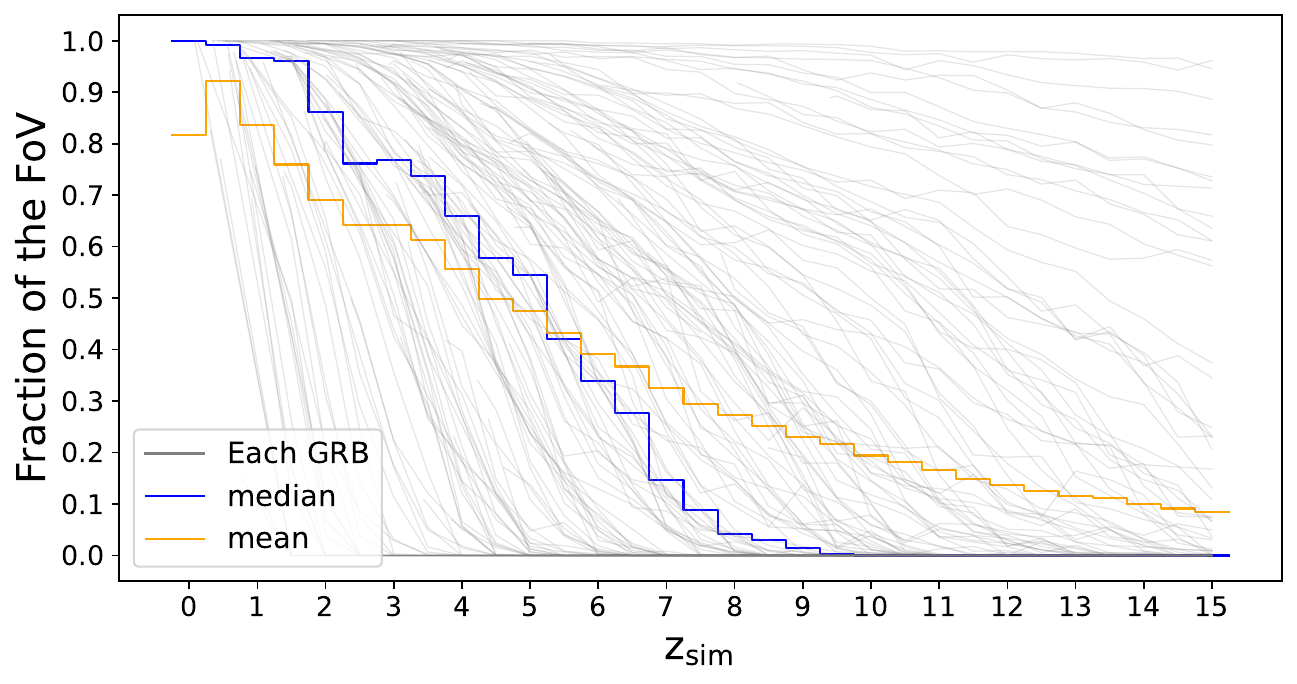}
    \caption{Fraction of the FoV with successful detection (\gls{isnr} $\geq$ 6.5) plotted against $\mathrm{z_{sim}}$ for each GRB in the sample (grey curves). The median and mean values for bins of 0.5 width in redshift are shown in blue and orange, respectively.}
    \label{fig:Fraction_fov_redshift}
\end{figure}

Figure \ref{fig:Fraction_fov_redshift} shows the evolution with increasing redshift of the fraction of \gls{fov} over which a GRB can be detected. The grey curves represent each simulated GRB, while the blue and orange histograms represent the median and mean values, respectively, for redshift bins of width 0.5.
At $\mathrm{z_{sim} \sim 5.5}$, nearly half of the GRBs in the sample (79 GRBs) remain detectable in more than 40\% of the \gls{fov}.
At $\mathrm{z_{sim} \sim 8}$ and $\mathrm{z_{sim} \sim 10}$, the number of GRBs would become 42 and 32, respectively.

At $\mathrm{z_{sim}}$ > 7.5, more than half of the GRBs exhibit an almost zero detectable fraction of the \gls{fov}, which is consistent with median value observed beyond this redshift range (see Figure \ref{fig:Fraction_fov_redshift}). 
However, a subset of bright GRBs remains detectable within a large fraction of the \gls{fov} at very high redshifts (e.g. 20 GRBs are detectable within more than 0.4 of the \gls{fov} fraction above $\mathrm{z_{sim} \sim 12}$). This explains the mean value diverging from the median value above redshift 6.
Consistently with the results over the \gls{fcfov}, the GRBs detectable up to high redshift values are mostly GRBs not contained within the lowest redshift bin in our sample [0.078 -- 1.33], but equally distributed within the other redshift bins.

\subsection{Sample $\mathrm{T_{90}}$ evolution with redshift}
\label{sec:t90_evolution_with_z}

\begin{figure*}
    \begin{subfigure}[b]{0.5\textwidth}
         \centering
         \includegraphics[width=\textwidth]{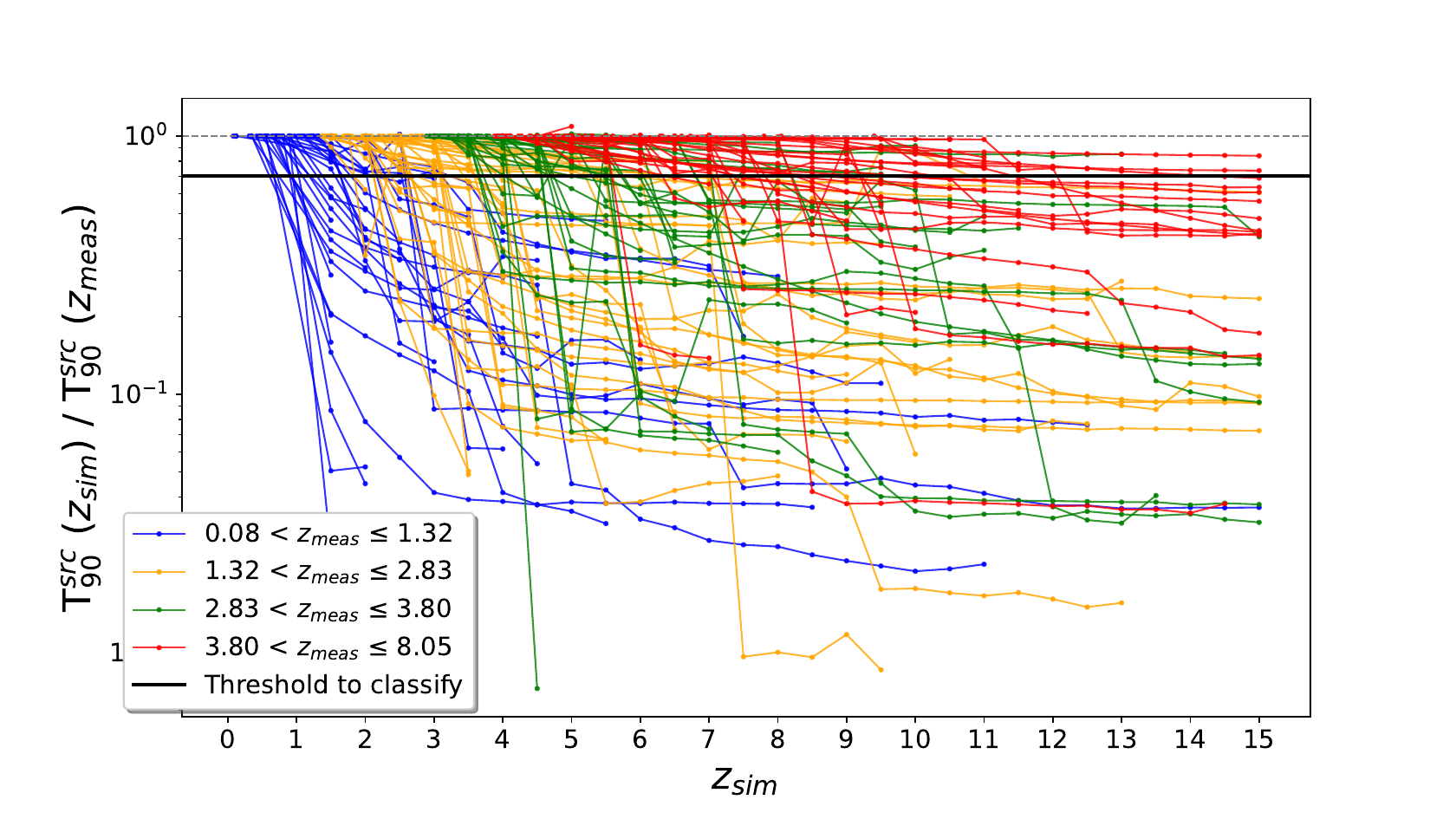}
        \label{fig:normalised_t90_ev_all_sample}
     \end{subfigure}
     \begin{subfigure}[b]{0.5\textwidth}
         \centering
         \includegraphics[width=\textwidth]{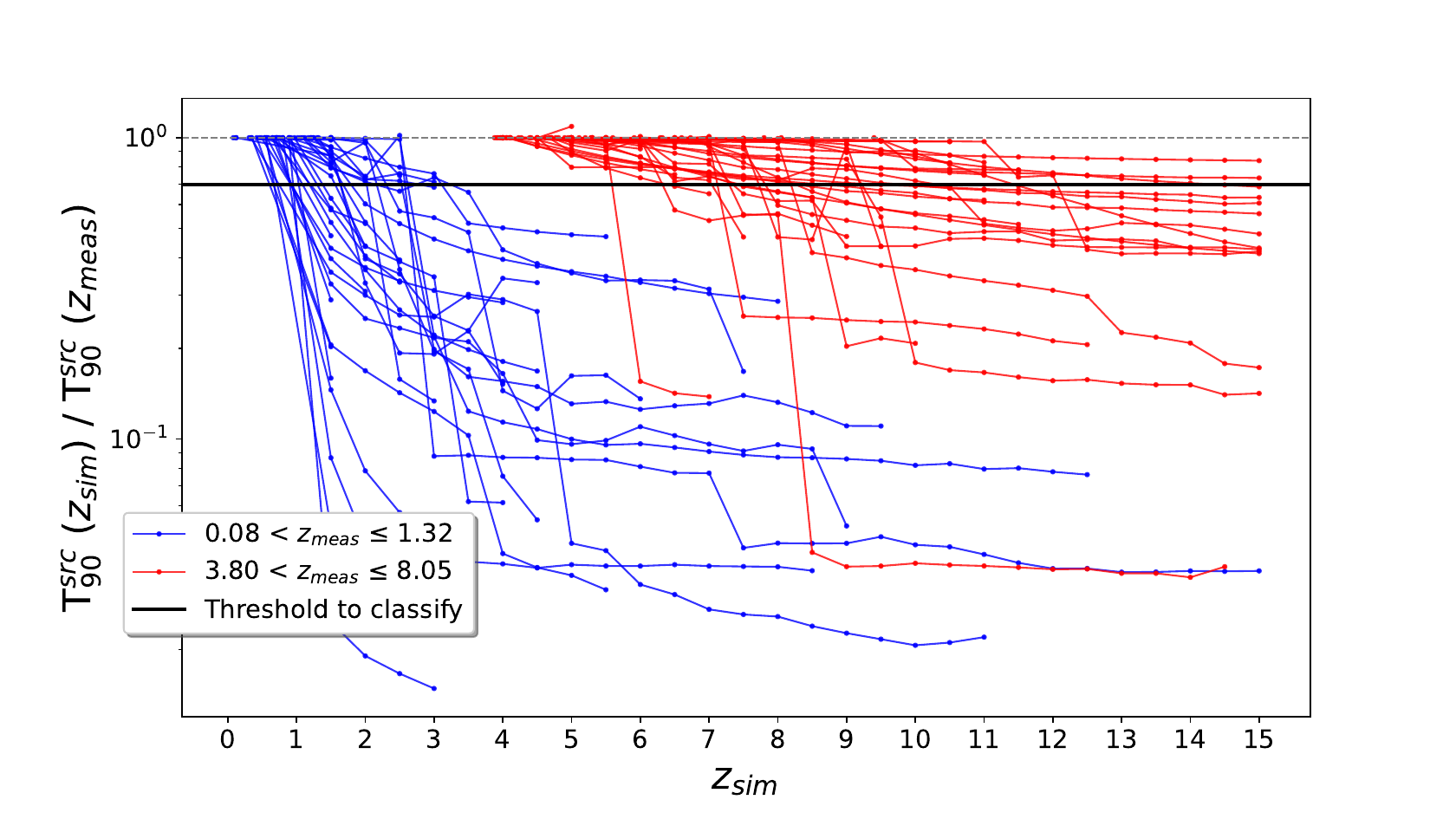}
         \label{fig:Normalisation_t90_red_and_blue}
    \end{subfigure}
    \caption{Evolution of $\mathrm{T_{90}^{src}}$ at each $\mathrm{z_{sim}}$ divided by the corresponding value at $\mathrm{z_{meas}}$, for all GRBs in the sample. The curves are colour-coded according to the subset to which the GRBs belong. The black horizontal line shows the value used to categorise the GRBs according to their $\mathrm{T_{90}}$ evolution with redshift. The left panel contains all GRBs in the sample, while the right panel contains only those in the lowest-z and highest-z subsets.}
    \label{fig:normalised_t90_ev}
\end{figure*}

The evolution with $\mathrm{z_{sim}}$ of $\mathrm{T_{90}^{src}}$ normalised by the derived $\mathrm{T_{90}^{src}}$ value at $\mathrm{z_{meas}}$, is depicted for the whole sample in the left panel of Figure \ref{fig:normalised_t90_ev}. 
The right panel shows only the GRBs from the lowest-z and highest-z subsets, for clarity.
The plots analogous to Figure \ref{fig:T90_ev_3GRBs} (i.e. without normalisation) by subsets of the entire sample can be found in Appendix \ref{appendix:t90_evolution}.

Each curve in Figure \ref{fig:normalised_t90_ev} represents the evolution of a given GRB, colour-coded according to the subset to which it belongs. The black horizontal line represents the threshold value used to classify the GRBs based on their $\mathrm{T_{90}}$ evolution with redshift (see below).

Many GRBs exhibit a behaviour similar to GRB 090424 and GRB 090516A, where a portion of the low-flux emission is buried into the background noise above a certain $\mathrm{z_{sim}}$, resulting in a noticeable decrease in the derived $\mathrm{T_{90}^{src}}$ value.
Some of these GRBs display an abrupt decrease in $\mathrm{T_{90}^{src}}$ at a specific redshift, followed by a plateau of relatively constant $\mathrm{T_{90}^{src}}$, as for GRB 090424, while others exhibit a more gradual or steady decrease, similar to GRB 090516A.

On the contrary, many other GRBs maintain a relatively constant $\mathrm{T_{90}^{src}}$ value across all $\mathrm{z_{sim}}$ values up to their $\mathrm{z_{hor}}$, indicating that the majority of their emission remains above the background noise level.
GRB 060927 and GRB 120521C, which belong to the highest-z bin, are special cases as they exhibit this behaviour, but feature a $\mathrm{z_{hor} \gtrsim 15}$. Therefore, it is possible that their $\mathrm{T_{90}^{src}}$ value would decrease at higher redshift values, once their low-flux emission is lost.

\glspl{grb} were categorised into two groups based on the ratio of $\mathrm{T_{90}^{src}}$ at $\mathrm{z_{hor}}$ to that at $\mathrm{z_{meas}}$:
$\mathrm{T_{90}^{src}(z_{hor}) \,/ \,T_{90}^{src}(z_{meas})}$.
This ratio provides insight into the extent of the $\mathrm{T_{90}^{src}}$ decrease between the first and last simulated redshift values.
A threshold value of 0.7 (i.e. a decrease of 30\%; see Figure \ref{fig:normalised_t90_ev}) was used to allocate the GRBs into these two groups.
This value was chosen based on the distribution of the ratio values for the whole sample.
GRBs with a ratio below 0.7 were allocated in the \texttt{Decrease} group indicating a significant decrease in $\mathrm{T_{90}^{src}}$, while those with higher ratio were allocated in the \texttt{Uniform} group indicating a rather constant $\mathrm{T_{90}^{src}}$ value.
Within the \texttt{Decrease} class, GRBs were further divided into those with an abrupt decrease ($\mathrm{\gtrsim 25\%}$ between two consecutive $\mathrm{z_{sim}}$) and those with a gradual decrease.
GRBs not detected above their $\mathrm{z_{meas}}$ were not categorised in any group (\texttt{N/C}).

Table \ref{tab:classification_GRBs_t90_ev} presents the number of GRBs allocated to each class, categorised by the subsets based on their $\mathrm{z_{meas}}$ in the sample. The table also indicates the number of GRBs in the \texttt{Uniform} class with $\mathrm{z_{hor} \gtrsim 15}$.


\begin{table*}[h]
\centering
\caption{GRBs categorised according to their $\mathrm{T_{90}^{src}}$ decrease with $\mathrm{z_{sim}}$.}
\label{tab:classification_GRBs_t90_ev}
\begin{tabular}{@{}cl||cccc||cccl||c@{}}
\toprule
\multirow{2}{*}{\textbf{}} & & \multicolumn{3}{c}{\textbf{Decrease (>30\%)}} & \multirow{2}{*}{} &
\multicolumn{3}{c}{\textbf{Uniform (<30\%)}} & & \multirow{2}{*}{\textbf{Not Categorized}} \\
& \textbf{\#} & Gradual   & Abrupt  & \textbf{Total}  & & $\mathrm{z_{hor} \gtrsim 15}$ &
$\mathrm{z_{hor} < 15}$ & \textbf{Total} &  & \\ \midrule
lowest-z subset & 41&  2  &  29  &  \textbf{31}  &&  0  & 3  &  \textbf{3} && 7\\ 
low-z subset & 40&  4  &  26  &  \textbf{30}  &&  0  & 10  &  \textbf{10} && 0\\ 
high-z subset & 40&  2  &  20  &  \textbf{22}  &&  0  & 18  &  \textbf{18} && 0\\ 
highest-z subset & 41&  9  &  9  &  \textbf{18}  &&  2  & 19  &  \textbf{21} && 2\\ \bottomrule
\end{tabular}
\end{table*}


There is an increasing trend in the number of GRBs categorised as \texttt{Uniform} as the $\mathrm{z_{meas}}$ value increases, while the subsets with lower $\mathrm{z_{meas}}$ contain a higher proportion of GRBs in the \texttt{Decrease} class.
Most of those categorised as decreasing show an abrupt decrease between two consecutive $\mathrm{z_{sim}}$, rather than a  gradual decrease.

This $\mathrm{T_{90}}$ bias may lead to significantly divergent measurements at different redshift values, dropping to less than 5\% of the value at $\mathrm{z_{meas}}$ for certain GRBs. 
A comparable loss of measured GRB counts arising from this $\mathrm{T_{90}}$ reduction could lead to a more pronounced impact on the measured properties (e.g. luminosity) of some bursts.
However, upon a preliminary examination of the count loss within the measured $\mathrm{T_{90}}$ interval, it appears to be generally less substantial than the reduction in $\mathrm{T_{90}}$.


\section{Discussion}
\label{sec:Discussion}

In this section we discuss the main findings of this study, outlined in Section \ref{sec:results}, after having simulated a carefully selected sample of GRBs (see Section \ref{sec:Sample}), emulating a realistic detection scenario by ECLAIRs, at different $\mathrm{z_{sim}}$ values, equal to and higher than their $\mathrm{z_{meas}}$ value (see Section \ref{sec:Method}).
Firstly, we discuss the main findings in terms of detection capabilities of ECLAIRs, and then we discuss the instrumental effects on the reconstruction of source measured properties, such as GRB duration.
While the majority of GRBs are spatially more likely to lie within the \gls{pcfov}, a significant part of the analysis presented in this study corresponds to the simulations within the \gls{fcfov} to serve as a reference.

The results of the simulations outlined in Section \ref{sec:detectability_fcfov} suggest that ECLAIRs has the potential to detect GRBs at extremely high redshifts ($\mathrm{z>15}$), significantly surpassing the redshift values of the highest-z GRBs known to date ($\mathrm{z \approx 9.4}$).
The ability to detect these high-z GRBs is particularly prominent when the GRBs are located within the \gls{fcfov}, which represents only a $\sim$ 7.24\% of the instrument's \gls{fov}.
However, even within the \gls{pcfov} (see Section \ref{sec:detectability_all_fov}), where more GRBs would lie based on spatial likelihood but where the detection is more challenging, a substantial number of bright GRBs, with $\mathrm{E_{iso} \gtrsim 10^{53}\,erg}$, can still be detected at very high-z values. Out of the 162 GRBs in the sample, 14 GRBs could be detected within more than 50\% of the \gls{fov} at $\mathrm{z\sim15}$ and 47 GRBs at $\mathrm{z\sim9}$.
On the other hand, the diminishing median value of the detectable fraction of the \gls{fov} beyond $\mathrm{z=7}$ (Figure \ref{fig:Fraction_fov_redshift}) suggests that standard GRBs are less likely to be detected outside the \gls{fcfov} at these redshift ranges.

The low-energy threshold ($\mathrm{E_{low}}$) of the ECLAIRs energy range, which extends down to 4\,keV, is one of the contributing factors to the good high-z GRB detection performance obtained from our simulations. 
Setting the low-energy threshold, $\mathrm{E_{low}}$ to 15\,keV instead, would deteriorate its detection capabilities in more cases than it would improve it. This is more significant at high-z values, reaching  $\mathrm{\gtrsim 10\%}$ loss at $\mathrm{z_{sim} > 10}$. It is therefore important to set the trigger $\mathrm{E_{low}}$ threshold as close as possible to that of the camera (i.e. 4 keV). To do so, specific care in the flight calibration should be devoted during the \textit{SVOM} commissioning phase. With this capability, ECLAIRs is expected to contribute to slightly augmenting the number of detected GRBs at high redshift.

Demonstrating that ECLAIRs will be able to detect GRBs possibly with $z>10$ does not mean that such GRBs will be identified as such, as there is a step to go  from detection to identification, which is challenging.
Accurate redshift measurements for high-z GRBs present challenges as they require near-infrared observations of the afterglow, which rely upon follow-up observations by other instruments.
The sensitivity and availability of multi-wavelength instruments for the follow-up are therefore limiting conditions to obtain the redshift measurements.
These limitations could play a role in the scarcity of known high-z GRBs in the current sample, thus contributing to the notable disparity in detectability at high-z between the simulations and the current GRB sample.
The detection of the afterglow is therefore crucial for increasing the sample of known high-z GRBs.
In order to address this challenge, and to fill the gap between the detection and the identification of GRBs, the \gls{svom} collaboration have put in place  a dedicated pointing strategy (i.e. quasi anti-solar pointing) coupled with a strong space and ground synergy to enhance the GRB follow-up on ground promptly and efficiently \citep[see][]{wei2016deep, atteia2022svom}.

We recall that we used a sample of already observed GRBs to perform our study, to stay as close to reality as possible, since there are no spectral nor temporal templates of GRB emission accounting for the diversity of their observational properties.
We shall therefore face that the instruments that have observed these GRBs have likely imprinted some instrumental biases on the source intrinsic properties. For instance, some low-flux temporal structures may have disappeared within the noise due to the observing conditions (e.g. angle of incidence; \citealt{moss2021_tip_iceberg}) and/or may have fallen outside of the energy range of the instruments.
Figure \ref{fig:Detectability_per_z_in_zbins}, considering the GRBs in the various redshift ranges subsets, clearly evidences the instrumental biases imprinted on the high-z GRB sample.
The lower detectability of the GRBs in the lowest-z subset ($\mathrm{z_{meas}}<1.33$, blue curve), is due to their lower intrinsic brightness.
This selection effect is also shown in the analysis of the $\mathrm{T_{90}}$ evolution with redshift, discussed below.

In this study, we estimated the detectability using the count-rate trigger with the same threshold as the onboard algorithm.
Some additional bursts may be found using the onboard image trigger, or using the on-ground ECLAIRs offline trigger featuring complementary algorithms or applying lower threshold values for the count-rate trigger.
In particular, considering that the emission of high-z GRBs in the observer’s frame may be significantly stretched, it would be beneficial to use the image trigger, which is well suited to detect GRBs featuring a very long emission \citep{dagoneau2020ultra}.
Therefore, one could expect a larger fraction of GRBs detected with the image trigger than with the count-rate trigger, consistently with \textit{Swift}/\gls{bat}  \citep[Figure 22]{Lien_2016_BAT_catalog}.
However, the $\mathrm{T_{90}}$ bias presented in Section \ref{sec:t90_evolution_with_z} might counteract this effect reducing the image trigger performance as the GRB emission that stands above the noise would be shorter. Further analyses need to be done in order to address this.

On the other hand, additional ECLAIRs instrumental effects \citep[see e.g.][]{arcier2022detection}, the evolution of the background on the orbit and over time, alongside the efficiency of background suppression on actual data (on the ground and on board) may reduce the overall efficiency in detecting this GRB population, compared to the simulations.
Likewise, even if significant effort has been put into making a realistic background simulator, accurately simulating the background is challenging, thus it could be underestimated in our simulations.
Therefore, the results of these simulations should be interpreted cautiously as they might lean towards an overly optimistic outlook.
The smaller effective area of ECLAIRs compared to that of \textit{Swift}/\gls{bat} should typically result in an overall lower detection efficiency.
It is worth emphasising that these results are not indicative of the expected rate of identified high-z GRBs with ECLAIRs, nor are  they  intended for direct comparison with observational results from past missions. 
Rather, their purpose is to estimate ECLAIRs' potential to detect GRBs at high redshift values, 
while considering the aforementioned caveats, and recalling that the detection of a GRB does not imply the identification of their redshift.
Moreover, the evolution of GRB progenitor populations with redshift cannot be ruled out.
Although the current study does not directly address this aspect, it aims to provide further insight by highlighting the potential biases involved in the detection and identification of high-z GRBs.

In Section \ref{sec:t90_evolution_with_z}, we showed the presence of detection biases in the GRB duration measurements as a function of redshift.
For some GRBs, such as GRB 090424 and GRB 090516A, the measured value of $\mathrm{T^{src}_{90}}$ decreases with increasing redshift (categorised as decreasing) because a part of the low-flux emission is lost within the noise.
On the other hand, the effect of this bias is not evident in other bursts like GRB 140311A, which display an almost non-decreasing $\mathrm{T^{src}_{90}}$ value (categorised as uniform).

GRBs with a lower $\mathrm{z_{meas}}$ are more likely to belong to the former category, while those with a higher $\mathrm{z_{meas}}$ are more likely to belong to the latter (see Figure \ref{fig:normalised_t90_ev} and Table \ref{tab:classification_GRBs_t90_ev}).
This is mainly because the low-flux emission of those with higher $\mathrm{z_{meas}}$ may have been lost for the detection by \textit{Swift}/\gls{bat}, and thus not present in the light curves used as input of our simulations.

Similarly to many other GRBs in the \texttt{Decreasing $\mathrm{T_{90}}$} class, the $\mathrm{T^{src}_{90}}$ value for GRB 090424 reaches a plateau (at $\mathrm{z_{sim}} > 3$) after an abrupt decrease (see Figure \ref{fig:T90_ev_3GRBs}).
This plateau resembles the curves exhibited by the GRBs in the \texttt{Uniform $\mathrm{T_{90}}$} class.
It is likely that many GRBs in this class do not show a significant decrease in $\mathrm{T^{src}_{90}}$ because the low-flux emission coming from these GRBs was not detected originally by \textit{Swift}/\gls{bat}.
If GRB 090424 had been detected at z$\sim$5, as for GRB 140311A, the low-flux emission would probably not have been detected, leading us to put it into the \texttt{Uniform} class since the $\mathrm{T^{src}_{90}}$ value would have remained constant throughout the simulations.

All of these pieces of evidence show the instrumental bias in the measurement of the GRB duration as a function of their redshift, in agreement with the results from \citet{Littlejohns_2013}. 
This effect may partially explain why the measured $\mathrm{T_{90}}$ values of currently known high-z GRBs tend to be lower than expected (see Figure \ref{fig:T90vsRedshiftSample}).
However, drawing definitive conclusions in this regard is complicated, primarily due to the limited size of the available sample.
The bias is highly dependent on the light curve shape, which in turn relies on the intrinsic characteristics of the GRB internal jet physics.
Therefore, discerning between scenarios where low-flux emissions remain undetected and those where the entire emission is captured proves to be challenging.

It is important to note that this bias is not specific to the $\mathrm{T_{90}}$ parameter alone, but to the GRB duration in general. One could naively think that the $\mathrm{T_{50}}$ would be less affected by this bias as its interval might not encompass the low-flux emission that is buried in noise at higher redshift values, unlike the $\mathrm{T_{90}}$ interval. However, the results from the simulations indicate that the $\mathrm{T_{50}}$ parameter is also notably affected by this bias.

Such biases may inadvertently lead to erroneous associations, potentially misclassifying core-collapse GRBs with a short measured duration as compact merger progenitor events. One should be cautious in using the GRB duration alone to infer the nature of their progenitors. 
Likewise, this effect may impact the reconstruction of other source properties.
For instance the derived fluence will be affected by the GRB duration, even if the derived flux was not highly impacted, as it is mainly the low-flux emission that is lost within the noise. 
This can directly impact the derivation of the $\mathrm{E_{iso}}$ value (see Eq. \ref{eq:eiso}), thus also impacting analyses related to energy and luminosity relations, which can be used as cosmological probes.
It is also likely that there will be a lack of useful \gls{grm} data for most high-z GRBs. Consequently, we will have to rely on the energy range covered by ECLAIRs to assess the burst energetics.
Due to its narrow energy range, similar to that of Swift/BAT, it is likely that our estimates on the prompt emission energetics of high-z GRBs may be biased.
Therefore, despite its complexity, it is essential to pursue a comprehensive understanding of the various biases (e.g. ECLAIRs detection, \gls{svom} observing strategy) that will affect the GRB populations observed by \gls{svom}.
A similar study, but including the potential detection with \gls{grm} could contribute to unravelling these biases.


\clearpage
\addcontentsline{toc}{section}{References}
\bibliographystyle{aa}

{\bibliography{references}}

\begin{thebibliography}{111}
\expandafter\ifx\csname natexlab\endcsname\relax\def\natexlab#1{#1}\fi

\bibitem[{Abbott {et~al.}(2017)}]{abbott2017ligo}
Abbott, B. {et~al.} 2017, Astrophys. J, 848

\bibitem[{Aghanim {et~al.}(2018)Aghanim, Akrami, Ashdown, Aumont, Baccigalupi,
  Ballardini, Banday, Barreiro, Bartolo, \& et~al.}]{planck_2018_results}
Aghanim, N., Akrami, Y., Ashdown, M., {et~al.} 2018, Astronomy \& Astrophysics,
  641, A6

\bibitem[{Antier-Farfar(2016)}]{antierfarfar_thesis_2016}
Antier-Farfar, S. 2016, Theses, {Universit{\'e} Paris Saclay}

\bibitem[{Arcier(2022)}]{arcier2022detection}
Arcier, B. 2022, PhD thesis, Universit{\'e} Paul Sabatier-Toulouse III

\bibitem[{Arcier {et~al.}(2020)Arcier, Atteia, Godet, Mate, Guillot, Dagoneau,
  Rodriguez, Götz, Schanne, \& Bernardini}]{Arcier_2020}
Arcier, B., Atteia, J.~L., Godet, O., {et~al.} 2020, Astrophysics and Space
  Science, 365

\bibitem[{Atteia {et~al.}(2022)Atteia, Cordier, \& Wei}]{atteia2022svom}
Atteia, J.-L., Cordier, B., \& Wei, J. 2022, International Journal of Modern
  Physics D, 31, 2230008

\bibitem[{Atteia {et~al.}(2017)Atteia, Heussaff, Dezalay, Klotz, Turpin,
  Tsvetkova, Frederiks, Zolnierowski, Daigne, \& Mochkovitch}]{Atteia_2017}
Atteia, J.-L., Heussaff, V., Dezalay, J.-P., {et~al.} 2017, The Astrophysical
  Journal, 837, 119

\bibitem[{{Band} {et~al.}(1993){Band}, {Matteson}, {Ford}, {Schaefer},
  {Palmer}, {Teegarden}, {Cline}, {Briggs}, {Paciesas}, {Pendleton}, {Fishman},
  {Kouveliotou}, {Meegan}, {Wilson}, \& {Lestrade}}]{Band}
{Band}, D., {Matteson}, J., {Ford}, L., {et~al.} 1993, The Astrophysical
  Journal, 413, 281

\bibitem[{Barthelmy {et~al.}(2005)Barthelmy, Barbier, Cummings, Fenimore,
  Gehrels, Hullinger, Krimm, Markwardt, Palmer, Parsons, \&
  et~al.}]{Barthelmy_2005}
Barthelmy, S.~D., Barbier, L.~M., Cummings, J.~R., {et~al.} 2005, Space Science
  Reviews, 120, 143–164

\bibitem[{Basa {et~al.}(2022)Basa, Lee, Dolon, Watson, Floriot, Atteia, Butler,
  Dornic, Lombardo, Ronayette, {et~al.}}]{basa2022colibri}
Basa, S., Lee, W.~H., Dolon, F., {et~al.} 2022, in Ground-based and Airborne
  Telescopes IX, Vol. 12182, SPIE, 602--613

\bibitem[{Bloom {et~al.}(2001)Bloom, Frail, \& Sari}]{bloom2001prompt}
Bloom, J.~S., Frail, D.~A., \& Sari, R. 2001, The Astronomical Journal, 121,
  2879

\bibitem[{Bromberg {et~al.}(2013)Bromberg, Nakar, Piran, \&
  Sari}]{Bromberg_2013}
Bromberg, O., Nakar, E., Piran, T., \& Sari, R. 2013, The Astrophysical
  Journal, 764, 179

\bibitem[{Bromm \& Loeb(2006)}]{bromm2006high}
Bromm, V. \& Loeb, A. 2006, The Astrophysical Journal, 642, 382

\bibitem[{Campana {et~al.}(2012)Campana, Salvaterra, Melandri, Vergani, Covino,
  D’Avanzo, Fugazza, Ghisellini, Sbarufatti, \& Tagliaferri}]{campana2012x}
Campana, S., Salvaterra, R., Melandri, A., {et~al.} 2012, Monthly Notices of
  the Royal Astronomical Society, 421, 1697

\bibitem[{Cannizzo {et~al.}(2009)Cannizzo, Barthelmy, Beardmore, Burrows,
  Curran, Evans, Gehrels, Gronwall, Guidorzi, Holland, Kennea, Krimm, Mao,
  Margutti, Marshall, O'Brien, Page, Romano, Rowlinson, Siegel, Tagliaferri, \&
  Vetere}]{Canizzo_GCN9223}
Cannizzo, J.~K., Barthelmy, S.~D., Beardmore, A.~P., {et~al.} 2009, GCN 9223

\bibitem[{Chaoul {et~al.}(2018)Chaoul, Mousset, \& Quenouille}]{chaoul2018svom}
Chaoul, L., Mousset, V., \& Quenouille, G. 2018, in 2018 SpaceOps Conference,
  2509

\bibitem[{Chornock {et~al.}(2014)Chornock, Fox, Tanvir, \&
  Berger}]{Chornock_GCN15966}
Chornock, R., Fox, D.~B., Tanvir, N.~R., \& Berger, E. 2014, GCN 15965

\bibitem[{Chornock {et~al.}(2009)Chornock, Perley, Cenko, \&
  Bloom}]{Chornock_GCN9243}
Chornock, R., Perley, D.~A., Cenko, S.~B., \& Bloom, J.~S. 2009, GCN 9243

\bibitem[{Christie {et~al.}(2009)Christie, de~Ugarte~Postigo, \&
  Natusch}]{Christie_GCN9396}
Christie, G.~W., de~Ugarte~Postigo, A., \& Natusch, T. 2009, GCN 9396

\bibitem[{Churazov {et~al.}(2007)Churazov, Sunyaev, {Revnivtsev, M.}, {Sazonov,
  S.}, {Molkov, S.}, {Grebenev, S.}, {Winkler, C.}, {Parmar, A.}, {Bazzano,
  A.}, {Falanga, M.}, {Gros, A.}, {Lebrun, F.}, {Natalucci, L.}, {Ubertini,
  P.}, {Roques, J.-P.}, {Bouchet, L.}, {Jourdain, E.}, {Kn\"odlseder, J.},
  {Diehl, R.}, {Budtz-Jorgensen, C.}, {Brandt, S.}, {Lund, N.}, {Westergaard,
  N. J.}, {Neronov, A.}, {T\"urler, M.}, {Chernyakova, M.}, {Walter, R.},
  {Produit, N.}, {Mowlavi, N.}, {Mas-Hesse, J. M.}, {Domingo, A.}, {Gehrels,
  N.}, {Kuulkers, E.}, {Kretschmar, P.}, \& {Schmidt, M.}}]{Churazov}
Churazov, E., Sunyaev, R., {Revnivtsev, M.}, {et~al.} 2007, A\&A, 467, 529

\bibitem[{Connaughton(2009)}]{Connaughton_GCN9230}
Connaughton, V. 2009, GCN 9230

\bibitem[{Corre {et~al.}(2018)Corre, Basa, Klotz, Watson, Ageron, Ambert,
  {\'A}ngeles, Atteia, Blanc, Boulade, {et~al.}}]{corre2018end}
Corre, D., Basa, S., Klotz, A., {et~al.} 2018, in Modeling, Systems
  Engineering, and Project Management for Astronomy VIII, Vol. 10705, SPIE,
  565--580

\bibitem[{Costa {et~al.}(1997)Costa, Frontera, Heise, Feroci,
  in~{\textquotesingle}t~Zand, Fiore, Cinti, Fiume, Nicastro, Orlandini,
  Palazzi, Rapisarda{\#}, Zavattini, Jager, Parmar, Owens, Molendi, Cusumano,
  Maccarone, Giarrusso, Coletta, Antonelli, Giommi, Muller, Piro, \&
  Butler}]{Costa_1997_Xray_afterglow}
Costa, E., Frontera, F., Heise, J., {et~al.} 1997, Nature, 387, 783

\bibitem[{Cucchiara {et~al.}(2011)Cucchiara, Levan, Fox, Tanvir, Ukwatta,
  Berger, Krühler, Yoldaş, Wu, Toma, \& et~al.}]{Cucchiara_2011}
Cucchiara, A., Levan, A.~J., Fox, D.~B., {et~al.} 2011, The Astrophysical
  Journal, 736, 7

\bibitem[{Dagoneau(2020)}]{dagoneau_thesis}
Dagoneau, N. 2020, Theses, {Universit{\'e} Paris-Saclay}

\bibitem[{Dagoneau {et~al.}(2020)Dagoneau, Schanne, Atteia, G{\"o}tz, \&
  Cordier}]{dagoneau2020ultra}
Dagoneau, N., Schanne, S., Atteia, J.-L., G{\"o}tz, D., \& Cordier, B. 2020,
  Experimental Astronomy, 50, 91

\bibitem[{Dagoneau {et~al.}(2018)Dagoneau, Schanne, Gros, \&
  Cordier}]{dagoneau2018ultralongGRBs}
Dagoneau, N., Schanne, S., Gros, A., \& Cordier, B. 2018, Detection capability
  of Ultra-Long Gamma-Ray Bursts with the ECLAIRs telescope aboard the SVOM
  mission

\bibitem[{Dagoneau {et~al.}(2021)Dagoneau, Schanne, Rodriguez, Atteia, \&
  Cordier}]{Dagoneau21}
Dagoneau, N., Schanne, S., Rodriguez, J., Atteia, J.-L., \& Cordier, B. 2021,
  A\&A, 645, A18

\bibitem[{Dai {et~al.}(2017)Dai, Daigne, \& M{\'e}sz{\'a}ros}]{dai2017theory}
Dai, Z., Daigne, F., \& M{\'e}sz{\'a}ros, P. 2017, Space Science Reviews, 212,
  409

\bibitem[{D'Avanzo {et~al.}(2014)D'Avanzo, D'Elia, Piranomonte, Melandri,
  Malesani, Pursimo, Baglio, \& Andreoni}]{Avanzo_GCN15964}
D'Avanzo, P., D'Elia, V., Piranomonte, S., {et~al.} 2014, GCN 15964

\bibitem[{de~Ugarte~Postigo {et~al.}(2009)de~Ugarte~Postigo, Gorosabel,
  Malesani, Fynbo, \& Levan}]{Deugarte_GCN9383}
de~Ugarte~Postigo, A., Gorosabel, J., Malesani, D., Fynbo, J., \& Levan, A.~J.
  2009, GCN 9383

\bibitem[{Dong {et~al.}(2009)}]{Dong_2009_GRM}
Dong, N.~R. {et~al.} 2009, in Science in China Series G: physics, Mechanics
  Astronomy, Vol.~53, 40

\bibitem[{Eichler {et~al.}(1989)Eichler, Livio, Piran, \&
  Schramm}]{Eichler1989}
Eichler, D., Livio, M., Piran, T., \& Schramm, D.~N. 1989, Nature, 340,
  126–128

\bibitem[{Fan {et~al.}(2020)Fan, Zou, Wei, Qiu, Gao, Wang, Yang, Zhang, Li,
  Zhao, {et~al.}}]{fan2020visible}
Fan, X., Zou, G., Wei, J., {et~al.} 2020, in Space Telescopes and
  Instrumentation 2020: Optical, Infrared, and Millimeter Wave, Vol. 11443,
  International Society for Optics and Photonics, 114430Q

\bibitem[{Fynbo {et~al.}(2009)Fynbo, Jakobsson, Prochaska, Malesani, Ledoux,
  de~Ugarte~Postigo, Nardini, Vreeswijk, Wiersema, Hjorth, \&
  et~al.}]{Fynbo_2009}
Fynbo, J. P.~U., Jakobsson, P., Prochaska, J.~X., {et~al.} 2009, The
  Astrophysical Journal Supplement Series, 185, 526–573

\bibitem[{Galama {et~al.}(1999)Galama, Vreeswijk, Van~Paradijs, Kouveliotou,
  Augusteijn, Patat, Heise, Groot, Wijers, Pian, {et~al.}}]{galama1999possible}
Galama, T., Vreeswijk, P., Van~Paradijs, J., {et~al.} 1999, Astronomy and
  Astrophysics Supplement Series, 138, 465

\bibitem[{Gardner {et~al.}(2006)Gardner, Mather, Clampin, Doyon, Greenhouse,
  Hammel, Hutchings, Jakobsen, Lilly, Long, {et~al.}}]{gardner2006james}
Gardner, J.~P., Mather, J.~C., Clampin, M., {et~al.} 2006, Space Science
  Reviews, 123, 485

\bibitem[{Gehrels(2004)}]{Gehrels_2004}
Gehrels, N. 2004, AIP Conference Proceedings

\bibitem[{Gehrels {et~al.}(2006)Gehrels, Norris, Barthelmy, Granot, Kaneko,
  Kouveliotou, Markwardt, M{\'e}sz{\'a}ros, Nakar, Nousek,
  {et~al.}}]{gehrels2006new}
Gehrels, N., Norris, J., Barthelmy, S., {et~al.} 2006, Nature, 444, 1044

\bibitem[{Godet {et~al.}(2022)Godet, Atteia, Amoros, Roger, Bouchet, Dezalay,
  Yassine, Arcier, Bordon, Lacombe, Lecomte, Llamas, Maestre, Marty, Papais,
  Ramon, Verdeil, Waegebaert, Schanne, Dagoneau, Chateau, Kestener, Provost,
  Tahoulan, Cordier, Tourrette, Daly, Triou, Coleiro, Goldwurm, Lachaud,
  Guillemot, Mouret, Charmeau, Perraud, Bousquet, Cervantes, Gasc, Pasquier,
  Perrin, Ruellan, Simonella, \& Yadallee}]{godet_2022_calibration}
Godet, O., Atteia, J.-L., Amoros, C., {et~al.} 2022, in Space Telescopes and
  Instrumentation 2022: Ultraviolet to Gamma Ray, ed. J.-W.~A. den Herder,
  S.~Nikzad, \& K.~Nakazawa, Vol. 12181, International Society for Optics and
  Photonics (SPIE), 121815O

\bibitem[{Godet {et~al.}(2014)Godet, Nasser, Atteia, Cordier, Mandrou, Barret,
  Triou, Pons, Amoros, Bordon, \& et~al.}]{Godet_2014}
Godet, O., Nasser, G., Atteia, J.-., {et~al.} 2014, Space Telescopes and
  Instrumentation 2014: Ultraviolet to Gamma Ray

\bibitem[{Godet {et~al.}(2009)Godet, Sizun, Barret, Mandrou, Cordier, Schanne,
  \& Remou{\'e}}]{godet2009monte}
Godet, O., Sizun, P., Barret, D., {et~al.} 2009, Nuclear Instruments and
  Methods in Physics Research Section A: Accelerators, Spectrometers, Detectors
  and Associated Equipment, 603, 365

\bibitem[{Goldstein {et~al.}(2012)Goldstein, Burgess, Preece, Briggs, Guiriec,
  van~der Horst, Connaughton, Wilson-Hodge, Paciesas, Meegan, \&
  et~al.}]{Goldstein_2012}
Goldstein, A., Burgess, J.~M., Preece, R.~D., {et~al.} 2012, The Astrophysical
  Journal Supplement Series, 199, 19

\bibitem[{Gottlieb {et~al.}(2023)Gottlieb, Metzger, Quataert, Issa, \&
  Foucart}]{gottlieb2023unified}
Gottlieb, O., Metzger, B.~D., Quataert, E., Issa, D., \& Foucart, F. 2023, A
  Unified Picture of Short and Long Gamma-ray Bursts from Compact Binary
  Mergers

\bibitem[{Greiner {et~al.}(2015)Greiner, Fox, Schady, Krühler, Trenti, Cikota,
  Bolmer, Elliott, Delvaux, Perna, \& et~al.}]{Greiner_2015}
Greiner, J., Fox, D.~B., Schady, P., {et~al.} 2015, The Astrophysical Journal,
  809, 76

\bibitem[{Greiner {et~al.}(2009)Greiner, Krühler, McBreen, Ajello, Giannios,
  Schwarz, Savaglio, Yolda{\c{s}}, Clemens, Stefanescu, Sala, Bertoldi,
  Szokoly, \& Klose}]{Greiner_2009}
Greiner, J., Krühler, T., McBreen, S., {et~al.} 2009, The Astrophysical
  Journal, 693, 1912

\bibitem[{Götz {et~al.}(2016)Götz, , {et~al.}}]{Gotz16}
Götz, D., , {et~al.} 2016, in Proc. of SPIE, Vol. 9905, Space Telescopes and
  Instrumentation 2016: Ultraviolet to Gamma Ray, 99054L

\bibitem[{Han {et~al.}(2021)Han, Xiao, Zhang, Turpin, Xin, Wu, Cai, Dong,
  Huang, Kang, {et~al.}}]{han2021automatic}
Han, X., Xiao, Y., Zhang, P., {et~al.} 2021, Publications of the Astronomical
  Society of the Pacific, 133, 065001

\bibitem[{Hjorth {et~al.}(2003)Hjorth, Sollerman, Møller, Fynbo, Woosley,
  Kouveliotou, Tanvir, Greiner, Andersen, Castro-Tirado, \&
  et~al.}]{Hjorth_2003}
Hjorth, J., Sollerman, J., Møller, P., {et~al.} 2003, Nature, 423, 847–850

\bibitem[{Ivezi{\'c} {et~al.}(2019)Ivezi{\'c}, Kahn, Tyson, Abel, Acosta,
  Allsman, Alonso, AlSayyad, Anderson, Andrew, {et~al.}}]{ivezic2019lsst}
Ivezi{\'c}, {\v{Z}}., Kahn, S.~M., Tyson, J.~A., {et~al.} 2019, The
  Astrophysical Journal, 873, 111

\bibitem[{{Kaneko} {et~al.}(2006){Kaneko}, {Preece}, {Briggs}, {Paciesas},
  {Meegan}, \& {Band}}]{Kaneko}
{Kaneko}, Y., {Preece}, R.~D., {Briggs}, M.~S., {et~al.} 2006, The
  Astrophysical Journal, 166, 298

\bibitem[{{Klebesadel} {et~al.}(1973){Klebesadel}, {Strong}, \&
  {Olson}}]{Klebesadel1973_def_GRBs}
{Klebesadel}, R.~W., {Strong}, I.~B., \& {Olson}, R.~A. 1973, apjl, 182, L85

\bibitem[{Krimm {et~al.}(2014)Krimm, Barthelmy, Baumgartner, Cummings, Gehrels,
  Lien, Markwardt, Palmer, Racusin, T.Sakamoto, Stamatikos, Tueller, \&
  Ukwatta}]{Krim_GCN15962}
Krimm, H.~A., Barthelmy, S.~D., Baumgartner, W.~H., {et~al.} 2014, GCN 15962

\bibitem[{Kumar \& Zhang(2015)}]{kumar2015physics}
Kumar, P. \& Zhang, B. 2015, Physics Reports, 561, 1

\bibitem[{Lacombe {et~al.}(2013)Lacombe, Nasser, Amoros, Atteia, Barret,
  Billot, Cordier, Gevin, Godet, Gonzalez, {et~al.}}]{lacombe2013development}
Lacombe, K., Nasser, G., Amoros, C., {et~al.} 2013, Nuclear Instruments and
  Methods in Physics Research Section A: Accelerators, Spectrometers, Detectors
  and Associated Equipment, 732, 122

\bibitem[{Laskar {et~al.}(2023)Laskar, Alexander, Margutti, Eftekhari,
  Chornock, Berger, Cendes, Duerr, Perley, Ravasio, {et~al.}}]{laskar2023radio}
Laskar, T., Alexander, K.~D., Margutti, R., {et~al.} 2023, The Astrophysical
  Journal Letters, 946, L23

\bibitem[{Levan {et~al.}(2023)Levan, Gompertz, Malesani, Tanvir, Burns,
  Salvaterra, Ackley, Lamb, Fynbo, Schneider, {et~al.}}]{levan2023grb}
Levan, A., Gompertz, B., Malesani, D., {et~al.} 2023, GRB Coordinates Network,
  33569, 1

\bibitem[{Lidz {et~al.}(2021)Lidz, Chang, Mas-Ribas, \& Sun}]{lidz2021future}
Lidz, A., Chang, T.-C., Mas-Ribas, L., \& Sun, G. 2021, The Astrophysical
  Journal, 917, 58

\bibitem[{Lien {et~al.}(2016)Lien, Sakamoto, Barthelmy, Baumgartner, Cannizzo,
  Chen, Collins, Cummings, Gehrels, Krimm, \& et~al.}]{Lien_2016_BAT_catalog}
Lien, A., Sakamoto, T., Barthelmy, S.~D., {et~al.} 2016, The Astrophysical
  Journal, 829, 7

\bibitem[{Littlejohns {et~al.}(2013)Littlejohns, Tanvir, Willingale, Evans,
  O’Brien, \& Levan}]{Littlejohns_2013}
Littlejohns, O.~M., Tanvir, N.~R., Willingale, R., {et~al.} 2013, Monthly
  Notices of the Royal Astronomical Society, 436, 3640–3655

\bibitem[{Macpherson {et~al.}(2013)Macpherson, Coward, \&
  Zadnik}]{macpherson2013potential}
Macpherson, D., Coward, D., \& Zadnik, M. 2013, The Astrophysical Journal, 779,
  73

\bibitem[{Mate(2021)}]{mate2021development}
Mate, S. 2021, PhD thesis, Universit{\'e} Paul Sabatier-Toulouse III

\bibitem[{Mate {et~al.}(2019)Mate, Bouchet, Atteia, Claret, Cordier, Dagoneau,
  Godet, Gros, Schanne, \& Triou}]{Sujay-bkg}
Mate, S., Bouchet, L., Atteia, J.-L., {et~al.} 2019, Experimental Astronomy,
  48, 171–198

\bibitem[{McBreen(2009)}]{McBreen_GCN9413}
McBreen, S. 2009, GCN 9413

\bibitem[{Meegan {et~al.}(2009)Meegan, Lichti, Bhat, Bissaldi, Briggs,
  Connaughton, Diehl, Fishman, Greiner, Hoover, \& et~al.}]{Meegan_2009}
Meegan, C., Lichti, G., Bhat, P.~N., {et~al.} 2009, The Astrophysical Journal,
  702, 791–804

\bibitem[{Melandri {et~al.}(2019)Melandri, Malesani, Izzo, Japelj, Vergani,
  Schady, Sagu{\'e}s~Carracedo, de~Ugarte~Postigo, Anderson, Barbarino,
  {et~al.}}]{melandri2019grb}
Melandri, A., Malesani, D., Izzo, L., {et~al.} 2019, Monthly Notices of the
  Royal Astronomical Society, 490, 5366

\bibitem[{M{\'e}sz{\'a}ros \& Rees(1997)}]{meszaros1997optical}
M{\'e}sz{\'a}ros, P. \& Rees, M.~J. 1997, The Astrophysical Journal, 476, 232

\bibitem[{Moss {et~al.}(2021)Moss, Lien, Guiriec, B., \&
  T}]{moss2021_tip_iceberg}
Moss, M., Lien, A., Guiriec, S., B., C.~S., \& T, S. 2021, Swift/BAT
  Tip-of-The-Iceberg Effects

\bibitem[{Nousek {et~al.}(2006)Nousek, Kouveliotou, Grupe, Page, Granot,
  Ramirez‐Ruiz, Patel, Burrows, Mangano, Barthelmy, \& et~al.}]{Nousek_2006}
Nousek, J.~A., Kouveliotou, C., Grupe, D., {et~al.} 2006, The Astrophysical
  Journal, 642, 389–400

\bibitem[{Nouvel de~la Fl{\`e}che {et~al.}(2023)Nouvel de~la Fl{\`e}che,
  Atteia, Boy, Klotz, Langlois, Larrieu, Mathon, Valentin, Ambert, Clemens,
  {et~al.}}]{nouvel2023cagire}
Nouvel de~la Fl{\`e}che, A., Atteia, J.-L., Boy, J., {et~al.} 2023,
  Experimental Astronomy, 1

\bibitem[{Palmerio \& Daigne(2021)}]{palmerio2021constraining}
Palmerio, J.~T. \& Daigne, F. 2021, Astronomy \& Astrophysics, 649, A166

\bibitem[{Pe’er(2015)}]{pe2015physics}
Pe’er, A. 2015, Advances in Astronomy, 2015

\bibitem[{Piran(2005)}]{piran2005physics}
Piran, T. 2005, Reviews of Modern Physics, 76, 1143

\bibitem[{Prochaska {et~al.}(2008)Prochaska, Dessauges-Zavadsky, Ramirez-Ruiz,
  \& Chen}]{prochaska2008survey}
Prochaska, J.~X., Dessauges-Zavadsky, M., Ramirez-Ruiz, E., \& Chen, H.-W.
  2008, The Astrophysical Journal, 685, 344

\bibitem[{Rastinejad {et~al.}(2022)Rastinejad, Gompertz, Levan, Fong, Nicholl,
  Lamb, Malesani, Nugent, Oates, Tanvir, {et~al.}}]{rastinejad2022kilonova}
Rastinejad, J.~C., Gompertz, B.~P., Levan, A.~J., {et~al.} 2022, Nature, 612,
  223

\bibitem[{Remoué {et~al.}(2010)Remoué, Barret, Godet, \& Mandrou}]{Remoue10}
Remoué, N., Barret, D., Godet, O., \& Mandrou, P. 2010, Nuclear Instruments
  and Methods in Physics Research A, 618, 199

\bibitem[{Robertson {et~al.}(2023)Robertson, Tacchella, Johnson, Hainline,
  Whitler, Eisenstein, Endsley, Rieke, Stark, Alberts,
  {et~al.}}]{robertson2023identification}
Robertson, B., Tacchella, S., Johnson, B., {et~al.} 2023, Nature Astronomy, 7,
  611

\bibitem[{Rossi {et~al.}(2009)Rossi, Afonso, \& J.Greiner}]{Rossi_GCN9382}
Rossi, A., Afonso, P., \& J.Greiner. 2009, GCN 9382

\bibitem[{Rowlinson {et~al.}(2009)Rowlinson, Beardmore, Evans, Guidorzi,
  Kennea, O'Brien, Page, Palmer, Romano, \& Sakamoto}]{Palmer_GCN9374}
Rowlinson, B.~A., Beardmore, A.~P., Evans, P.~A., {et~al.} 2009, GCN 9374

\bibitem[{Saccardi {et~al.}(2023)Saccardi, Vergani, De~Cia, D’Elia, Heintz,
  Izzo, Palmerio, Petitjean, Rossi, de~Ugarte~Postigo,
  {et~al.}}]{Saccardi_2023_ISM}
Saccardi, A., Vergani, S., De~Cia, A., {et~al.} 2023, Astronomy \&
  Astrophysics, 671, A84

\bibitem[{Salvaterra {et~al.}(2009)Salvaterra, Valle, Campana, Chincarini,
  Covino, D’avanzo, Fern{\'a}ndez-Soto, Guidorzi, Mannucci, Margutti,
  {et~al.}}]{salvaterra2009grb}
Salvaterra, R., Valle, M.~D., Campana, S., {et~al.} 2009, Nature, 461, 1258

\bibitem[{Sazonov {et~al.}(2007)Sazonov, Churazov, Sunyaev, \&
  Revnivtsev}]{sazonov_2007}
Sazonov, S., Churazov, E., Sunyaev, R., \& Revnivtsev, M. 2007, Monthly Notices
  of the Royal Astronomical Society, 377, 1726

\bibitem[{Scargle(1998)}]{Scargle_1998}
Scargle, J.~D. 1998, The Astrophysical Journal, 504, 405

\bibitem[{Schanne {et~al.}(2013)Schanne, Le~Provost, Kestener, Gros, Cortial,
  Götz, Sizun, Château, \& Cordier}]{schanne_2013_trigger_detailed}
Schanne, S., Le~Provost, H., Kestener, P., {et~al.} 2013, in 2013 IEEE Nuclear
  Science Symposium and Medical Imaging Conference (2013 NSS/MIC), 1--5

\bibitem[{Schulze {et~al.}(2015)Schulze, Chapman, Hjorth, Levan, Jakobsson,
  Bj{\"o}rnsson, Perley, Kr{\"u}hler, Gorosabel, Tanvir, de~Ugarte~Postigo,
  Fynbo, Milvang-Jensen, M{\o}ller, \& Watson}]{Schulze_2015}
Schulze, S., Chapman, R., Hjorth, J., {et~al.} 2015, The Astrophysical Journal,
  808, 73

\bibitem[{Sparre {et~al.}(2014)Sparre, Hartoog, Kr{\"u}hler, Fynbo, Watson,
  Wiersema, d'Elia, Zafar, Afonso, Covino, {et~al.}}]{sparre2014metallicity}
Sparre, M., Hartoog, O., Kr{\"u}hler, T., {et~al.} 2014, The Astrophysical
  Journal, 785, 150

\bibitem[{Sun {et~al.}(2023)Sun, Wang, Yang, Zhang, Xiong, Yin, Liu, Li, Xue,
  Yan, {et~al.}}]{sun2023magnetar}
Sun, H., Wang, C.-W., Yang, J., {et~al.} 2023, arXiv preprint arXiv:2307.05689

\bibitem[{Tanvir {et~al.}(2019)Tanvir, Grindlay, Berger, Metzger, Gezari,
  Ivezic, Jencson, Kasliwal, Kutyrev, Macleod,
  {et~al.}}]{tanvir2019TSO_early_universe}
Tanvir, N., Grindlay, J., Berger, E., {et~al.} 2019, Bulletin of the the AAS,
  51

\bibitem[{Tanvir {et~al.}(2013)Tanvir, Levan, Fruchter, Hjorth, Hounsell,
  Wiersema, \& Tunnicliffe}]{tanvir2013kilonova}
Tanvir, N., Levan, A.~J., Fruchter, A., {et~al.} 2013, Nature, 500, 547

\bibitem[{Tanvir {et~al.}(2009)Tanvir, Fox, Levan, Berger, Wiersema, Fynbo,
  Cucchiara, Krühler, Gehrels, Bloom, \& et~al.}]{Tanvir_2009}
Tanvir, N.~R., Fox, D.~B., Levan, A.~J., {et~al.} 2009, Nature, 461,
  1254–1257

\bibitem[{Tanvir {et~al.}(2018)Tanvir, Laskar, Levan, Perley, Zabl, Fynbo,
  Rhoads, Cenko, Greiner, Wiersema, \& et~al.}]{Tanvir_2018}
Tanvir, N.~R., Laskar, T., Levan, A.~J., {et~al.} 2018, The Astrophysical
  Journal, 865, 107

\bibitem[{Tanvir {et~al.}(2014)Tanvir, Levan, Wiersema, \&
  Cucchiara}]{Tanvir_GCN15961}
Tanvir, N.~R., Levan, A.~J., Wiersema, K., \& Cucchiara, A. 2014, GCN 15961

\bibitem[{Toma {et~al.}(2016)Toma, Yoon, \& Bromm}]{toma2016_GRB_pop3}
Toma, K., Yoon, S.-C., \& Bromm, V. 2016, Space Science Reviews, 202, 159

\bibitem[{Totani {et~al.}(2006)Totani, Kawai, Kosugi, Aoki, Yamada, Iye, Ohta,
  \& Hattori}]{totani2006reionization_implications_050904}
Totani, T., Kawai, N., Kosugi, G., {et~al.} 2006, Publications of the
  Astronomical Society of Japan, 58, 485

\bibitem[{Valle {et~al.}(2006)Valle, Chincarini, Panagia, Tagliaferri,
  Malesani, Testa, Fugazza, Campana, Covino, Mangano,
  {et~al.}}]{valle2006enigmatic}
Valle, M.~D., Chincarini, G., Panagia, N., {et~al.} 2006, Nature, 444, 1050

\bibitem[{Van~Paradijs {et~al.}(1997)Van~Paradijs, Groot, Galama, Kouveliotou,
  Strom, Telting, Rutten, Fishman, Meegan, Pettini,
  {et~al.}}]{vanParadijs1997_Optical_afterglow}
Van~Paradijs, J., Groot, P., Galama, T., {et~al.} 1997, Nature, 386, 686

\bibitem[{Vedrenne \& Atteia(2009)}]{vedrenne_and_atteia_2009gamma}
Vedrenne, G. \& Atteia, J.-L. 2009, Gamma-ray bursts: The brightest explosions
  in the universe (Springer Science \& Business Media)

\bibitem[{Vergani {et~al.}(2009)Vergani, Petitjean, Ledoux, Vreeswijk, Smette,
  \& Meurs}]{vergani2009statistics}
Vergani, S.~D., Petitjean, P., Ledoux, C., {et~al.} 2009, Astronomy \&
  Astrophysics, 503, 771

\bibitem[{von Kienlin {et~al.}(2020)von Kienlin, Meegan, Paciesas, Bhat,
  Bissaldi, Briggs, Burns, Cleveland, Gibby, Giles, Goldstein, Hamburg, Hui,
  Kocevski, Mailyan, Malacaria, Poolakkil, Preece, Roberts, Veres, \&
  Wilson-Hodge}]{von_Kienlin_2020}
von Kienlin, A., Meegan, C.~A., Paciesas, W.~S., {et~al.} 2020, The
  Astrophysical Journal, 893, 46

\bibitem[{Wei {et~al.}(2016)Wei, Cordier, Antier, Antilogus, Atteia, Bajat,
  Basa, Beckmann, Bernardini, Boissier, {et~al.}}]{wei2016deep}
Wei, J., Cordier, B., Antier, S., {et~al.} 2016, arXiv preprint
  arXiv:1610.06892

\bibitem[{Wen {et~al.}(2021)Wen, Sun, He, Song, Wang, Zhou, Dong, Liu, Liu, Li,
  {et~al.}}]{wen2021calibrationGRM}
Wen, X., Sun, J., He, J., {et~al.} 2021, Nuclear Instruments and Methods in
  Physics Research Section A: Accelerators, Spectrometers, Detectors and
  Associated Equipment, 1003, 165301

\bibitem[{Wiersema {et~al.}(2009)Wiersema, de~Ugarte~Postigo, \&
  Levan}]{Wiersema_GCN9250}
Wiersema, K., de~Ugarte~Postigo, A., \& Levan, A. 2009, GCN 9250

\bibitem[{Woosley \& Bloom(2006)}]{Woosley_2006}
Woosley, S. \& Bloom, J. 2006, Annual Review of Astronomy and Astrophysics, 44,
  507–556

\bibitem[{Woosley(1993)}]{woosley1993gamma}
Woosley, S.~E. 1993, Astrophysical Journal, Part 1 (ISSN 0004-637X), vol. 405,
  no. 1, p. 273-277., 405, 273

\bibitem[{Yoshida {et~al.}(2003)Yoshida, Sokasian, Hernquist, \&
  Springel}]{yoshida2003early}
Yoshida, N., Sokasian, A., Hernquist, L., \& Springel, V. 2003, The
  Astrophysical Journal Letters, 591, L1

\bibitem[{Yu {et~al.}(2016)Yu, Preece, Greiner, Bhat, Bissaldi, Briggs,
  Cleveland, Connaughton, Goldstein, von Kienlin, {et~al.}}]{Yu_TR_spectra}
Yu, H.-F., Preece, R.~D., Greiner, J., {et~al.} 2016, A\&A, 588, A135

\bibitem[{Zhang(2007)}]{zhang2007gamma}
Zhang, B. 2007, Chinese Journal of Astronomy and Astrophysics, 7, 1

\bibitem[{Zhang(2018)}]{zhang2018physics}
Zhang, B. 2018, The physics of gamma-ray bursts (Cambridge University Press)

\bibitem[{{Zhang} {et~al.}(2006){Zhang}, {Fan}, {Dyks}, {Kobayashi},
  {M{\'e}sz{\'a}ros}, {Burrows}, {Nousek}, \&
  {Gehrels}}]{zhang_2006_afterglow_LC}
{Zhang}, B., {Fan}, Y.~Z., {Dyks}, J., {et~al.} 2006, apj, 642, 354

\bibitem[{Zhang {et~al.}(2013)Zhang, Fan, Shao, \& Wei}]{Zhang_2013}
Zhang, F.-W., Fan, Y.-Z., Shao, L., \& Wei, D.-M. 2013, The Astrophysical
  Journal Letters, 778, L11

\bibitem[{Zitouni {et~al.}(2014)Zitouni, Guessoum, \& Azzam}]{Zitouni_2014}
Zitouni, H., Guessoum, N., \& Azzam, W.~J. 2014, Astrophysics and Space
  Science, 351, 267–279

\end{thebibliography}

\clearpage
\begin{appendices}

\setcounter{table}{0}
\setcounter{figure}{0}
\renewcommand{\thetable}{\Alph{section}.\arabic{table}}
\renewcommand{\thefigure}{\Alph{section}.\arabic{figure}}

\begin{table*}[t]

\section{GRB sample table}
\label{appendix:grb_table}

\caption{\textit{Swift}/\gls{bat} GRBs analysed in this study.}

\begin{tabular}{lccc cc lccc cc lccc} 
\label{tab:GRB_sample}
GRB  & $\mathrm{z_{meas}}$ &  $\mathrm{T_{90}^{obs}}$ (s) & \textit{Fermi} & & &  GRB  & $\mathrm{z_{meas}}$ &  $\mathrm{T_{90}^{obs}}$ (s) & \textit{Fermi} & & &  GRB  & $\mathrm{z_{meas}}$ &  $\mathrm{T_{90}^{obs}}$ (s) & \textit{Fermi}\\  
\cmidrule{1-4}\cmidrule(lr){7-10} \cmidrule{13-16}  
&&&&&&&&&&&&\\ 

050319 & 3.24 & 151.6 &  &&& 090927 & 1.37 & 2.2 & Yes &&& 140304A & 5.28 & 14.8 & Yes	\\  
 050502B & 5.20 & 17.7 &  &&& 091020 & 1.71 & 38.9 & Yes &&& 140311A & 4.95 & 70.5 & 	\\  
 050505 & 4.27 & 58.9 &  &&& 091024 & 1.09 & 112.3 & Yes &&& 140419A & 3.96 & 80.1 & 	\\  
 050730 & 3.97 & 154.6 &  &&& 091109A & 3.29 & 48.0 &  &&& 140423A & 3.26 & 134.1 & Yes	\\  
 050814 & 5.30 & 142.9 &  &&& 091127 & 0.49 & 7.0 & Yes &&& 140428A & 4.70 & 17.4 & 	\\  
 050904 & 6.29 & 181.6 &  &&& 091208B & 1.06 & 14.8 & Yes &&& 140506A & 0.89 & 111.1 & Yes	\\  
 050908 & 3.35 & 18.3 &  &&& 100219A & 4.62 & 27.6 &  &&& 140512A & 0.72 & 154.1 & Yes	\\  
 060115 & 3.53 & 139.1 &  &&& 100302A & 4.81 & 17.9 &  &&& 140515A & 6.32 & 23.4 & 	\\  
 060116 & 6.60 & 104.8 &  &&& 100513A & 4.77 & 83.5 &  &&& 140518A & 4.71 & 60.5 & 	\\  
 060206 & 4.05 & 7.6 &  &&& 100615A & 1.40 & 38.8 & Yes &&& 140614A & 4.23 & 77.4 & 	\\  
 060210 & 3.91 & 288.0 &  &&& 100728A & 1.57 & 193.4 & Yes &&& 140703A & 3.14 & 68.6 & Yes	\\  
 060223A & 4.41 & 11.3 &  &&& 100728B & 2.45 & 12.1 & Yes &&& 140907A & 1.21 & 80.0 & Yes	\\  
 060510B & 4.90 & 262.9 &  &&& 100814A & 1.44 & 177.3 & Yes &&& 141004A & 0.57 & 3.9 & Yes	\\  
 060522 & 5.11 & 69.1 &  &&& 100816A & 0.80 & 2.9 & Yes &&& 141026A & 3.35 & 139.5 & 	\\  
 060526 & 3.21 & 298.0 &  &&& 100906A & 1.73 & 114.6 & Yes &&& 141220A & 1.32 & 7.2 & Yes	\\  
 060605 & 3.76 & 79.8 &  &&& 110128A & 2.34 & 14.2 & Yes &&& 141225A & 0.92 & 86.1 & Yes	\\  
 060607A & 3.08 & 103.0 &  &&& 110213A & 1.46 & 48.0 & Yes &&& 150120A & 0.46 & 1.2 & Yes	\\  
 060707 & 3.43 & 66.6 &  &&& 110731A & 2.83 & 40.9 & Yes &&& 150301B & 1.52 & 17.1 & Yes	\\  
 060906 & 3.69 & 44.6 &  &&& 110818A & 3.36 & 102.8 & Yes &&& 150314A & 1.76 & 14.8 & Yes	\\  
 060926 & 3.21 & 8.8 &  &&& 111008A & 4.99 & 62.8 &  &&& 150403A & 2.06 & 37.3 & Yes	\\  
 060927 & 5.60 & 22.4 &  &&& 111107A & 2.89 & 31.1 & Yes &&& 150413A & 3.17 & 243.6 & 	\\  
 061110B & 3.44 & 135.2 &  &&& 111123A & 3.15 & 290.0 &  &&& 150727A & 0.31 & 88.0 & Yes	\\  
 061222B & 3.35 & 37.2 &  &&& 111228A & 0.71 & 101.2 & Yes &&& 150821A & 0.76 & 168.9 & Yes	\\  
 070721B & 3.63 & 336.9 &  &&& 120118B & 2.94 & 20.3 & Yes &&& 151027B & 4.06 & 80.0 & 	\\  
 071025 & 5.30 & 241.3 &  &&& 120119A & 1.73 & 68.0 & Yes &&& 160203A & 3.52 & 17.4 & 	\\  
 080129 & 4.39 & 50.2 &  &&& 120326A & 1.80 & 69.5 & Yes &&& 160327A & 4.99 & 33.7 & 	\\  
 080516 & 3.20 & 5.8 &  &&& 120521C & 6.00 & 27.1 &  &&& 160624A & 0.48 & 0.2 & Yes	\\  
 080607 & 3.04 & 79.0 &  &&& 120712A & 4.11 & 14.8 & Yes &&& 160804A & 0.74 & 152.7 & Yes	\\  
 080804 & 2.20 & 37.9 & Yes &&& 120729A & 0.80 & 93.9 & Yes &&& 161014A & 2.82 & 23.0 & Yes	\\  
 080810 & 3.35 & 107.7 & Yes &&& 120802A & 3.80 & 50.3 &  &&& 161017A & 2.01 & 217.0 & Yes	\\  
 080905A & 0.12 & 1.0 & Yes &&& 120805A & 3.10 & 48.0 &  &&& 161117A & 1.55 & 125.7 & Yes	\\  
 080905B & 2.37 & 120.9 & Yes &&& 120811C & 2.67 & 24.3 & Yes &&& 161129A & 0.65 & 35.5 & Yes	\\  
 080913 & 6.73 & 7.5 &  &&& 120907A & 0.97 & 6.1 & Yes &&& 170113A & 1.97 & 20.3 & Yes	\\  
 080916A & 0.69 & 61.3 & Yes &&& 120909A & 3.93 & 220.6 & Yes &&& 170202A & 3.65 & 37.8 & 	\\  
 080928 & 1.69 & 233.7 & Yes &&& 120922A & 3.10 & 168.2 & Yes &&& 170405A & 3.51 & 165.3 & Yes	\\  
 081008 & 1.97 & 187.8 & Yes &&& 120923A & 7.80 & 26.1 &  &&& 170607A & 0.56 & 320.0 & Yes	\\  
 081028A & 3.04 & 284.4 &  &&& 121128A & 2.20 & 23.4 & Yes &&& 170705A & 2.01 & 223.2 & Yes	\\  
 081029 & 3.85 & 275.1 &  &&& 121201A & 3.38 & 38.0 &  &&& 170903A & 0.89 & 27.7 & Yes	\\  
 081121 & 2.51 & 17.5 & Yes &&& 121211A & 1.02 & 182.7 & Yes &&& 171222A & 2.41 & 173.9 & Yes	\\  
 081221 & 2.26 & 33.9 & Yes &&& 130215A & 0.59 & 66.2 & Yes &&& 180205A & 1.41 & 15.5 & Yes	\\  
 081222 & 2.77 & 33.0 & Yes &&& 130408A & 3.76 & 4.2 &  &&& 180314A & 1.45 & 50.5 & Yes	\\  
 090102 & 1.55 & 28.3 & Yes &&& 130420A & 1.30 & 121.1 & Yes &&& 180620B & 1.12 & 224.0 & Yes	\\  
 090113 & 1.75 & 9.1 & Yes &&& 130427A & 0.34 & 244.3 & Yes &&& 180720B & 0.65 & 108.4 & Yes	\\  
 090205 & 4.67 & 8.8 &  &&& 130514A & 3.60 & 214.2 &  &&& 180728A & 0.12 & 8.7 & 	\\  
 090313 & 3.37 & 83.0 &  &&& 130606A & 5.91 & 276.7 &  &&& 181010A & 1.39 & 15.6 & 	\\  
 090423 & 8.05 & 10.3 & Yes &&& 130610A & 2.09 & 47.7 & Yes &&& 181020A & 2.94 & 238.0 & 	\\  
 090424 & 0.54 & 49.5 & Yes &&& 130612A & 2.01 & 4.0 & Yes &&& 190114A & 3.38 & 67.1 & 	\\  
 090429B & 9.40 & 5.6 &  &&& 130925A & 0.35 & 160.3 & Yes &&& 190114C & 0.42 & 361.5 & 	\\  
 090510 & 0.90 & 5.7 & Yes &&& 131004A & 0.71 & 1.5 & Yes &&& 190324A & 1.17 & 22.8 & 	\\  
 090516A & 4.10 & 181.0 & Yes &&& 131117A & 4.11 & 10.9 &  &&& 190719C & 2.47 & 185.8 & 	\\  
 090519 & 3.88 & 58.0 & Yes &&& 131227A & 5.30 & 18.0 &  &&& 190829A & 0.08 & 56.9 & 	\\  
 090618 & 0.54 & 113.3 & Yes &&& 140114A & 3.00 & 139.9 &  &&& 191004B & 3.50 & 300.1 & 	\\  
 090715B & 3.00 & 266.4 &  &&& 140206A & 2.74 & 94.2 & Yes &&& 191011A & 1.72 & 7.4 & 	\\  
 090926B & 1.24 & 99.3 & Yes &&& 140213A & 1.21 & 59.9 & Yes &&& 200829A & 1.25 & 13.1 & 	\\  
 \end{tabular} 
 \caption*{The rightmost column indicates whether a corresponding \textit{Fermi}/\gls{gbm} detection was found and therefore its derived spectrum used.}
\end{table*}
\setcounter{table}{0}
\setcounter{figure}{0}
\section{Use of time resolved spectral models}
\label{appendix:TR_spectra}

We used the \gls{ti} spectral data available in the \textit{Fermi}/\gls{gbm} and \textit{Swift}/\gls{bat} catalogues for our GRB sample in this study. However, it is important to note that \gls{ti} data lack temporal spectral evolution details, unlike \gls{tr} spectral data. This simplification may have an impact on the results, likely more prominent when redshifting the GRBs.

To address this, we referred to the GRB TR spectral catalogue derived by \citet{Yu_TR_spectra}, which covers the brightest GRBs detected by \textit{Fermi}/\gls{gbm} during its four first years of operation. This catalogue includes three GRBs from our sample (GRB 090424, GRB 110731A, GRB 120119A). 
Simulations were performed using \gls{tr} spectra for these GRBs and their results were compared to those with TI spectra in order to assess the impact of using the TI-spectra simplification.
The \gls{ti} spectral model was used for the intervals for which no specific spectral model was given (intervals outside of the time interval accounted for in \citealt{Yu_TR_spectra}, or intervals with no model defined in the catalogue).

\begin{figure}
    \centering
    \includegraphics[width = \hsize]{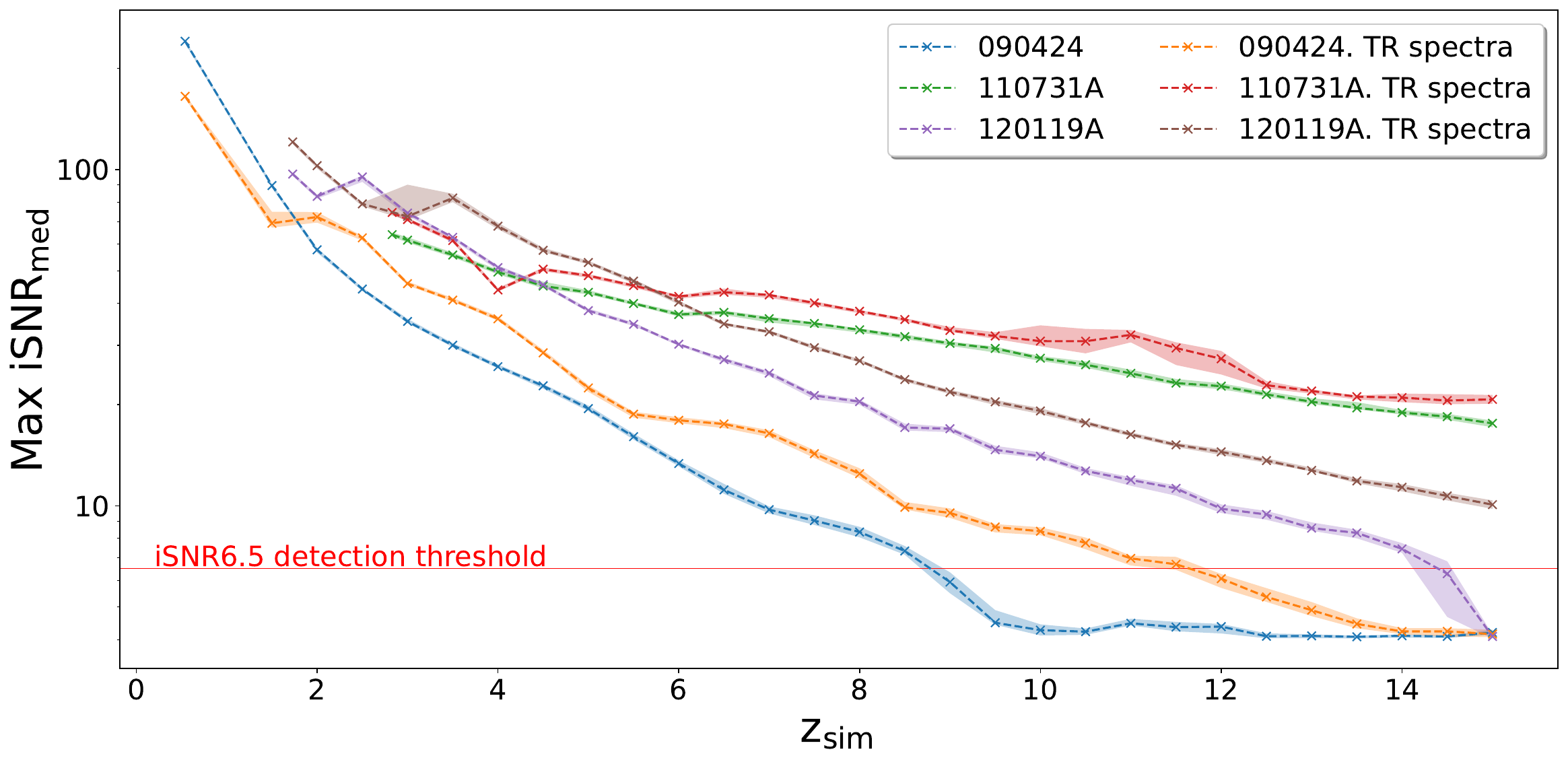}
    \caption{Evolution of the maximum \gls{isnr}$\mathrm{_{med}}$ with increasing $\mathrm{z_{sim}}$ for GRB 090424, GRB 110731A, and GRB 120119A. The shaded area corresponds to the 90\% confidence interval for the median. The red horizontal line shows the \gls{isnr} threshold of 6.5.}
    \label{fig:iSNR_ev_TRspectra}
\end{figure}

\begin{figure}
    \centering
    \includegraphics[width = \hsize]{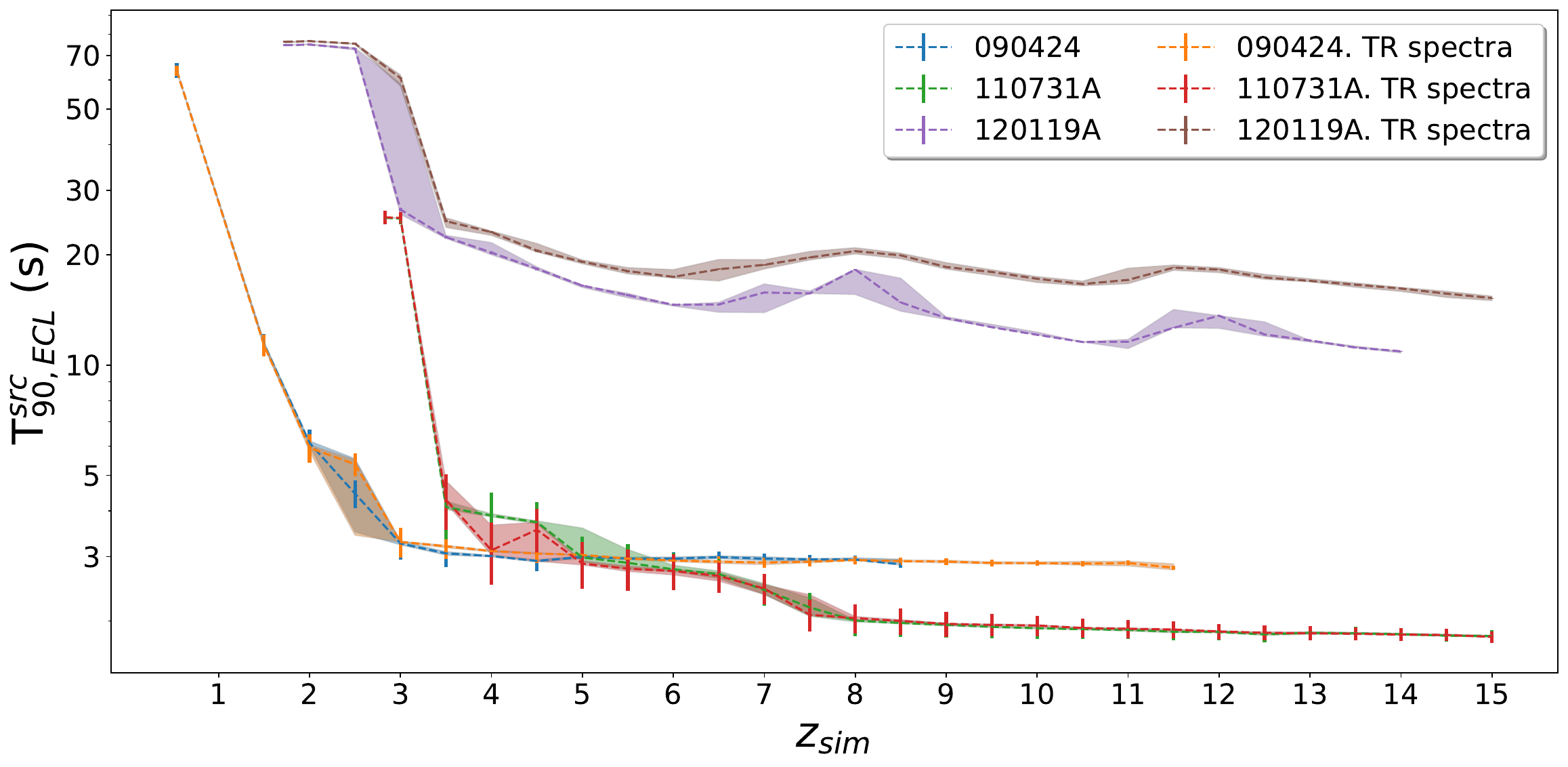}
    \caption{Similar to Figure \ref{fig:LC_evolution}, but for GRB 090424, GRB 110731A, and GRB 120119A and including also the \gls{tr} spectral model analysis along with the \gls{ti} spectral model for comparison.}
    \label{fig:T90_ev_TRspectra}
\end{figure}

Figures \ref{fig:iSNR_ev_TRspectra} and \ref{fig:T90_ev_TRspectra} present the evolution of the maximum \gls{isnr}$\mathrm{_{med}}$ and the $\mathrm{T_{90}^{src}}$ values, respectively, with increasing redshift, for both the \gls{ti} and \gls{tr} scenarios. 
One hundred trials were performed to obtain these results.

In Figure \ref{fig:iSNR_ev_TRspectra}, the \gls{isnr}$\mathrm{_{med}}$ curves for GRB 110731A show minor differences between the two scenarios. 
For GRB 090424 and GRB 120119A the disparities are more noticeable, though the general patterns remain similar.
Above $\mathrm{z_{sim}\sim2}$ for GRB 090424 and $\mathrm{z_{sim}\sim3}$ for GRB 120119A, the \gls{isnr} values slightly favour the \gls{tr} scenario, leading to a larger $\mathrm{z_{hor}}$ value.
In the case of GRB 090424, this leads to a noticeable difference in the $\mathrm{z_{hor}}$ value.

In Figure \ref{fig:T90_ev_TRspectra} the curves for GRB 090424 and GRB 110731A exhibit negligible differences between the two scenarios, and consistent with the derived confidence intervals and the uncertainties obtained from \texttt{battblocks}.
The curves corresponding to GRB 120119A exhibit slightly more disparities between the scenarios, generally with lower $\mathrm{T_{90}}$ values in the \gls{ti} scenario, although the general trends remain consistent.

Although this comparison is limited to only three GRBs and may not fully represent the entire sample, its purpose is to demonstrate that while there might be quantitative impacts on the results, the qualitative conclusions remain valid. 
We recall that there is a limited availability of \gls{tr} spectra, and hence the adoption of the \gls{ti} spectra simplification, assuming their potential caveats.
\setcounter{table}{0}
\setcounter{figure}{0}
\section{Potential bias in the analysis}
\label{appendix:LC_bias_in_t90}
The comparison of the $\mathrm{T_{90}^{obs}}$ median values obtained for each GRB in the sample at $\mathrm{z_{meas}}$ within ECLAIRs \gls{fcfov}, with the measured values from \textit{Swift}/\gls{bat} (hereafter $\mathrm{T_{90}^{ECL}}$ and $\mathrm{T_{90}^{BAT}}$, respectively), revealed a potential bias in the analysis. Specifically, for certain GRBs $\mathrm{T_{90}^{ECL}}$ was found to be significantly larger than $\mathrm{T_{90}^{BAT}}$. These GRBs exhibit a common characteristic in their light curves, characterised by a prolonged period over which the count rate remains nearly at the background noise level after the main peak (i.e. long low-flux tail emission).

\begin{figure}[h!]
    \centering
    \includegraphics[width = \hsize]{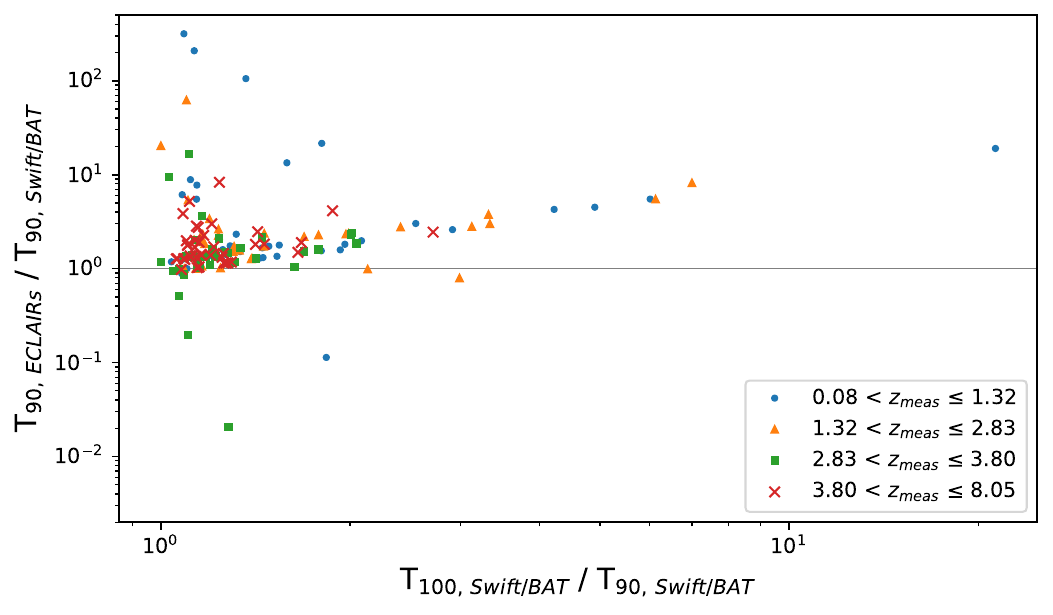}
    \caption{Comparison between the values of $\mathrm{T_{90}^{obs}}$ obtained from the simulations within ECLAIRs \gls{fcfov} and the measured values from \textit{Swift}/\gls{bat}, plotted against the $\mathrm{T_{100} / T_{90}}$ ratio from \textit{Swift}/\gls{bat} measurements.}
    \label{fig:bias_t90_t100_BAT_LC}
\end{figure}

Figure \ref{fig:bias_t90_t100_BAT_LC} illustrates the ratio $\mathrm{T_{90}^{ECL}/T_{90}^{BAT}}$ plotted against $\mathrm{T_{100}^{BAT} / T_{90}^{BAT}}$ ($\mathrm{T_{100}}$ being the time interval that encompasses 100\% of the burst counts).
The clear trend observed between these two ratios suggests that the inclusion of the long tail emission in the calculated $\mathrm{T_{90}^{ECL}}$ values contributes to the observed differences.

A potential explanation for this might be related to the background subtracted light curves from \textit{Swift}/\gls{bat} used as input for the simulations.
When the raw light curve exhibits a prolonged time interval with only a few GRB counts above the noise level, the background subtraction procedure may lead to the positive parts of the background noise fluctuations being interpreted as GRB photons, while the negative parts may be subtracted below the estimated background curve. In other words, during the background subtraction, only the positive fluctuations would be retained.

In addition, selecting the time interval and the binning applied to the light curve passed to the \texttt{battblocks} tool may influence the results. In this study, we specifically opted for 80\,ms binning and selected the events within 25\% before and after the GRB events.
This selection might introduce an additional impact on the results, potentially leading to a $\mathrm{T_{90}}$ that is more closely related to the initial $\mathrm{T_{100}^{BAT}}$ time interval.

However, it is important to note that this bias should not significantly affect the qualitative results concerning the evolution of $\mathrm{T_{90}}$ with redshift, as the input light curve was the same for all $\mathrm{z_{sim}}$ values. Additionally, the \gls{snr} computation is unlikely to be influenced since the maximum \gls{isnr} primarily depends on the intensity of the peak and not on the low-flux tail emission.
\setcounter{table}{0}
\setcounter{figure}{0}

\begin{figure*}
\section{Evolution of T90 for each GRB in the sample}
\label{appendix:t90_evolution}
     \centering
     \begin{subfigure}[b]{0.95\textwidth}
         \centering
         \includegraphics[width=\textwidth]{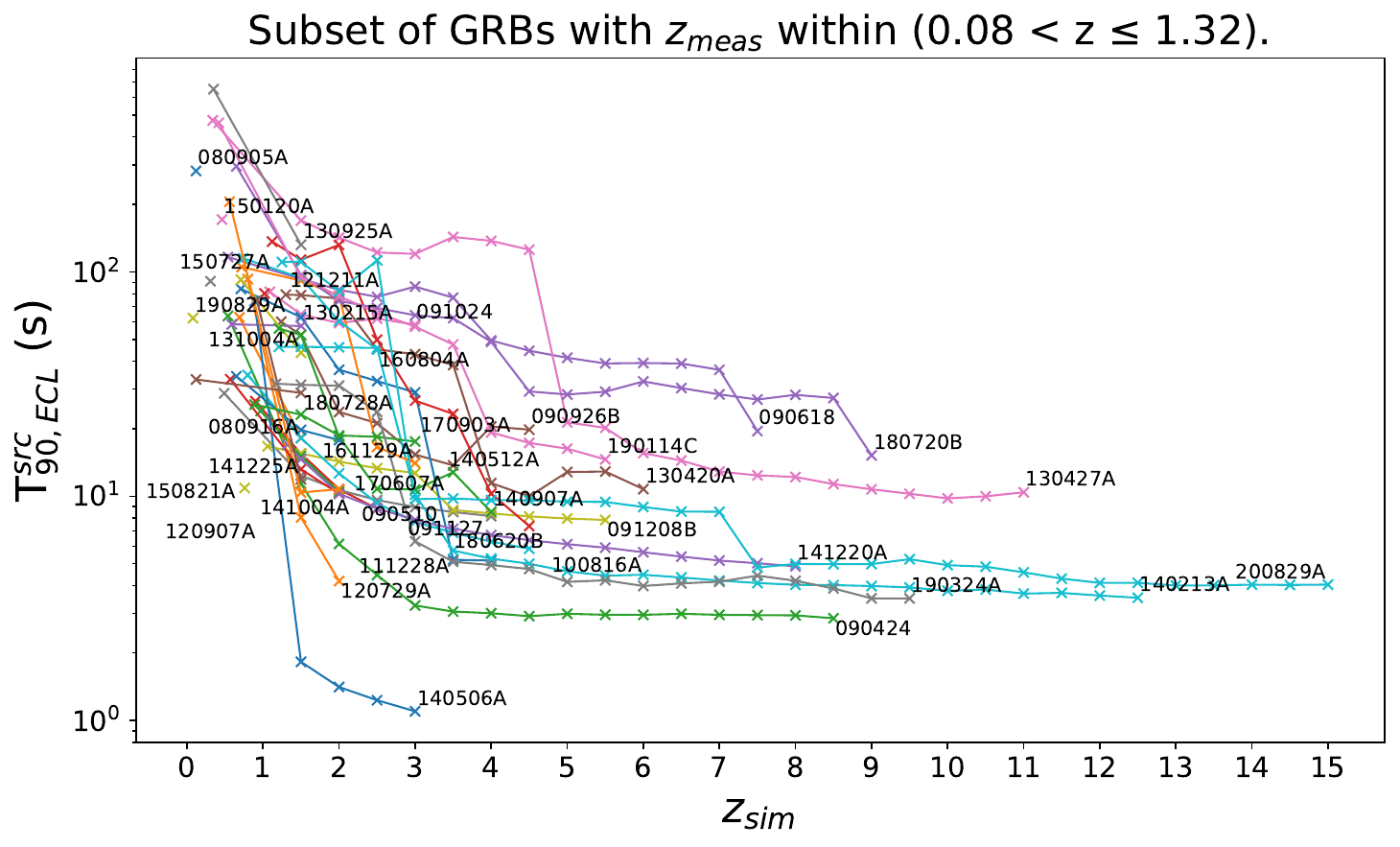}
         \caption{Lowest-z subset (0.078 > $\mathrm{z_{meas}}$ > 1.33)}
         \label{fig:T90_ev_lowest}
     \end{subfigure}
     \hfill
     \begin{subfigure}[b]{0.95\textwidth}
         \centering
         \includegraphics[width=\textwidth]{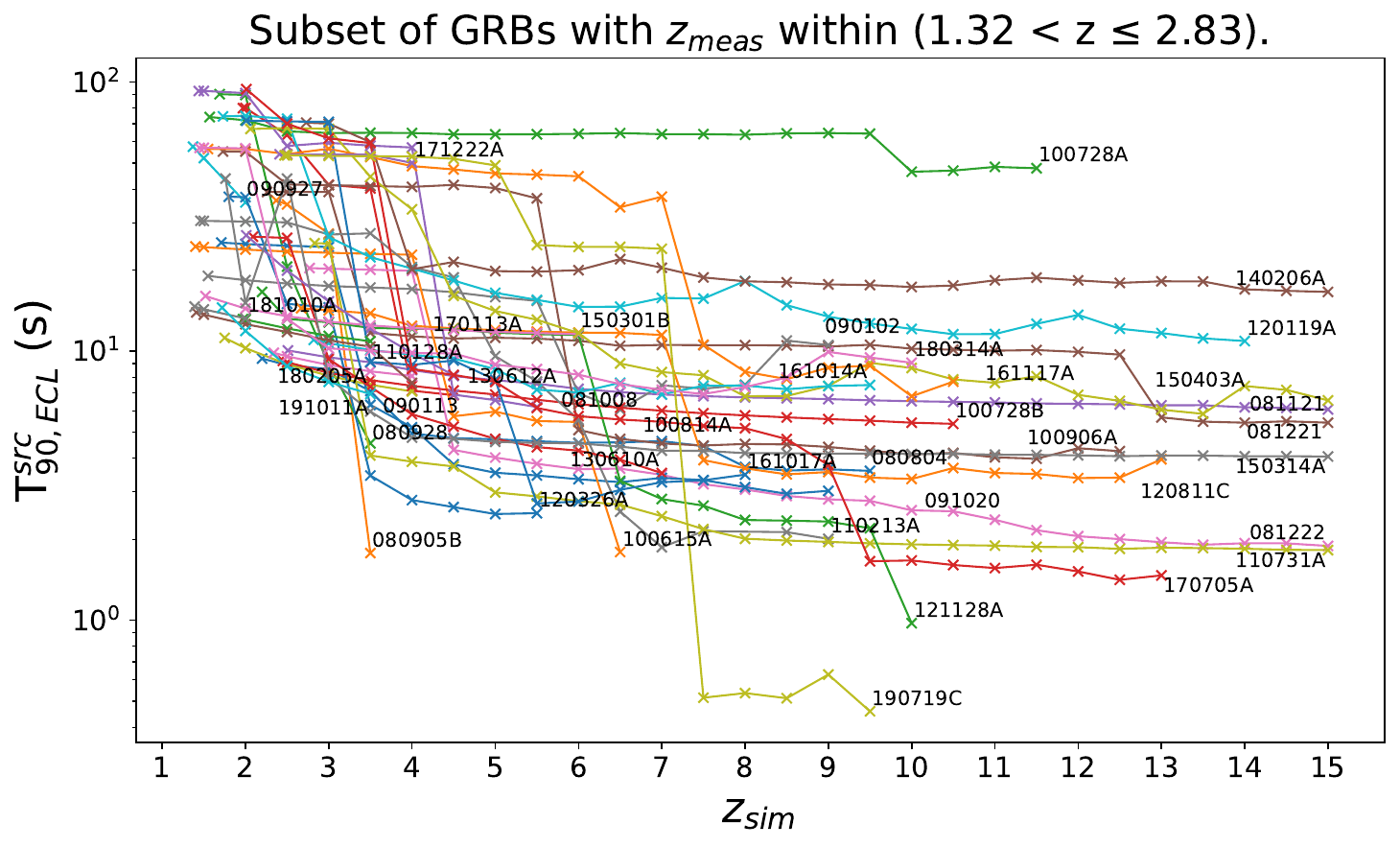}
         \caption{Low-z subset (1.33 > $\mathrm{z_{meas}}$ > 2.86)}
         \label{fig:T90_ev_low}
     \end{subfigure}
     \hfill
     \caption{Analogous to Figure \ref{fig:T90_ev_3GRBs} (evolution of $\mathrm{T_{90}^{src}}$ at each $\mathrm{z_{sim}}$ within the \gls{fcfov}), for all GRBs in the sample, but divided into four subsets according to their $\mathrm{z_{meas}}$.}
\end{figure*}
\begin{figure*}
     \ContinuedFloat
     \begin{subfigure}[b]{0.95\textwidth}
         \centering
         \includegraphics[width=\textwidth]{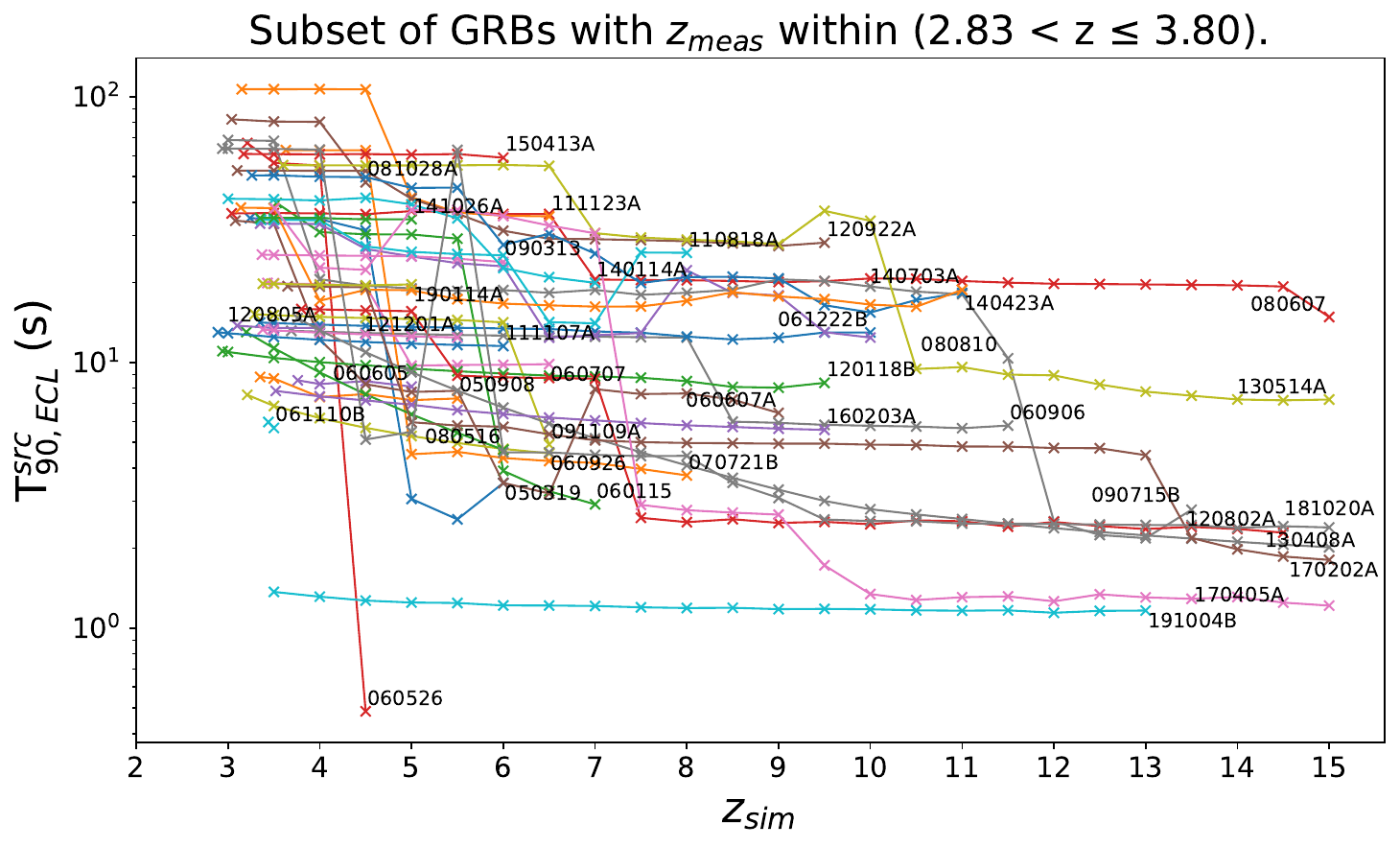}
         \caption{High-z subset (2.86 > $\mathrm{z_{meas}}$ > 3.83)}
         \label{fig:T90_ev_high}
     \end{subfigure}
    \begin{subfigure}[b]{0.95\textwidth}
         \centering
         \includegraphics[width=\textwidth]{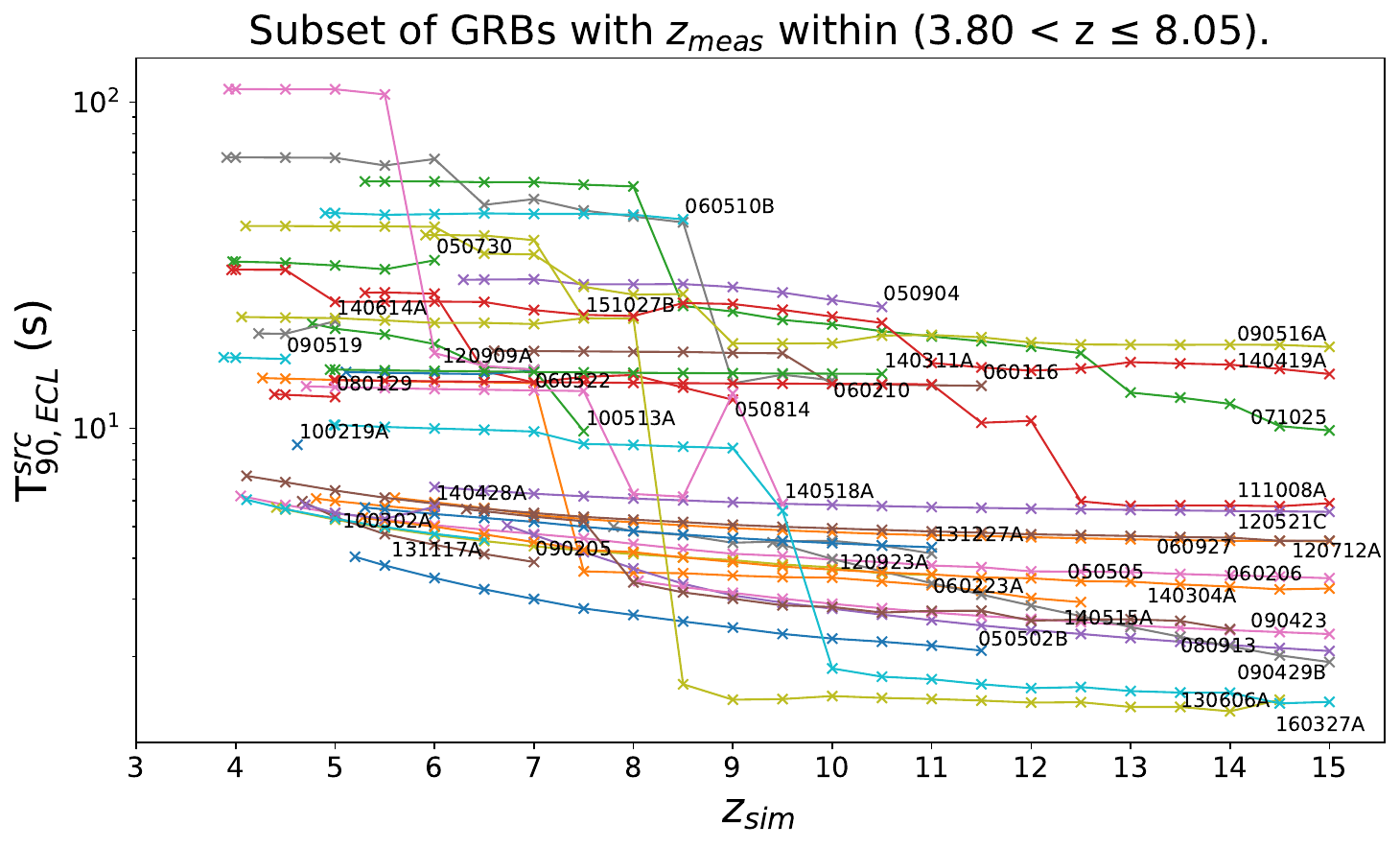}
         \caption{Highest-z subset (3.83 > $\mathrm{z_{meas}}$ > 9.4)}
         \label{fig:T90_ev_highest}
     \end{subfigure}
     \hfill
    \caption{continued.}
    \label{fig:t90_ev_all_sample}
\end{figure*}
\setcounter{table}{0}
\setcounter{figure}{0}
\begin{table*}[t]

\section{Table of resulting values from the simulations}
\label{appendix:results_table}

\caption{General results for each GRB in the sample.}

\begin{tabular}{lccccc cc lccccc} 
GRB & \gls{isnr}$\mathrm{_{med}}$ & $\mathrm{T_{90}^{obs}}$ (s) & FFoV (\%)& $\mathrm{z_{hor}}$ & $\mathrm{\frac{T_{90}^{src}(z_{hor})}{T_{90}^{src}(z_{meas})}}$ &&& GRB & \gls{isnr}$\mathrm{_{med}}$ & $\mathrm{T_{90}^{obs}}$ (s) & FFoV (\%)& $\mathrm{z_{hor}}$ & $\mathrm{\frac{T_{90}^{src}(z_{hor})}{T_{90}^{src}(z_{meas})}}$ \\  
\cmidrule{1-6}\cmidrule(lr){9-14}\\ 
\label{tab:results_table}

050319 & 20.1 & 147.9 & 75.6 & 6.0 & 0.10 &&&090927 & 15.7 & 136.4 & 73.6 & 2.5 & 0.62\ \\  
 050502B & 18.0 & 25.1 & 74.7 & 11.5 & 0.52 &&&091020 & 67.2 & 68.6 & 97.7 & 9.0 & 0.12\ \\  
 050505 & 35.1 & 75.2 & 87.8 & 13.5 & 0.21 &&&091024 & 45.8 & 170.5 & 95.7 & 3.0 & 0.71\ \\  
 050730 & 16.1 & 161.2 & 65.5 & 6.0 & 1.01 &&&091109A & 20.4 & 65.1 & 74.8 & 7.0 & 0.32\ \\  
 050814 & 19.9 & 164.1 & 74.1 & 9.0 & 0.47 &&&091127 & 175.3 & 42.9 & 100.0 & 4.0 & 0.28\ \\  
 050904 & 16.0 & 208.1 & 69.3 & 10.5 & 0.83 &&&091208B & 74.1 & 34.4 & 99.4 & 5.5 & 0.47\ \\  
 050908 & 21.1 & 38.3 & 76.7 & 6.5 & 0.83 &&&100219A & 7.1 & 50.1 & 17.1 & 5.0 & --\ \\  
 060115 & 23.0 & 180.3 & 77.3 & 7.0 & 0.07 &&&100302A & 13.6 & 35.4 & 55.6 & 7.0 & 0.92\ \\  
 060116 & 16.3 & 131.4 & 68.2 & 11.5 & 0.78 &&&100513A & 15.3 & 120.9 & 63.8 & 7.5 & 0.47\ \\  
 060206 & 45.4 & 31.4 & 95.7 & 15.0 & 0.56 &&&100615A & 83.8 & 58.7 & 98.1 & 6.5 & 0.07\ \\  
 060210 & 34.4 & 332.1 & 91.6 & 10.0 & 0.21 &&&100728A & 63.1 & 190.5 & 99.2 & 11.5 & 0.65\ \\  
 060223A & 29.5 & 31.0 & 87.0 & 12.5 & 0.62 &&&100728B & 42.8 & 31.7 & 95.5 & 11.5 & 0.58\ \\  
 060510B & 18.6 & 269.5 & 75.1 & 8.5 & 0.96 &&&100814A & 43.6 & 226.1 & 96.5 & 5.5 & 0.07\ \\  
 060522 & 11.3 & 90.9 & 49.9 & 7.0 & 1.01 &&&100816A & 42.6 & 62.5 & 98.1 & 4.5 & 0.17\ \\  
 060526 & 13.4 & 281.5 & 60.9 & 4.5 & 0.01 &&&100906A & 143.1 & 150.7 & 99.3 & 12.5 & 0.08\ \\  
 060605 & 10.3 & 40.7 & 41.8 & 5.0 & 0.95 &&&110128A & 21.0 & 32.9 & 76.2 & 4.5 & 0.82\ \\  
 060607A & 32.3 & 139.2 & 89.0 & 9.0 & 0.19 &&&110213A & 111.2 & 75.1 & 99.4 & 9.0 & 0.07\ \\  
 060707 & 16.4 & 88.1 & 70.6 & 6.5 & 0.50 &&&110731A & 64.1 & 96.5 & 98.6 & 15.0 & 0.07\ \\  
 060906 & 72.7 & 61.6 & 98.2 & 11.5 & 0.44 &&&110818A & 27.5 & 149.8 & 85.9 & 8.0 & 0.75\ \\  
 060926 & 37.8 & 31.7 & 92.6 & 7.0 & 0.60 &&&111008A & 105.0 & 84.2 & 99.4 & 15.0 & 0.42\ \\  
 060927 & 28.7 & 40.5 & 82.6 & 15.0 & 0.74 &&&111107A & 20.5 & 50.5 & 70.1 & 6.0 & 0.89\ \\  
 061110B & 6.5 & 26.5 & 10.8 & 3.5 & 0.95 &&&111123A & 31.0 & 442.3 & 88.6 & 6.5 & 0.33\ \\  
 061222B & 63.7 & 61.4 & 98.0 & 10.0 & 0.92 &&&111228A & 155.2 & 143.8 & 100.0 & 4.0 & 0.06\ \\  
 070721B & 19.5 & 291.1 & 72.4 & 8.0 & 0.06 &&&120118B & 50.2 & 43.4 & 95.5 & 10.5 & 0.76\ \\  
 071025 & 55.6 & 359.8 & 97.2 & 15.0 & 0.17 &&&120119A & 96.9 & 203.7 & 100.0 & 14.0 & 0.15\ \\  
 080129 & 8.4 & 68.5 & 26.5 & 5.0 & 0.98 &&&120326A & 67.1 & 105.0 & 98.1 & 5.5 & 0.07\ \\  
 080516 & 15.6 & 54.5 & 73.4 & 6.0 & 0.36 &&&120521C & 37.6 & 46.3 & 91.2 & 15.0 & 0.84\ \\  
 080607 & 198.9 & 146.8 & 100.0 & 15.0 & 0.41 &&&120712A & 30.7 & 36.5 & 93.9 & 15.0 & 0.63\ \\  
 080804 & 28.7 & 30.0 & 89.8 & 9.5 & 0.38 &&&120729A & 42.1 & 167.2 & 96.1 & 2.0 & 0.05\ \\  
 080810 & 25.5 & 144.8 & 88.7 & 10.0 & 0.37 &&&120802A & 56.0 & 76.2 & 96.3 & 14.5 & 0.14\ \\  
 080905A & 11.6 & 315.8 & 45.1 & 0.1 & -- &&&120805A & 9.8 & 56.3 & 42.4 & 4.0 & 0.98\ \\  
 080905B & 10.7 & 122.4 & 50.0 & 3.5 & 0.05 &&&120811C & 76.4 & 53.0 & 98.9 & 14.5 & 0.27\ \\  
 080913 & 14.2 & 39.0 & 58.8 & 15.0 & 0.41 &&&120907A & 15.5 & 47.2 & 75.6 & 2.0 & 0.44\ \\  
 080916A & 38.6 & 106.0 & 96.0 & 1.5 & 0.29 &&&120909A & 16.7 & 539.9 & 69.7 & 7.0 & 0.14\ \\  
 080928 & 27.0 & 242.5 & 86.2 & 3.5 & 0.05 &&&120922A & 46.5 & 216.0 & 97.6 & 10.0 & 0.54\ \\  
 081008 & 24.6 & 237.7 & 78.4 & 4.5 & 0.09 &&&120923A & 15.1 & 44.0 & 58.6 & 11.0 & 0.83\ \\  
 081028A & 18.0 & 331.8 & 72.3 & 4.5 & 0.58 &&&121128A & 97.9 & 53.1 & 99.3 & 10.0 & 0.06\ \\  
 081029 & 5.9 & -- & 7.0 & -- & -- &&&121201A & 20.0 & 58.2 & 72.6 & 5.5 & 0.93\ \\  
 081121 & 64.1 & 35.3 & 99.3 & 15.0 & 0.60 &&&121211A & 6.8 & 161.9 & 11.8 & 1.0 & --\ \\  
 081221 & 193.2 & 127.2 & 100.0 & 15.0 & 0.14 &&&130215A & 42.9 & 92.9 & 94.1 & 1.5 & 0.98\ \\  
 081222 & 55.2 & 76.8 & 99.2 & 15.0 & 0.09 &&&130408A & 28.0 & 69.8 & 94.7 & 15.0 & 0.14\ \\  
 090102 & 60.9 & 48.5 & 99.1 & 9.0 & 0.55 &&&130420A & 103.4 & 182.1 & 100.0 & 6.0 & 0.14\ \\  
 090113 & 26.0 & 30.8 & 82.1 & 4.0 & 0.63 &&&130427A & 475.5 & 634.8 & 100.0 & 11.0 & 0.02\ \\  
 090205 & 16.3 & 33.9 & 67.6 & 8.0 & 0.65 &&&130514A & 89.5 & 254.0 & 98.9 & 15.0 & 0.13\ \\  
 090313 & 22.1 & 111.1 & 79.6 & 6.0 & 0.94 &&&130606A & 22.8 & 270.3 & 79.8 & 14.5 & 0.04\ \\  
 090423 & 30.5 & 30.9 & 91.8 & 15.0 & 0.69 &&&130610A & 50.8 & 82.3 & 96.9 & 7.0 & 0.13\ \\  
 090424 & 237.5 & 98.0 & 100.0 & 8.5 & 0.04 &&&130612A & 27.2 & 81.1 & 87.8 & 4.5 & 0.30\ \\  
 090429B & 31.0 & 46.5 & 88.9 & 15.0 & 0.43 &&&130925A & 202.6 & 881.3 & 100.0 & 1.5 & 0.20\ \\  
 090510 & 16.6 & 50.4 & 74.7 & 2.5 & 0.33 &&&131004A & 25.1 & 157.9 & 89.3 & 1.5 & 0.47\ \\  
 090516A & 66.2 & 212.7 & 97.9 & 15.0 & 0.43 &&&131117A & 15.8 & 30.9 & 68.3 & 7.5 & 0.75\ \\  
 090519 & 8.6 & 80.6 & 23.2 & 4.5 & 0.99 &&&131227A & 21.9 & 36.1 & 73.2 & 11.0 & 0.76\ \\  
 090618 & 398.4 & 179.5 & 100.0 & 7.5 & 0.17 &&&140114A & 44.7 & 165.0 & 95.4 & 7.0 & 0.48\ \\  
 090715B & 63.6 & 274.5 & 97.5 & 13.5 & 0.04 &&&140206A & 94.8 & 263.6 & 100.0 & 15.0 & 0.23\ \\  
 090926B & 48.6 & 134.3 & 98.1 & 4.5 & 0.33 &&&140213A & 183.8 & 102.4 & 100.0 & 12.5 & 0.08\ \\  
 \end{tabular} 
\caption*{Note: The values of $\mathrm{T_{90}^{obs}}$, \gls{isnr} and the detectable Fraction of the FoV (FFoV) correspond to those at $\mathrm{z_{meas}}$. All values, except for FFoV, refer to the results within the \gls{fcfov}.}
\end{table*}

\begin{table*}[t]
\caption{continued.}
\begin{tabular}{lccccc cc lccccc} 
GRB & \gls{isnr}$\mathrm{_{med}}$ & $\mathrm{T_{90}^{obs}}$ (s) & FFoV (\%)& $\mathrm{z_{hor}}$ & $\mathrm{\frac{T_{90}^{src}(z_{hor})}{T_{90}^{src}(z_{meas})}}$ &&& GRB & \gls{isnr}$\mathrm{_{med}}$ & $\mathrm{T_{90}^{obs}}$ (s) & FFoV (\%)& $\mathrm{z_{hor}}$ & $\mathrm{\frac{T_{90}^{src}(z_{hor})}{T_{90}^{src}(z_{meas})}}$ \\  
\cmidrule{1-6}\cmidrule(lr){9-14}\\ 

140304A & 22.7 & 33.5 & 83.9 & 15.0 & 0.61 &&&160804A & 83.5 & 200.2 & 98.3 & 2.5 & 0.39\ \\  
 140311A & 27.1 & 90.1 & 80.5 & 10.5 & 0.97 &&&161014A & 30.9 & 42.1 & 92.8 & 10.0 & 0.68\ \\  
 140419A & 69.7 & 151.8 & 99.5 & 15.0 & 0.48 &&&161017A & 42.3 & 216.5 & 93.4 & 8.0 & 0.05\ \\  
 140423A & 33.8 & 215.2 & 96.0 & 11.0 & 0.36 &&&161117A & 136.6 & 143.8 & 99.5 & 10.5 & 0.14\ \\  
 140428A & 11.8 & 33.2 & 41.1 & 7.0 & 0.99 &&&161129A & 56.9 & 56.6 & 97.7 & 2.0 & 0.52\ \\  
 140506A & 40.9 & 140.0 & 94.5 & 3.0 & 0.01 &&&170113A & 30.6 & 39.1 & 84.8 & 4.0 & 0.83\ \\  
 140512A & 96.4 & 180.9 & 99.1 & 3.0 & 0.13 &&&170202A & 69.2 & 89.9 & 98.7 & 15.0 & 0.09\ \\  
 140515A & 25.7 & 41.4 & 84.0 & 15.0 & 0.43 &&&170405A & 25.5 & 171.3 & 90.7 & 15.0 & 0.03\ \\  
 140518A & 28.8 & 76.8 & 80.2 & 9.5 & 0.44 &&&170607A & 76.3 & 320.5 & 99.2 & 2.0 & 0.05\ \\  
 140614A & 8.9 & 102.2 & 34.1 & 5.0 & 1.09 &&&170705A & 95.1 & 283.4 & 99.3 & 13.0 & 0.02\ \\  
 140703A & 42.1 & 158.1 & 93.6 & 11.0 & 0.49 &&&170903A & 57.2 & 48.4 & 96.7 & 3.0 & 0.68\ \\  
 140907A & 58.1 & 123.8 & 97.9 & 4.0 & 0.15 &&&171222A & 22.8 & 183.8 & 76.5 & 4.0 & 0.93\ \\  
 141004A & 42.5 & 52.2 & 97.2 & 2.0 & 0.31 &&&180205A & 52.1 & 33.6 & 96.3 & 4.0 & 0.55\ \\  
 141026A & 20.3 & 151.9 & 78.6 & 5.0 & 0.99 &&&180314A & 82.3 & 139.4 & 100.0 & 10.0 & 0.16\ \\  
 141220A & 38.1 & 39.5 & 96.6 & 8.0 & 0.29 &&&180620B & 70.9 & 289.6 & 97.9 & 4.5 & 0.05\ \\  
 141225A & 15.8 & 143.4 & 69.3 & 1.5 & 0.16 &&&180720B & 382.7 & 488.1 & 100.0 & 9.0 & 0.05\ \\  
 150120A & 16.8 & 250.0 & 77.0 & 0.5 & -- &&&180728A & 431.7 & 37.1 & 100.0 & 1.5 & 0.87\ \\  
 150301B & 44.4 & 40.4 & 96.1 & 6.0 & 0.72 &&&181010A & 22.4 & 35.0 & 74.5 & 2.0 & 0.90\ \\  
 150314A & 112.2 & 120.9 & 100.0 & 15.0 & 0.09 &&&181020A & 88.4 & 251.1 & 99.0 & 15.0 & 0.04\ \\  
 150403A & 104.4 & 204.8 & 100.0 & 15.0 & 0.10 &&&190114A & 17.6 & 86.5 & 71.8 & 5.0 & 0.99\ \\  
 150413A & 23.2 & 253.7 & 80.8 & 6.0 & 0.97 &&&190114C & 352.5 & 656.2 & 100.0 & 5.5 & 0.03\ \\  
 150727A & 23.6 & 119.1 & 82.6 & 0.3 & -- &&&190324A & 131.6 & 68.9 & 100.0 & 9.5 & 0.11\ \\  
 150821A & 7.9 & 19.2 & 15.1 & 0.8 & -- &&&190719C & 39.0 & 185.4 & 95.7 & 9.5 & 0.01\ \\  
 151027B & 25.4 & 111.1 & 80.6 & 7.5 & 0.99 &&&190829A & 276.2 & 67.3 & 100.0 & 0.1 & --\ \\  
 160203A & 52.6 & 35.3 & 96.2 & 10.5 & 0.71 &&&191004B & 19.5 & 6.2 & 83.1 & 13.0 & 0.85\ \\  
 160327A & 43.8 & 61.4 & 95.7 & 15.0 & 0.14 &&&191011A & 27.2 & 39.5 & 88.4 & 4.0 & 0.48\ \\  
 160624A & 7.6 & -- & 20.5 & 0.5 & -- &&&200829A & 184.9 & 249.0 & 100.0 & 15.0 & 0.04\ \\  
 \end{tabular} 
\end{table*}

\end{appendices}

\end{document}